\documentclass[3p,times]{elsarticle}

\usepackage{xcolor}
\usepackage{amssymb,amsmath,mathtools,wrapfig,subcaption,booktabs,multirow,float,xcolor,pdfpages,bbm,colortbl,url}
\usepackage{physics,bm,tikz}
\usetikzlibrary{arrows.meta}
\usepackage{amsthm}
\usepackage{amsfonts}
\usepackage[normalem]{ulem}
\usepackage{lipsum}
\allowdisplaybreaks
\usepackage{comment} 

\usepackage{framed}
\definecolor{shadecolor}{gray}{0.90}

\usepackage[
pagebackref=false,
colorlinks=true,
linkcolor=blue,
urlcolor=blue,
filecolor=black,
citecolor=red,
pdfstartview=FitV,
pdftitle={},
pdfauthor={},
pdfsubject={},
pdfkeywords={},
pdfpagemode=None,
bookmarksopen=true
]{hyperref}

\definecolor{refkey}{rgb}{0.9451,0.2706,0.4941}
\definecolor{labelkey}{rgb}{0.9451,0.2706,0.4941}

\theoremstyle{plain}
\newtheorem{thm}{Theorem}

\makeatletter
\renewcommand{\p@enumii}{}
\makeatother

\newcommand{\ii}{\mathrm{i}}
\newcommand{\ee}{\mathrm{e}}

\newcommand{\ctext}[1]{\raise0.2ex\hbox{\textcircled{\scriptsize{#1}}}}

\newcommand{\RR}{\mathrm{R}}
\newcommand{\LL}{\mathrm{L}}

\newcommand{\up}{\uparrow}
\newcommand{\down}{\downarrow}

\newcommand{\Hm}{H_\mathrm{m}}

\newcommand{\hatPsi}{\hat{\Psi}}

\DeclareMathOperator{\sign}{sgn}

\usepackage[final]{showkeys}

\begin{document}
\begin{frontmatter}




\title{Introducing a Kondo-type interaction to the model of quantum walkers}


\author[label1,label2]{Manami Yamagishi\corref{cor1}}
\ead{manami@iis.u-tokyo.ac.jp}
\author[label3]{Naomichi Hatano\corref{cor2}}
\ead{hatano@iis.u-tokyo.ac.jp}
\author[label4]{Kohei Kawabata}
\ead{kawabata@issp.u-tokyo.ac.jp}
\author[label5]{Chusei Kiumi}
\ead{c.kiumi.qiqb@osaka-u.ac.jp}
\author[label6]{Akinori Nishino}
\ead{nishino@kanagawa-u.ac.jp}
\author[label2,label7]{Franco Nori}
\ead{fnori@riken.jp}
\author[label8,label3,label4,label9]{Hideaki Obuse}
\ead{hideaki.obuse@eng.hokudai.ac.jp}

\address[label1]{Department of Physics, The University of Tokyo,
5-1-5 Kashiwanoha, Kashiwa, Chiba 277-8574, Japan}
\address[label2]{RIKEN Center for Quantum Computing, RIKEN,
2-1 Hirosawa, Wako, Saitama 351-0198, Japan}
\address[label3]{Institute of Industrial Science, The University of Tokyo,
5-1-5 Kashiwanoha, Kashiwa, Chiba 277-8574, Japan}
\address[label4]{Institute for Solid State Physics, University of Tokyo,
Kashiwa, Chiba 277-8581, Japan}
\address[label5]{Center for Quantum Information and Quantum Biology, The University of Osaka,
1-2 Machikaneyamacho, Toyonaka, Osaka 560-0043, Japan}
\address[label6]{Department of Applied Physics, Kanagawa University,
3-27-1 Rokkakubashi, Kanagawa, Yokohama, Kanagawa 221-8686, Japan}
\address[label7]{Physics Department, The University of Michigan,
450 Church St., Ann Arbor, Michigan 48109-1040, U.S.A.}
\address[label8]{Department of Applied Physics, Hokkaido University,
Kita13 Nishi8, Kita, Sapporo, Hokkaido 060-8628, Japan}
\address[label9]{Institute for Frontier Education and Research on Semiconductors, Hokkaido University,
Kita8 Nishi5, Kita, Sapporo, Hokkaido 060-0808, Japan}

\cortext[cor1]{First corresponding author}
\cortext[cor2]{Second corresponding author}

\begin{abstract}
The discrete-time quantum walk is a quantum model useful for studying quantum information science and computing, as well as biophysics, statistical physics, and condensed-matter physics.
The quantum walk has also been experimentally realized in various systems, including laser systems, cold atom systems, photonic systems, and superconducting qubits, thereby making the model more valuable.
Despite its significance across various fields of research, the model has been studied primarily as a one-body system.
We here introduce a model of quantum walkers interacting with a magnetic impurity localized at the origin. 
Each quantum walker interacts with the spin degree of freedom of a magnetic impurity, through which multiple quantum walkers indirectly interact with each other, as in the Kondo model.
We first identify a discrete-time quantum walker as a massless Dirac particle propagating in continuous space via a series of Dirac's delta potentials.
Based on the identification, we add a spin-$1/2$ degree of freedom to Dirac's potential at the origin.
We present an algorithm for the unitary time evolution of quantum walkers, in which all scattering matrices for massless Dirac particles arising from the magnetic impurity are derived.
To demonstrate its utility, we then analytically study a bound state of one quantum walker interacting with the localized magnetic impurity and numerically simulate the dynamics of two quantum walkers interacting indirectly through the magnetic impurity, the latter of which is a nontrivial three-body problem.
First, for a simple set of parameter values, we analytically obtain the eigenvalues and eigenvectors of the bound states, in which a quantum walker is bound to the magnetic impurity.
Second, we study two quantum walkers indirectly interacting with each other via the magnetic impurity.
We numerically simulate the collision dynamics in one dimension when the spin-spin interaction at the origin is of the XX type and the SU(2) Heisenberg type.
In the case of the XX interaction, we calculate the entanglement negativity to quantify how much the two quantum walkers are entangled with each other, and find that the negativity increases drastically upon the collision of the two walkers.
In the case of the SU(2) Heisenberg interaction, we simulate the dynamics starting from the initial state in which one fermionic walker is in a bound eigenstate around the origin and the other fermionic walker is a delta function colliding with the first walker. 
We find that a bound eigenstate closest to the singlet state of the first walker and the magnetic impurity is least perturbed by the collision of the second walker.
We speculate that this finding may be related to Kondo screening-like behavior at the lowest level of the real-space renormalization-group procedure.
\end{abstract}


\end{frontmatter}

\tableofcontents

\section{Introduction}\label{sec1}

The quantum walk~\cite{Kempe03,Venegas-Andraca12,Kadian21} is a quantum analogue of a random walk.
Nonetheless, the quantum walk exhibits deterministic quantum dynamics without any stochasticity.
Instead of the stochastic fluctuations of a classical random walker, a quantum walker moves under the interference of quantum fluctuations at each site, which deterministically governs the walker's dynamics.
Quantum interference gives the quantum walker a probability density distinct from the Gaussian distribution of the classical random walk.
While many paths starting from the origin and ending around the origin add up to make a Gaussian peak for the classical random walk, the paths quantum-mechanically interfere with each other in the quantum walk and deplete the probability around the origin.
Instead, sharp peaks emerge at the edges of the probability distribution, because of less quantum interference of the paths.

The original idea of the quantum walk was first introduced in a book by Gudder~\cite{Gudder88}.
Later, Aharonov \textit{et al.}~\cite{Aharonov93}, in their paper entitled ``Quantum random walks'', studied quantum walks with coin operators, comparing properties to classical random walk.
Meyer~\cite{Meyer96} developed a systematic model of the quantum walk and revealed its correspondence to the Feynman path integral~\cite{Feynman65} for the Dirac equation.
Beginning with Farhi and Gutmann~\cite{Farhi98}, quantum walks have been well studied in quantum information science and computing~\cite{Ambainis12,Asaka21,Zhou21,Qiang24}.
To date, studies of quantum walks have become even more interdisciplinary and have extended to a variety of research fields, such as biophysics~\cite{Engel07, Dudhe22}, statistical physics~\cite{Mukai20,Meng24}, and condensed-matter physics~\cite{Oka05}, particularly topological physics~\cite{Kitagawa10,Obuse11,Kitagawa12,Asboth13,Tong25}.
Quantum walks have been experimentally realized in various systems such as in laser systems~\cite{Schreiber10}, cold atom systems~\cite{Mugel16}, photon systems~\cite{Xiao17,Su19}, and superconducting qubits~\cite{Yan19}.
The quantum walk is now an indispensable model across various scientific fields interdisciplinarily.

The term ``quantum walk" often refers to two types of time evolution, 
continuous-time quantum walks and discrete-time quantum walks.
In the present paper, we focus on the latter, in which the space and time are both discrete.
The standard quantum walk is defined on a one-dimensional lattice; see Ref.~\cite{Yamagishi23} for an extension to multidimensional lattices.
In the simplest case, each lattice point $x$ (with the lattice constant $\Delta_x$) has a two-dimensional internal degree of freedom spanned by two states often called the left-going state and the right-going state, which are here denoted by $\ket{x,\LL}$ and $\ket{x,\RR}$.
The state takes the form
$\ket{\Psi(t)}=\sum_x\qty(\ell_x\ket{x,\LL}+r_x\ket{x,\RR})$,
where $\ell_x$ and $r_x$ denote the wave function amplitudes at site $x$.

The unitary time evolution of the state is given by the product of the coin and shift operators.
The coin operator shuffles the internal degree of freedom at each site separately.
A simple version that we use here is
\begin{align}\label{eq1001}
\hat{C}\coloneqq\sum_{x} \dyad{x}\otimes C_{\varphi},
\end{align}
where
\begin{align}
\label{eq1001-1}
C_{\varphi} \coloneqq \ee^{-\ii\sigma^y \varphi}
=\begin{pmatrix}
\cos\varphi & -\sin\varphi \\
\sin\varphi & \cos\varphi
\end{pmatrix},
\end{align}
with $\sigma^\mu$ denoting the $\mu$th component of the Pauli matrices.
On the other hand, the shift operator moves each state depending on its direction:
\begin{equation}\label{eq1002}
\hat{S}\coloneqq\sum_x\qty(\dyad{x-\Delta_x,\LL}{x,\LL} +\dyad{x+\Delta_x,\RR}{x,\RR}).
\end{equation}
The initial state $\ket{\psi(0)}$ evolves in time with the repeated multiplication of the shift and coin operators as in
\begin{equation}\label{eq1003}
\ket{\psi(t)} = \qty(\hat{S}\hat{C})^t\ket{\psi(0)}.
\end{equation}
Our primary purpose of the present paper is to introduce a new type of interaction between two quantum walkers.

The original quantum walk, as introduced above, is a one-body model.
Research on multiparticle discrete-time quantum walks was initiated by Omar \textit{et al.}~\cite{Omar06}, who studied two non-interacting discrete-time quantum walks on a line, and has been put forward both theoretically~\cite{Pathak07,Rohde11,Goyal10,Venegas-Andraca09} and experimentally~\cite{Peruzzo10,Owens11,Sansoni12}, aiming to study entanglement generation for possible realization of exponential speed up in quantum computation.
The meeting problem of two non-interacting quantum walkers was first studied by  \v{S}tefa\v{n}\'{a}k \textit{et al.}~\cite{Stefanak06} and later by Rigovacca and Di Franco with percolation induced by missing links~\cite{Rigovacca16}.
Interplay between disorder-induced localization and entanglement is studied in Refs.~\cite{Rigovacca16,Crespi13}.

For most of the above works on non-interacting multiparticle quantum walks, operators for the total system are given by the direct product of operators for each walker.
\v{S}tefa\v{n}\'{a}k \textit{et al.}~\cite{Stefanak11} first introduced ``$\delta$-interaction" between two walkers.
They introduced a coin operator that cannot be expressed as the product of two coin operators for each walker when the two walkers are at the same site, and studied the probability of finding both on the same side of the system.
This kind of $\delta$-interaction was further studied through Fourier analysis in Ref.~\cite{Ampadu11} and from the perspective of localization~\cite{Malishava20}.
Berry and Wang~\cite{Berry11} studied similar interacting discrete-time quantum walkers but with a different interaction, which are the ``$\mathbbm{1}$ interaction" operating the identity operator instead of usual coin operator or the ``$\pi$-phase interaction" adding a phase $\pi$ to the usual coin operator when the two walkers are at the same site, and applied the model to graph isomorphism problems.
With the same on-site phase interaction, effects of statistics~\cite{Wang16}, interaction and its strength on entanglement dynamics~\cite{Gan15,Carson15,Verga18} and localization properties~\cite{Sun18,Toikka20}, and interplay between interaction and robustness of topological edge states~\cite{Verga18} were also investigated.
One notable work with the same interaction scheme was carried out by Ahlbrecht \textit{et al.}~\cite{Ahlbrecht12}, who 
found that the interaction can make two walkers bound to each other as one quasi-particle, as if they are forming a molecule.
They theoretically analyzed this molecular formation and proposed how to realize it experimentally using atoms in optical lattice potentials.
The molecular formation and Bloch oscillation predicted in Ref.~\cite{Ahlbrecht12} were later observed in experiments of two-dimensional quantum walk with photons~\cite{Schreiber12} and with ultracold atoms in optical lattices~\cite{Preiss15}. 
Mittal and Sowi\'{n}ski~\cite{Mittal25} generalized the local phase interactions so that the interaction framework that they introduced can recover several previous works~\cite{Ahlbrecht12,Wang16,Verga18,Costa19} in some types of limit.
There are studies on other types of interacting models~\cite{Bisio18,Bisio18-2,Costa19,Alonso-Lobo18,Badhani21,Asaka25}, but we do not go into their details here.

As yet another type of study on interacting models that is more relevant to the present paper, Lockhart and Paternostro~\cite{Lockhart16} studied a model of a single-particle discrete-time quantum walk interacting with qubits at each lattice site (or vertices of a graph) via a phase-shift interaction.
Verga~\cite{Verga19} studied a quantum walker in a graph of interacting spins, aiming to investigate relaxation properties.
A similar yet distinct interaction that may be closest to our work was studied by Sellapillay and Verga~\cite{Sellapillay21}.
In their model, a quantum walker on a one-dimensional lattice interacts with localized non-interacting spins sitting on each edge of the graph.

In the present paper, we introduce a Kondo-like interaction between quantum walkers. 
Each quantum walker interacts with a localized magnetic impurity at the origin; hence, they indirectly interact with each other via this impurity but do not interact directly.
The Kondo model~\cite{Kondo64} describes conduction electrons in a non-magnetic metal interacting with a localized magnetic impurity spin at the origin with the following Hamiltonian:
\begin{equation}\label{eq1004}
H_{\mathrm{K}}=\sum_{\bm{k},\sigma}\epsilon_{\bm{k}}c^\dag_{\bm{k}\sigma}c_{\bm{k}\sigma}+J\vb{s}_0^{\mathrm{c}}\cdot\vb{s}_0^{\mathrm{l}}
\end{equation}
with the one-electron energy $\epsilon_{\bm{k}}$ of the conduction electrons.
The first term describes the kinetic energy due to free hopping of electrons 
with the annihilation and creation operators, $c_{\bm{k}\sigma}$ and $c^\dag_{\bm{k}\sigma}$, with the wave number $\bm{k}$ and the electron spin $\sigma$. 
The second term describes the s-d exchange interaction, \textit{i.e.}, the spin-spin interaction between each conduction electron (s electron) superscripted by $\mathrm{c}$ and the localized magnetic impurity (d electron) superscripted by~$\mathrm{l}$.
The coupling constant $J$ is positive, and hence the coupling is antiferromagnetic.
The Kondo Hamiltonian can also be seen as an effective Hamiltonian of the Anderson model~\cite{Anderson61} in the limit of large Coulomb interaction~\cite{Hewson93,Coleman07}.
We introduce a similar interaction to the Kondo model between quantum walkers as we will see in Eq.~\eqref{eq2201} in Sec.~\ref{sec2}.
Note, however, that the Hamiltonian~\eqref{eq2201} for our quantum walk not only has the spin-spin interaction between quantum walkers and the magnetic impurity, but also has the spin-orbit coupling for each walker.
This fact leads to a different behavior of our quantum walk from that of the Kondo model, as we will see in Sec.~\ref{sec4}.

The present paper is composed of the following contents.
Section~\ref{sec2} describes the main point of the present paper. We first identify the discrete-time quantum walk as a massless Dirac particle scattering through a series of Dirac's delta potentials. Based on this identification, we introduce a model in which one quantum walker interacts with a localized magnetic impurity at the origin, and two quantum walkers interact through the impurity.
We derive all scattering matrices for the one and two quantum walkers interacting with the impurity.
We thereby present the algorithm for the unitary dynamics for one and two quantum walkers under the interaction with the impurity.
Introducing the interaction to the model of quantum walkers is the main point of the present paper.
In the following sections, we demonstrate the model's utility.
In Sec.~\ref{sec3}, using a transfer-matrix method, we analytically obtain the eigenvalues and eigenvectors of bound states, in which the quantum walker is bound to the magnetic impurity.
In Sec.~\ref{sec4}, we numerically simulate collision dynamics of the two quantum walkers interacting via the impurity.
We present examples with different types of interactions, namely, the XX case and the Heisenberg case, as well as different statistics, namely, fermions, bosons, and distinguishable particles.

\section{Quantum-walk model of impurity interaction}\label{sec2}

We present our main point in the present section.
We first identify the quantum walker on a discrete lattice as a massless Dirac particle scattering through a series of delta potentials in continuum space.
So far, the quantum walk has been identified as a massive Dirac particle based on the Trotter formula~\cite{Strauch06, Yamagishi23},
in which the limit $\Delta_t\to0$ is taken, and thereby the limit $\Delta_x\to0$ is also taken.
From our point of view given in Subsec.~\ref{sec2.1}, this is the limit of dense potentials. 
Based on this identification, we introduce the Kondo-type interaction by adding a spin-$1/2$ degree of freedom to one of the delta potentials.
We then derive the scattering matrices due to the magnetic impurity in the cases of one and two massless Dirac particles.
These scattering matrices of the massless Dirac particles translate into the coin operators of the quantum walkers.
We present the details of the computation of the scattering matrices in \ref{appA}, \ref{appB}, and \ref{appC}, 
and focus on the algorithm for the unitary-time dynamics of the quantum walkers in the present section.

\subsection{Identification of a quantum walk as propagation of a massless Dirac particle in continuum space}\label{sec2.1}

The identification of the quantum walker with a massless Dirac particle is schematically shown in Fig.~\ref{fig01}.
\begin{figure}
\centering
\begin{tikzpicture}
\draw[line width = 0.5, arrows = {-{Stealth[length=3mm]}}] (0,3) -- (11,3);
\draw[thick] (1,0)--(1,6);
\draw[thick] (4,0)--(4,6);
\draw[thick] (7,0)--(7,6);
\draw[thick] (10,0)--(10,6);
\draw[line width = 2, arrows = {-{Stealth[length=5mm]}}] (1.5,3.75) -- (3.5,3.75);
\draw (2.5,4)node[above]{$A\ee^{\ii kx}\mqty(0 \\ 1)$};
\draw[line width = 2, arrows = {-{Stealth[length=5mm]}}] (4.5,3.75) -- (6.5,3.75);
\draw (5.5,4)node[above]{$C\ee^{\ii kx}\mqty(0 \\ 1)$};
\draw[line width = 2, arrows = {-{Stealth[length=5mm]}}] (7.5,3.75) -- (9.5,3.75);
\draw (8.5,4)node[above]{$E\ee^{\ii kx}\mqty(0 \\ 1)$};
\draw[line width = 2, arrows = {-{Stealth[length=5mm]}}] (3.5,2.25) -- (1.5,2.25);
\draw (2.5,2)node[below]{$B\ee^{-\ii kx}\mqty(1 \\ 0)$};
\draw[line width = 2, arrows = {-{Stealth[length=5mm]}}] (6.5,2.25) -- (4.5,2.25);
\draw (5.5,2)node[below]{$D\ee^{-\ii kx}\mqty(1 \\ 0)$};
\draw[line width = 2, arrows = {-{Stealth[length=5mm]}}] (9.5,2.25) -- (7.5,2.25);
\draw (8.5,2)node[below]{$F\ee^{-\ii kx}\mqty(1 \\ 0)$};
\draw (0,4.5)node[font=\Huge]{$\cdots$};
\draw (0,1.5)node[font=\Huge]{$\cdots$};
\draw (11,4.5)node[font=\Huge]{$\cdots$};
\draw (11,1.5)node[font=\Huge]{$\cdots$};
\draw (1,0)node[below]{$-\Delta_x$};
\draw (4,0)node[below]{$0$};
\draw (7,0)node[below]{$+\Delta_x$};
\draw (10,0)node[below]{$+2\Delta_x$};
\draw (11,3)node[right]{$x$};
\end{tikzpicture}
\caption{Scattering of a Dirac particle in a series of delta potentials.}
\label{fig01}
\end{figure}
The total one-body Hamiltonian is given by 
\begin{equation}\label{eq2114}
H_{\mathrm{QW}}=\epsilon p_x\sigma^z+\sum_nm\sigma^y\delta(x-n\Delta_x),
\end{equation}
where the first term represents the kinetic energy with $\epsilon$ being a positive constant for the velocity of the massless Dirac particle, while the second term represents a series of Dirac's delta potentials at $x=n\Delta_x$ with $m$ being the strength of the potentials.
We use the branch of the negative energy $-\epsilon k$ for the massless Dirac particle because we use the convention that the upper component $\begin{pmatrix} 1 & 0 \end{pmatrix}^T$ is the left-going state and the lower component $\begin{pmatrix} 0 & 1 \end{pmatrix}^T$ is the right-going one; see \ref{appA} for details.

The Dirac wave function is scattered at each delta potential in the second term of the Hamiltonian~\eqref{eq2114}.
At the origin $x=0$ in Fig.~\ref{fig01}, for example, 
the right-going state of the amplitude $A$ and the left-going state of the amplitude $D$
are scattered to the left-going state of the amplitude $B$ and the right-going state of the amplitude $C$.
We identify the scattering matrix due to the delta potential [Eq.~\eqref{eq2111} in \ref{appA}] as the coin operator~\eqref{eq1001-1} with the parametrization
\begin{subequations}\label{eq2113}
\begin{align}
\cos\varphi&=\frac{\epsilon^2-\frac{m^2}{4}}{\epsilon^2+\frac{m^2}{4}},\label{eq2113a}
\\
\sin\varphi&=-\frac{\epsilon m}{\epsilon^2+\frac{m^2}{4}}.\label{eq2113b}
\end{align}
\end{subequations}
Each delta potential at $x=n\Delta_x$ in the Hamiltonian~\eqref{eq2114} thus produces the coin operator at the corresponding lattice point $n$.

We also identify the shift operation~\eqref{eq1002} as the propagation of a massless Dirac particle from one delta potential to the neighboring one over the distance $\Delta_x$.
Since the velocity $\epsilon$ is constant, it takes the wave in every segment of Fig.~\ref{fig01} a constant time $\Delta_t:=\Delta_x/\epsilon$ to propagate after the scattering at one potential until the scattering at the neighboring one.
This also justifies the discretization of the time steps.

%

\subsection{Hamiltonian and the scattering operator for one quantum walker with a magnetic impurity}\label{sec2.2}

The above identification motivates us to introduce the Kondo-type interaction between two quantum walkers.
We add a localized magnetic impurity to the delta potential at the origin.
Each quantum walker interacts directly with the magnetic impurity, which allows quantum walkers to interact indirectly with each other via the magnetic impurity, just as in the Kondo model~\eqref{eq1004}.
In the present and next subsections~\ref{sec2.2} and~\ref{sec2.3}, we begin with the analysis of the case of one quantum walker interacting with a magnetic impurity, which is a two-body problem.
We derive the scattering matrix for one quantum walker with the impurity and present the algorithm for the unitary dynamics.
In subsection~\ref{sec4.1}, we explore the case of two quantum walkers interacting indirectly with each other via the magnetic impurity, which is a nontrivial three-body problem.
We also derive the scattering matrix for two quantum walkers with the impurity and present the algorithm for the unitary dynamics.

As we show in the previous subsection~\ref{sec2.1}, the scattering matrix due to Dirac's delta potential is a $2\times2$ matrix, which corresponds to the coin operator of the quantum walker at each site.
The scattering matrix due to the spin-$1/2$ impurity of one quantum walker is a $4\times4$ matrix.
The matrix for two quantum walkers is an $8\times8$ matrix. 
This is schematically illustrated in Fig.~\ref{fig02}.
\begin{figure}
\centering
\begin{subfigure}[c]{0.33\textwidth}
\centering
\begin{tikzpicture}[baseline={(0,0)}]
\draw[line width = 0.5, arrows = {-{Stealth[length=3mm]}}] (0,3) -- (4,3);
\draw[thick] (2,1)--(2,5);
\draw[line width = 2, arrows = {-{Stealth[length=5mm]}}] (0.3,3.75) -- (1.7,3.75);
\draw[line width = 2, arrows = {-{Stealth[length=5mm]}}] (3.7,3.75) -- (2.3,3.75);
\draw[line width = 2, arrows = {-{Stealth[length=5mm]}}] (1.7,2.25) -- (0.3,2.25);
\draw[line width = 2, arrows = {-{Stealth[length=5mm]}}] (2.3,2.25) -- (3.7,2.25);
\draw (2,0.5)node{$m_0\sigma^y\delta(x)$};
\draw (4,2.5)node[right]{$x$};
\end{tikzpicture}
\vspace{0.5\baselineskip}
\caption{}
\end{subfigure}
\hfill
\begin{subfigure}[c]{0.33\textwidth}
\centering
\begin{tikzpicture}[baseline={(0,0)}]
\draw[line width = 0.5, arrows = {-{Stealth[length=3mm]}}] (0,3) -- (4,3);
\draw[thick] (2,1)--(2,5);
\draw[line width = 2, arrows = {-{Stealth[length=5mm]}}] (0.3,3.75) -- (1.7,3.75);
\draw[line width = 2, arrows = {-{Stealth[length=5mm]}}] (3.7,3.75) -- (2.3,3.75);
\draw[line width = 2, arrows = {-{Stealth[length=5mm]}}] (1.7,2.25) -- (0.3,2.25);
\draw[line width = 2, arrows = {-{Stealth[length=5mm]}}] (2.3,2.25) -- (3.7,2.25);
\draw (2,0.5)node{$m_0\sigma^y\delta(x)+H_\mathrm{m}\delta(x)$};
\draw[line width = 3, arrows = {-{Stealth[length=3mm]}}] (1.9,2.5) -- (1.9,3.5);
\draw[line width = 3, arrows = {-{Stealth[length=3mm]}}] (2.1,3.5) -- (2.1,2.5);
\draw (4,3)node[right]{$x$};
\end{tikzpicture}
\vspace{0.5\baselineskip}
\caption{}
\end{subfigure}
\hfill
\begin{subfigure}[c]{0.33\textwidth}
\centering
\begin{tikzpicture}[baseline={(0,0)}]
\draw[line width = 0.5, arrows = {-{Stealth[length=3mm]}}] (0,3) -- (4,3);
\draw[thick] (2,1)--(2,5);
\draw[line width = 2, arrows = {-{Stealth[length=5mm]}}] (0.3,3.7) -- (1.7,3.7);
\draw[line width = 2, arrows = {-{Stealth[length=5mm]}}] (0.3,4.4) -- (1.7,4.4);
\draw[line width = 2, arrows = {-{Stealth[length=5mm]}}] (3.7,3.7) -- (2.3,3.7);
\draw[line width = 2, arrows = {-{Stealth[length=5mm]}}] (3.7,4.4) -- (2.3,4.4);
\draw[line width = 2, arrows = {-{Stealth[length=5mm]}}] (1.7,2.3) -- (0.3,2.3);
\draw[line width = 2, arrows = {-{Stealth[length=5mm]}}] (1.7,1.6) -- (0.3,1.6);
\draw[line width = 2, arrows = {-{Stealth[length=5mm]}}] (2.3,2.3) -- (3.7,2.3);
\draw[line width = 2, arrows = {-{Stealth[length=5mm]}}] (2.3,1.6) -- (3.7,1.6);
\draw (2,0.5)node{$m_0\sigma^y\delta(x_1)+H_\mathrm{m1}\delta(x_1)$};
\draw (2,0.0)node{$+m_0\sigma^y\delta(x_2)+H_\mathrm{m2}\delta(x_2)$};
\draw[line width = 3, arrows = {-{Stealth[length=3mm]}}] (1.9,2.5) -- (1.9,3.5);
\draw[line width = 3, arrows = {-{Stealth[length=3mm]}}] (2.1,3.5) -- (2.1,2.5);
\draw (4,3)node[right]{$x$};
\end{tikzpicture}
\vspace{0.5\baselineskip}
\caption{}
\end{subfigure}
\caption{Schematic illustrations of (a) a $2\times2$ scattering matrix of a Dirac particle due to a delta potential without the spin-$1/2$ impurity, (b) a $4\times4$ scattering matrix of a Dirac particle due to the impurity, and (c) an $8\times8$ scattering matrix of two Dirac particles due to the impurity. The horizontal arrows represent the incoming and outgoing waves of the scattering event, while the vertical arrow at the origin represents the two states of the magnetic impurity.}
\label{fig02}
\end{figure}
The scattering matrix in Fig.~\ref{fig02}(b) gives a $4\times4$ coin operator at the origin for one quantum walker with the magnetic impurity.
The one in Fig.~\ref{fig02}(c) gives an $8\times8$ coin operator for two quantum walkers with the magnetic impurity.

We start from the following Hamiltonian of one massless Dirac particle interacting with a magnetic impurity:
\begin{equation}\label{eq2201}
H_\mathrm{K}^{\mathrm{1w}}\coloneqq%
\epsilon p (\sigma^z \otimes s^0)+ \sum_{n}m(\sigma^y\otimes s^0)\delta(x-n)+\Hm\delta(x),
\end{equation}
where the first and second terms are the same as the Hamiltonian~\eqref{eq2114} with $s^0$ being the identity operator for the degree of freedom of the impurity spin introduced below. We set $\Delta_x$ to unity for brevity.
The third, new term represents a magnetic interaction between the Dirac particle and the impurity spin at the origin,
\begin{equation}\label{eq2202}
\Hm\coloneqq J_x\sigma^x \otimes s^x+J_y\sigma^y \otimes s^y+J_z\sigma^z \otimes s^z,
\end{equation}
where $\{J_x,J_y,J_z\}$ are real parameters.
The new degree of freedom in Eq.~\eqref{eq2202} represented by a set of Pauli matrices $\{s^x,s^y,s^z\}$ is a magnetic impurity localized at $x=0$. 

In Sec.~\ref{sec2.1}, we showed that the kinetic term of the massless Dirac particle produces the standard shift operation of a quantum walker, 
\begin{equation}\label{eq2203}
\hat{S}^{\mathrm{1w}}\coloneqq\sum_{x,S_0}\qty(\dyad{x-1,\LL}{x,\LL} +\dyad{x+1,\RR}{x,\RR})\otimes \dyad{S_0},
\end{equation}
where $S_0$ denotes the states $\up$ and $\down$ of the magnetic impurity at the origin.
In Sec.~\ref{sec2.1}, we also showed that the scattering of the Dirac particle due to the potential at each lattice point $x$ produces the standard coin operator of a quantum walker,
\begin{equation}\label{eq2204}
\hat{C}^{\mathrm{1w}}\coloneqq\sum_{x,S_0} \dyad{x}\otimes C_{\varphi}\otimes\dyad{S_0},
\end{equation}
where $C_{\varphi}$ is given as $C_{\varphi}\coloneqq\ee^{-\ii\sigma^y \varphi}$ [Eq.~\eqref{eq1001-1}] with a real parameter $\varphi$ for $S_0=\up,\down$,
as in Eq.~\eqref{eq2113}. 
This coin operator shuffles the internal states $\LL$ and $\RR$ of the quantum walker at each lattice site and for each $S_0=\up,\down$.

Meanwhile, in \ref{appB}, we show that the scattering of the massless Dirac particle due to the magnetic impurity at the origin $x=0$ is given by 
\begin{equation}\label{eq2205}
\begin{pmatrix}
\psi_{\LL\up}(0-)\\
\psi_{\RR\up}(0+)\\
\psi_{\LL\down}(0-)\\
\psi_{\RR\down}(0+)
\end{pmatrix}
=
S_{\mathrm{imp}}^{\mathrm{1w}}
\begin{pmatrix}
\psi_{\LL\up}(0+)\\
\psi_{\RR\up}(0-)\\
\psi_{\LL\down}(0+)\\
\psi_{\RR\down}(0-)
\end{pmatrix},
\end{equation}
where the scattering matrix is given as
\begin{equation}\label{eq2206}
S_{\mathrm{imp}}^{\mathrm{1w}}\coloneqq
\begin{pmatrix}
\alpha^{\mathrm{1w}} & 0 & 0 & \beta^{\mathrm{1w}} \\
 0 & \gamma^{\mathrm{1w}} & \delta^{\mathrm{1w}} & 0 \\
 0 & \delta^{\mathrm{1w}} & \gamma^{\mathrm{1w}} & 0 \\
\beta^{\mathrm{1w}} & 0 & 0 & \alpha^{\mathrm{1w}} \\
\end{pmatrix},
\end{equation}
with
\begin{subequations}\label{eq2207}
\begin{align}\label{eq2207-a}
\alpha^{\mathrm{1w}}
&\coloneqq-\frac{4\epsilon^2-\qty(J_x-J_y)^2+{J_z}^2}%
{(2\ii\epsilon+J_z)^2-\qty(J_x-J_y)^2}, \\
\beta^{\mathrm{1w}}
&\coloneqq-\frac{4\ii\epsilon\qty(J_x-J_y)}%
{(2\ii\epsilon+J_z)^2-\qty(J_x-J_y)^2} , \\
\gamma^{\mathrm{1w}}&\coloneqq-\frac{4\epsilon^2-\qty(J_x+J_y)^2+{J_z}^2}%
{(2\ii\epsilon-J_z)^2-\qty(J_x+J_y)^2}, \\
\label{eq2207-d}
\delta^{\mathrm{1w}}&\coloneqq-\frac{4\ii\epsilon\qty(J_x+J_y)}%
{(2\ii\epsilon-J_z)^2-\qty(J_x+J_y)^2}.
\end{align}
\end{subequations}
This gives the special coin operator for one quantum walker and the magnetic impurity at the origin in the following form:
\begin{equation}\label{eq2302}
\hat{C}_0^{\mathrm{1w}}\coloneqq%
\sum_{x(\ne0)}\dyad{x}{x}\otimes\mathbb{I}_{4\times4}+\dyad{x=0}{x=0}\otimes S_{\mathrm{imp}}^{\mathrm{1w}}.
\end{equation}

\subsection{Time evolution with chiral symmetry}\label{sec2.3}

The wave function of the system of one quantum walker and one magnetic impurity is given in the form of $\Psi(t;x,\sigma;S_0)$,
where $x$ denotes the lattice point for the walker, $\sigma$ denotes the states $\LL$ and $\RR$ of the walker, and $S_0$ denotes the states $\up$ and $\down$ of the magnetic impurity at the origin.
We carry out the time-evolution update of the state with the standard coin operator $\hat{C}^{\mathrm{1w}}$ in Eq.~\eqref{eq2204},
the special coin operator $\hat{C}_0^{\mathrm{1w}}$ in Eq.~\eqref{eq2302},
and the shift operator $\hat{S}^{\mathrm{1w}}$ in Eq.~\eqref{eq2203}.

To make the time-evolution operator 
take a symmetric form,
we follow the approach introduced in Ref.~\cite{Asboth13} and arrange the operators so that the unitary time-evolution operator $\hat{U}_{\mathrm{K}}^{\mathrm{1w}}$ of one time step $\Delta_t$ is expressed as
\begin{equation}\label{eq2301}
\hat{U}_{\mathrm{K}}^{\mathrm{1w}}\coloneqq
\sqrt{\hat{C}_0^{\mathrm{1w}}}\sqrt{\hat{C}^{\mathrm{1w}}}\hat{S}^{\mathrm{1w}}%
\sqrt{\hat{C}^{\mathrm{1w}}}\sqrt{\hat{C}_0^{\mathrm{1w}}}.
\end{equation}
We define the time-evolution operator as above so that it can satisfy chiral symmetry.

More specifically, chiral symmetry of a Hamiltonian $H$ is defined by
\begin{equation}
    \Gamma H\Gamma^{-1}=-H,
        \label{eq2306a}
\end{equation}
with a unitary operator $\Gamma$,
which also leads to pseudo-Hermiticity of a time-evolution operator $U$:
\begin{equation}
    \Gamma U \Gamma^{-1} =U^{\dag}.
        \label{eq2306b}
\end{equation}
When one of the coupling constants $\{J_x,J_y,J_z\}$ is equal to zero, both the total Hamiltonian $H_{\mathrm{K}}^{\mathrm{1w}}$ in Eq.~\eqref{eq2201} and the time-evolution operator in Eq.~\eqref{eq2301} respect chiral symmetry. 
In the following, we will focus on the following two cases under the following names:
\begin{subequations}
\begin{align}
\label{eq-xx}
\mbox{XX case: } &\quad J_x=J_y=J,\quad J_z=0,  \\
\label{eq-su2}
\mbox{Heisenberg case: } & \quad J_x=J_y=J_z=J.
\end{align}
\end{subequations}
Note that the former case is chiral symmetric, while the latter SU(2) case is not.

The square roots of the coin operators in Eq.~\eqref{eq2301} are explicitly defined as follows. 
The square root of $\sqrt{\hat{C}^{\mathrm{1w}}}$ contains $\sqrt{C_{\varphi}}$, which takes exactly the same form as $C_{\varphi}$ in Eq.~\eqref{eq1001-1} but with a half of the argument, \textit{i.e.},
\begin{equation}\label{eq2305}
\sqrt{C_{\varphi}}=\ee^{-\ii\sigma^y (\varphi/2)}.
\end{equation}
The operator $\sqrt{\hat{C}_0^{\mathrm{1w}}}$, on the other hand, contains $\sqrt{S_{\mathrm{imp}}^{\mathrm{1w}}}$, which is given by
\begin{equation}\label{eq2303}
\sqrt{S_{\mathrm{imp}}^{\mathrm{1w}}}=\frac{1}{2}
\begin{pmatrix}
\tilde{\alpha}^{\mathrm{1w}} & & & \tilde{\beta}^{\mathrm{1w}} \\
 & \tilde{\gamma}^{\mathrm{1w}} & \tilde{\delta}^{\mathrm{1w}} & \\
 & \tilde{\delta}^{\mathrm{1w}} & \tilde{\gamma}^{\mathrm{1w}} & \\
\tilde{\beta}^{\mathrm{1w}} & & & \tilde{\alpha}^{\mathrm{1w}}
\end{pmatrix},
\end{equation}
where
\begin{align}\label{eq2304}
\begin{matrix}
\tilde{\alpha}^{\mathrm{1w}}%
\coloneqq\sqrt{\alpha^{\mathrm{1w}}+\beta^{\mathrm{1w}}}+\sqrt{\alpha^{\mathrm{1w}}-\beta^{\mathrm{1w}}}, &
\tilde{\beta}^{\mathrm{1w}}%
\coloneqq\sqrt{\alpha^{\mathrm{1w}}+\beta^{\mathrm{1w}}}-\sqrt{\alpha^{\mathrm{1w}}-\beta^{\mathrm{1w}}}, \\
\tilde{\gamma}^{\mathrm{1w}}%
\coloneqq\sqrt{\gamma^{\mathrm{1w}}+\delta^{\mathrm{1w}}}+\sqrt{\gamma^{\mathrm{1w}}-\delta^{\mathrm{1w}}}, &
\tilde{\delta}^{\mathrm{1w}}%
\coloneqq\sqrt{\gamma^{\mathrm{1w}}+\delta^{\mathrm{1w}}}-\sqrt{\gamma^{\mathrm{1w}}-\delta^{\mathrm{1w}}}.
\end{matrix}
\end{align}
Indeed, we confirm that the square of Eq.~\eqref{eq2303} yields Eq.~\eqref{eq2206}.
We can thus explicitly write down the time evolution operator $\hat{U}_\mathrm{K}^{\mathrm{1w}}$ defined in Eq.~\eqref{eq2301} as a matrix of the size $2L_x\times 2L_x$, where $L_x$ denotes the lattice size.

\begin{figure}[h]
\centering
\includegraphics[width=0.5\textwidth]{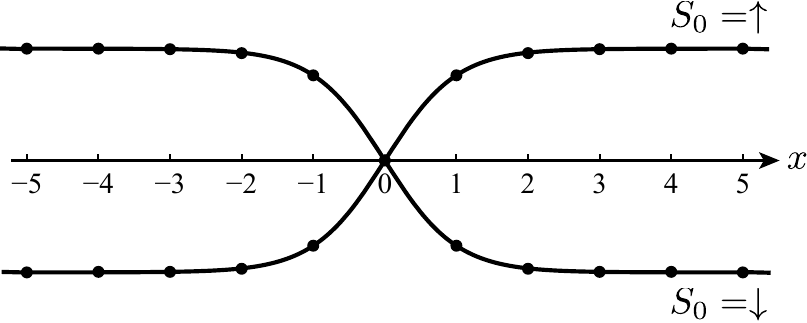}
\caption{
Visualization of the entire Hilbert space of the quantum walker, which consists
of the sectors of $S_0 =\uparrow$ and $S_0=\downarrow$. Each sector of $S_0$ constitutes a one-dimensional lattice with $x = -L_x/2, -L_x/2+1, \cdots, 0, \cdots, L_x/2-1, L_x/2$. For the explanation of the internal states of each lattice point, see the main text.
}
\label{fig03}
\end{figure}

We carry out the following algorithm for the time evolution of the walker with the impurity spin. Refer to Fig.~\ref{fig03} for the explanation.
\begin{itemize}
\item 
Each point $(x,S_0)$ with $x\neq0$ and $S_0=\uparrow,\downarrow$ has the internal degrees of freedom of $\sigma=\LL,\RR$. The state $\sigma$ at each site is separately updated by the standard coin operator $C_{\varphi}$ in Eq.~\eqref{eq1001-1}, and also independently of the state of $S_0$.

\item 
The origin $x=0$ has the internal degrees of freedom of $\sigma=\LL,\RR$ and $S_0=\uparrow,\downarrow$. The combined state $\sigma\otimes S_0$ is updated by the new coin operator~\eqref{eq2206}. In addition, the state $\sigma$ of each site is separately updated by the standard coin operator $C_{\varphi}$ in Eq.~\eqref{eq1001-1}, and also independently of the state of $S_0$.

\item 
The shift operator~\eqref{eq2203} moves the quantum walker according to the internal state, independently of the state of $S_0$.

\end{itemize}

\begin{figure}
\centering
\includegraphics[width=\textwidth]{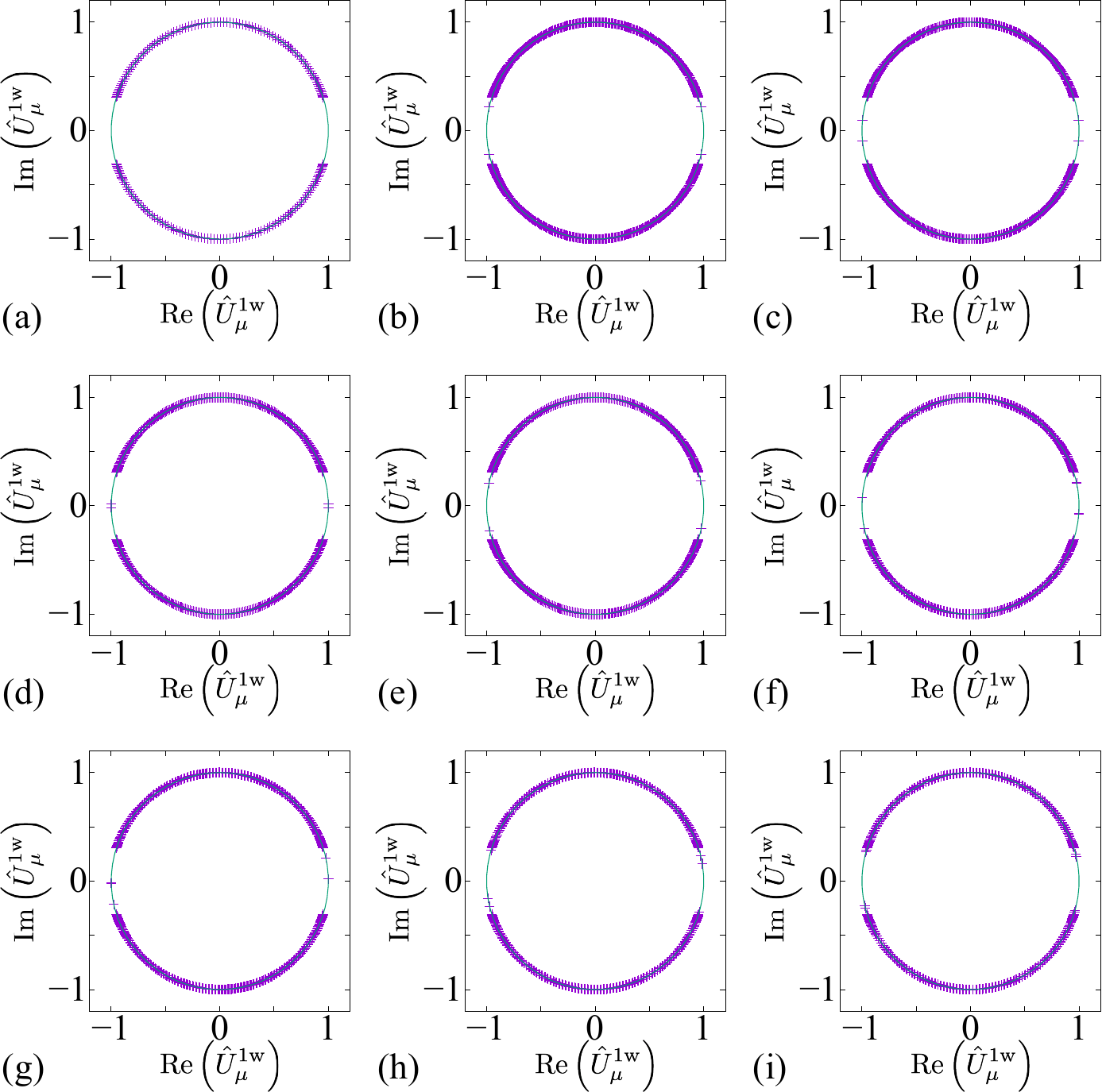}
\vspace{-\baselineskip}
\caption{Eigenvalues of the time-evolution operators on the unit circle (green circle) in the XX case [(a)--(d)] and in the Heisenberg case [(e)--(i)]. For (a)--(d), we set the XX interaction $J_x=J_y=J, J_z=0$ with (a) $J=0$, (b) $J=1$, (c) $J=3$, and (d) $J=20$. For (e)--(i), we set the Heisenberg interaction $J_x=J_y=J_z=J$ with (e) $J=1$, (f) $J=2$, (g) $J=3$, (h) $J=7$, and (i) $J=20$. We set $\varphi=\pi/10$ and the system is of size $L_x=201$ under the periodic boundary conditions.}
\label{fig04}
\end{figure}

Figure~\ref{fig04} shows the eigenvalues of the time-evolution operator $\hat{U}_\mathrm{K}^{\mathrm{1w}}$ defined in Eq.~\eqref{eq2301} on the complex plane, which we obtained by numerical diagonalization with $\varphi=\pi/10$ and the system of size $L_x=201\,(-100\le x\le 100)$ under the periodic boundary conditions.
%
Since the time-evolution operator is unitary, all eigenvalues $\{U_\mu^{\mathrm{1w}}\}$ are on the unit circle, as in $U_\mu^{\mathrm{1w}}=\ee^{\ii\lambda_\mu}$ with all psuedo-eigenvalues $\lambda_\mu$ being real.

We first observe two bulks of eigenvalues, whose eigenvectors we find are extended in the positional space.
We presume that they constitute continuous spectra in the infinite-size limit $L_z\to\infty$.
We also observe several isolated eigenvalues, whose eigenvectors we find are exponentially localized around the magnetic impurity.
We hereafter refer to them as bound states, whose eigenvalues and eigenvectors
we analytically obtain in Sec.~\ref{sec3}.

In the XX case in Eq.~\eqref{eq-xx} with chiral symmetry [Fig.~\ref{fig04}(a)--(d)], we find four isolated eigenvalues, which depart from the edges of the two bulks and approach the points $U_\mu^{\mathrm{1w}}=\pm1$ as we increase $J$.
The isolated eigenvalues become almost unity up to sign with $J=20$, as shown in Fig.~\ref{fig04}(d).
All the eigenvalues are distributed symmetrically with respect to the real axis in the XX case with chiral symmetry. 
This is because chiral symmetry of the Hamiltonian in Eq.~\eqref{eq2306a} leads to pseudo-Hermiticity of the time-evolution operator in Eq.~\eqref{eq2306b},
which implies that the eigenvalues of the time-evolution operator are either real or complex-conjugate pairs.

Introducing a finite value of $J_z$ generally breaks chiral symmetry, tilting the locations of the isolated eigenvalues. In Fig.~\ref{fig04}(e)--(i), we set the SU(2) Heisenberg interaction $J_x=J_y=J_z=J$, as in Eq.~\eqref{eq-su2}.
For the parameter range of $J\lesssim3$, as in Fig.~\ref{fig04}(e)--(g), there exist four isolated eigenvalues with corresponding localized eigenvectors around the magnetic impurity for each of them.
As we increase the coupling strength $J$, as in Fig.~\ref{fig04}(h)--(i), we obtain more isolated eigenvalues; for the parameter range $3\lesssim J\lesssim7$, there exist six of them, and for the parameter range $7\lesssim J$, there exist eight of them.
As we further increase the coupling strength $J$, the isolated eigenvalues get closer to the edge of the bulks as shown in Fig.~\ref{fig04}(i), but they are never absorbed to the bulks and remain isolated.

For each of the isolated states bound to the magnetic impurity, the expectation value of the angular momentum squared ${\vb{J}_{10}}^2=(\vb{s}_1+\vb{s}_0)^2$ is nonzero. 
In other words, none of them is a singlet state.
Here, $\vb{s}_1\coloneqq 
1/2\times(\sigma^x\otimes s^0, \sigma^y\otimes s^0, \sigma^z\otimes s^0)^\top$ and $\vb{s}_0\coloneqq
1/2\times(\sigma^0\otimes s^x, \sigma^0\otimes s^y, \sigma^0\otimes s^z)^\top$.
Indeed, the singlet and triplet states cannot be eigenstates of the Hamiltonian~\eqref{eq2201}, because its kinetic term has a spin-orbit coupling $p\sigma^z$.
This fact affects our numerical results in Sec.~\ref{sec4}.
See Sec.~\ref{sec3} for the explicit forms of bound states. 

\subsection{Hamiltonian and the scattering operator for two quantum walkers with a magnetic impurity}\label{sec4.1}

Let us now consider the non-trivial two-walker problem, in which two quantum walkers interact with each other through a magnetic impurity in the form of the Kondo problem.
In other words, this is a three-body problem.

We start from the Hamiltonian of two massless Dirac particles
\begin{align}\label{eq4101}
&H_\mathrm{K}^{\mathrm{2w}}\coloneqq%
\ \epsilon p_1 (\sigma_1^z\otimes \sigma_2^0 \otimes s^0)+\epsilon p_2  (\sigma_1^0\otimes \sigma_2^z \otimes s^0) \notag \\
&\qquad+\sum_{n}m (\sigma_1^y\otimes \sigma_2^0\otimes s^0)\delta(x_1-n)+%
\sum_{n}m (\sigma_1^0\otimes \sigma_2^y\otimes s^0)\delta(x_2-n) \notag \\
&\qquad+{\Hm}_1\delta(x_1)+{\Hm}_2\delta(x_2),
\end{align}
where the first and second terms represent the kinetic energy of each Dirac particle, the third and fourth terms represent series of scattering delta potentials at lattice points $n$, and the fifth and sixth terms represent an interaction of each walker with the impurity spin at the origin:
\begin{subequations}
\begin{align}
{\Hm}_1&\coloneqq J_x\sigma_1^x \otimes \sigma_2^0\otimes s^x+J_y\sigma_1^y \otimes \sigma_2^0\otimes s^y+J_z\sigma_1^z \otimes \sigma_2^0\otimes s^z,\label{eq4102} \\
{\Hm}_2&\coloneqq J_x\sigma_1^0 \otimes \sigma_2^x \otimes s^x+J_y\sigma_1^0\otimes \sigma_2^y \otimes s^y+J_z\sigma_1^0\otimes \sigma_2^z \otimes s^z.\label{eq4103}
\end{align}
\end{subequations}
Here, each set of Pauli matrices $\{\sigma_1^x,\sigma_1^y,\sigma_1^z\}$ and $\{\sigma_2^x,\sigma_2^y,\sigma_2^z\}$ is defined on the Hilbert spaces of the corresponding quantum walker, and $\sigma_1^0$ and $\sigma_2^0$ are the identity operators in the respective space.

The kinetic term of each massless Dirac particle yields the standard shift operation of the respective quantum walkers, 
\begin{subequations}\label{eq4108-0}
\begin{align}
&\hat{S}_1^{\mathrm{2w}}\coloneqq%
\sum_{x_1,x_2,\sigma_2,S_0}\qty(\dyad{x_1-1,\sigma_1=\LL}{x_1,\sigma_1=\LL}+\dyad{x_1+1,\sigma_1=\RR}{x_1,\sigma_1=\RR})%
\otimes\dyad{x_2,\sigma_2}\otimes\dyad{S_0}, \label{eq4108}\\
&\hat{S}_2^{\mathrm{2w}}\coloneqq%
\sum_{x_1,\sigma_1,x_2,S_0}\dyad{x_1,\sigma_1}\otimes%
\qty(\dyad{x_2-1,\sigma_2=\LL}{x_2,\sigma_2=\LL}+\dyad{x_2+1,\sigma_2=\RR}{x_2,\sigma_2=\RR})%
\otimes\dyad{S_0}, \label{eq4109}
\end{align}
\end{subequations}
The scattering of each Dirac particle due to the delta potential at each lattice point $x$ produces the standard coin operator of the respective quantum walker,
\begin{subequations}
\begin{align}
&\hat{C}_1^{\mathrm{2w}}\coloneqq%
\sum_{x_1,x_2,\sigma_2,S_0}\dyad{x_1}\otimes C_{x_1,S_0}\otimes\dyad{x_2,\sigma_2}\otimes\dyad{S_0}, \label{eq4110}\\
&\hat{C}_2^{\mathrm{2w}}\coloneqq%
\sum_{x_1,\sigma_1,x_2,S_0}\dyad{x_1,\sigma_1}\otimes\dyad{x_2}\otimes C_{x_2,S_0}\otimes\dyad{S_0}, \label{eq4111}
\end{align}
\end{subequations}
These coin operators shuffle the internal states $\LL$ and $\RR$ of the quantum walkers at each lattice site and for each $S_0=\up,\down$.

Finally, we derive the scattering matrix of two Dirac particles with the magnetic impurity.
Analytic calculation in \ref{appC} produces the following $8\times8$ scattering matrix:
\begin{equation}
S_{\mathrm{imp}}^{\mathrm{2w}}
=\qty(\begin{array}{cccc|cccc}
\alpha_+^{\mathrm{2w}} & & & \beta^{\mathrm{2w}} & & \epsilon_-^{\mathrm{2w}} & \epsilon_-^{\mathrm{2w}} & \\
& \gamma^{\mathrm{2w}} & \delta^{\mathrm{2w}} & & \epsilon_+^{\mathrm{2w}} & & & \epsilon_-^{\mathrm{2w}} \\
& \delta^{\mathrm{2w}} & \gamma^{\mathrm{2w}} & & \epsilon_+^{\mathrm{2w}} & & & \epsilon_-^{\mathrm{2w}} \\
\beta^{\mathrm{2w}} & & & \alpha_-^{\mathrm{2w}} & & \epsilon_+^{\mathrm{2w}} & \epsilon_+^{\mathrm{2w}} & \\ \hline
& \epsilon_+^{\mathrm{2w}} & \epsilon_+^{\mathrm{2w}} & & \alpha_-^{\mathrm{2w}} & & & \beta^{\mathrm{2w}} \\
\epsilon_-^{\mathrm{2w}} & & & \epsilon_+^{\mathrm{2w}} & & \gamma^{\mathrm{2w}} & \delta^{\mathrm{2w}} & \\
\epsilon_-^{\mathrm{2w}} & & & \epsilon_+^{\mathrm{2w}} & & \delta^{\mathrm{2w}} & \gamma^{\mathrm{2w}} & \\
& \epsilon_-^{\mathrm{2w}} & \epsilon_- ^{\mathrm{2w}}& & \beta^{\mathrm{2w}} & & & \alpha_+^{\mathrm{2w}}
\end{array})\label{eq4105}
\end{equation}
with
\begin{subequations}\label{eq4106}
\begin{align}
&\alpha_\pm^{\mathrm{2w}}=\frac{2\ii J_xJ_yJ_z-{J_z}^2\epsilon+\epsilon^3\pm(2\ii J_z\epsilon^2+2J_xJ_y\epsilon)}%
                             {-2\ii J_xJ_yJ_z+({J_x}^2+{J_y}^2+{J_z}^2)\epsilon+\epsilon^3}, \\
&\beta^{\mathrm{2w}}=\frac{(-{J_x}^2+{J_y}^2)\epsilon}{-2\ii J_xJ_yJ_z+({J_x}^2+{J_y}^2+{J_z}^2)\epsilon+\epsilon^3}, \\
&\gamma^{\mathrm{2w}}=\frac{{J_z}^2\epsilon+\epsilon^3}{-2\ii J_xJ_yJ_z+({J_x}^2+{J_y}^2+{J_z}^2)\epsilon+\epsilon^3}, \\
&\delta^{\mathrm{2w}}=\frac{2\ii J_xJ_yJ_z-({J_x}^2+{J_y}^2)\epsilon}{-2\ii J_xJ_yJ_z+({J_x}^2+{J_y}^2+{J_z}^2)\epsilon+\epsilon^3}, \\
&\epsilon_\pm^{\mathrm{2w}}=\pm\frac{(J_x\pm J_y)(J_z\pm\ii\epsilon)\epsilon}{-2\ii J_xJ_yJ_z+({J_x}^2+{J_y}^2+{J_z}^2)\epsilon+\epsilon^3}.
\end{align}
\end{subequations}
The corresponding coin operator for the two quantum walkers with the impurity is given by
\begin{align}
&\hat{C}_0^{\mathrm{2w}}\coloneqq%
\sum_{x_1(\neq0),x_2(\neq0)}\dyad{x_1,x_2}\otimes\mathbb{I}_{8\times8}%
+\dyad{x_1=0,x_2=0}\otimes S_{\mathrm{imp}}^{\mathrm{2w}}\notag \\%
&\qquad+\sum_{x_2(\neq0),\sigma_2}\dyad{x_1=0}\otimes S_{\mathrm{imp}}^{\mathrm{1w}}\otimes\dyad{x_2,\sigma_2}\notag \\
&\qquad+\sum_{x_1(\neq0),\sigma_1}\dyad{x_1,\sigma_1}\otimes\dyad{x_2=0}\otimes S_{\mathrm{imp}}^{\mathrm{1w}}.\label{eq4112}
\end{align}

\begin{figure}
\centering
\includegraphics[width=0.6\textwidth]{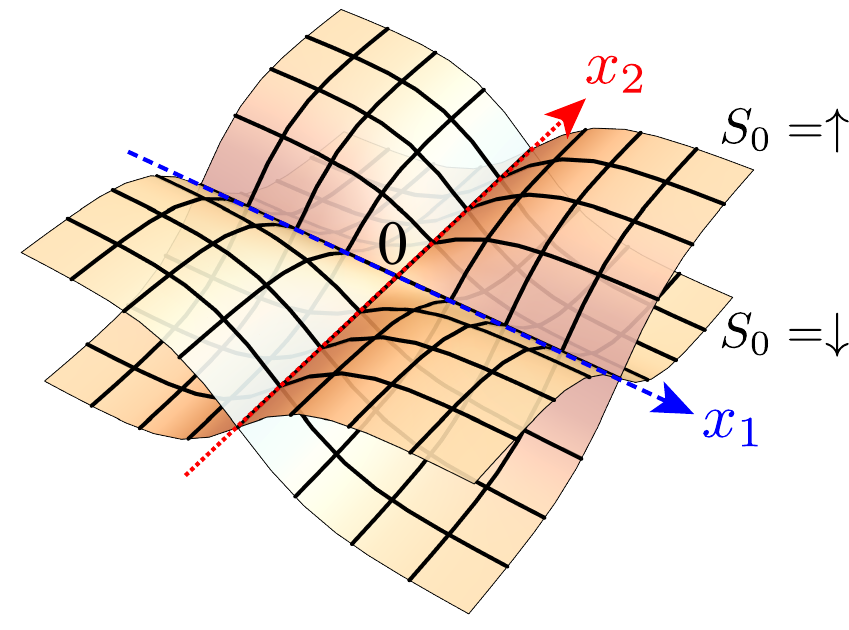}
\caption{
Visualization of the entire Hilbert space of the two quantum walkers, which consists
of the sectors of $S_0 =\uparrow$ and $S_0=\downarrow$. Each sector of $S_0$ constitutes a two-dimensional lattice $(x_1, x_2)$ with $x_1=-L_x/2, -L_x/2+1, \cdots, 0, \cdots, L_x/2-1, L_x/2$ and $x_2=-L_x/2, -L_x/2+1, \cdots, 0, \cdots, L_x/2-1, L_x/2$. For the explanation of the internal states of each lattice point, see the main text.
}
\label{fig05}
\end{figure}

The wave function in the two-walker sector is given by
$\Psi(t;x_1,\sigma_1;x_2,\sigma_2;S_0)$,
where $x_1$ and $x_2$ denote the lattice points for the two walkers, $\sigma_1$ and $\sigma_2$ denote the states $\LL$ and $\RR$ of each walker, and $S_0$ denotes the states $\up$ and $\down$ of the magnetic impurity at the origin.
We visualize the Hilbert space in Fig.~\ref{fig05}.
Note that the wave function of the standard quantum walker is given by
$\psi_i(t;x,\sigma)$,
where the subscript $i$ differentiates the wave functions starting from different initial conditions.
The wave function of two independent quantum walkers is then given by
\begin{equation}\label{eq4104}
\Psi(t;x_1,\sigma_1;x_2,\sigma_2)=\psi_1(t;x_1,\sigma_1)\psi_2(t;x_2,\sigma_2)
\end{equation}
for distinguishable particles and
\begin{equation}\label{eq4105-1}
\Psi(t;x_1,\sigma_1;x_2,\sigma_2)=\frac{1}{\sqrt{2}}
\qty[\psi_1(t;x_1,\sigma_1)\psi_2(t;x_2,\sigma_2)\mp\psi_1(t;x_2\sigma_2)\psi_2(t;x_1,\sigma_1)]
\end{equation}
for fermions (negative sign) and bosons (positive sign).
The wave function $\Psi(t;x_1,\sigma_1;x_2,\sigma_2;S_0)$ 
 has an extra argument that specifies the state of the magnetic impurity at the origin; hence, this is a three-body problem.

We carry out the time evolution of the two walkers with the impurity spin using the updates listed below. Refer to Fig.~\ref{fig05} for the following description.
\begin{itemize}
\item 
Each lattice point $(x_1,x_2,S_0)$ on the solid black lines with $x_1\neq0$, $x_2\neq0$, and $S_0=\uparrow,\downarrow$ has the internal degrees of freedom of $\sigma_1=\LL_1,\RR_1$ and $\sigma_2=\LL_2,\RR_2$. Each of the states $\sigma_1$ and $\sigma_2$ is separately updated by the standard coin operator $C_{\varphi}$ in Eq.~\eqref{eq1001-1}, and also independently of the state of $S_0$.

\item 
Each lattice point $(x_1,x_2)$ on the dotted red line with $x_1=0$ and $x_2\neq0$ has the internal degrees of freedom of $\sigma_1=\LL_1,\RR_1$, $\sigma_2=\LL_2,\RR_2$, and $S_0=\uparrow,\downarrow$. The combined state $\sigma_1\otimes S_0$ is updated by the new coin operator~\eqref{eq2206}. In addition, each of the states $\sigma_1$ and $\sigma_2$ is separately updated by the standard coin operator $C_{\varphi}$ in Eq.~\eqref{eq1001-1}, and also independently of the state of $S_0$.

\item 
Each lattice point $(x_1,x_2)$ on the dashed blue line with $x_1\neq0$ and $x_2=0$ has the internal degrees of freedom of $\sigma_1=\LL_1,\RR_1$, $\sigma_2=\LL_2,\RR_2$, and $S_0=\uparrow,\downarrow$. The combined state $\sigma_2\otimes S_0$ is updated by the new coin operator~\eqref{eq2206}. In addition, each of the states $\sigma_1$ and $\sigma_2$ is separately updated by the standard coin operator $C_{\varphi}$ in Eq.~\eqref{eq1001-1}, and also independently of the state of $S_0$.

\item
The origin $(x_1,x_2)=(0,0)$ has the internal degrees of freedom of $\sigma_1=\LL_1,\RR_1$, $\sigma_2=\LL_2,\RR_2$, and $S_0=\uparrow,\downarrow$. The combined state $\sigma_1\otimes\sigma_2\otimes  S_0$ is updated by the newer coin operator~\eqref{eq4105}. 
In addition, each of the states $\sigma_1$ and $\sigma_2$ is separately updated by the standard coin operator~\eqref{eq2204}, and also independently of the state of $S_0$. 

\item 
The shift operators~\eqref{eq4108-0} shift each of the two walkers according to the internal states $\sigma_1$ and $\sigma_2$.
\end{itemize}

As in the case in Eq.~\eqref{eq2301} with one quantum walker and one impurity spin, we introduce chiral symmetry to the time evolution of our model.
Reordering the coin operators listed above, we define the unitary time-evolution operator $\hat{U}_{\mathrm{K}}^{\mathrm{2w}}$ of one time step $\Delta_t$ as
\begin{equation}\label{eq4107}
\hat{U}_{\mathrm{K}}^{\mathrm{2w}}=%
\sqrt{\hat{C}_0^{\mathrm{2w}}}\sqrt{\hat{C}_2^{\mathrm{2w}}\hat{C}_1^{\mathrm{2w}}}%
\hat{S}_2^{\mathrm{2w}}\hat{S}_1^{\mathrm{2w}}%
\sqrt{\hat{C}_2^{\mathrm{2w}}\hat{C}_1^{\mathrm{2w}}}\sqrt{\hat{C}_0^{\mathrm{2w}}}.
\end{equation}
In Eq.~\eqref{eq4107}, $\sqrt{\hat{C}_0^{\mathrm{2w}}}$ contains $\sqrt{S_{\mathrm{imp}}^{\mathrm{2w}}}$, which is given by
\begin{equation}\label{eq4113}
\sqrt{S_{\mathrm{imp}}^{\mathrm{2w}}}
=\qty(\begin{array}{cccc|cccc}
\tilde{\alpha}_+^{\mathrm{2w}} & & & & & & & \\
& \tilde{\gamma}^{\mathrm{2w}} & \tilde{\delta}^{\mathrm{2w}} & & \tilde{\epsilon}_+^{\mathrm{2w}} & & & \\
& \tilde{\delta}^{\mathrm{2w}} & \tilde{\gamma}^{\mathrm{2w}} & & \tilde{\epsilon}_+^{\mathrm{2w}} & & & \\
& & & \tilde{\alpha}_-^{\mathrm{2w}} & & \tilde{\epsilon}_+^{\mathrm{2w}} & \tilde{\epsilon}_+^{\mathrm{2w}} & \\ \hline
& \tilde{\epsilon}_+^{\mathrm{2w}} & \tilde{\epsilon}_+^{\mathrm{2w}} & & \tilde{\alpha}_-^{\mathrm{2w}} & & & \\
& & & \tilde{\epsilon}_+^{\mathrm{2w}} & & \tilde{\gamma}^{\mathrm{2w}} & \tilde{\delta}^{\mathrm{2w}} & \\
& & & \tilde{\epsilon}_+^{\mathrm{2w}} & & \tilde{\delta}^{\mathrm{2w}} & \tilde{\gamma}^{\mathrm{2w}} & \\
& & & & & & & \tilde{\alpha}_+^{\mathrm{2w}}
\end{array}),
\end{equation}
with 
\begin{subequations}\label{eq4114}
\begin{align}
\tilde{\alpha}_+^{\mathrm{2w}}&\coloneqq1, \label{eq4114a} \\
\tilde{\alpha}_-^{\mathrm{2w}}&\coloneqq\frac{\epsilon}{\sqrt{2J^2+\epsilon^2}}, \label{eq4114b} \\
\tilde{\gamma}^{\mathrm{2w}}&\coloneqq\frac{1}{2}\qty[+1+\frac{\epsilon}{\sqrt{2J^2+\epsilon^2}}], \label{eq4114c} \\
\tilde{\delta}^{\mathrm{2w}}&\coloneqq\frac{1}{2}\qty[-1+\frac{\epsilon}{\sqrt{2J^2+\epsilon^2}}], \label{eq4114d} \\
\tilde{\epsilon}_+^{\mathrm{2w}}&\coloneqq\frac{\ii J}{\sqrt{2J^2+\epsilon^2}}, \label{eq4114e}
\end{align}
\end{subequations}
in the XX case of the coupling parameter $J_x=J_y=J$ and $J_z=0$, and
\begin{subequations}\label{eq4115}
\begin{align}
\tilde{\alpha}_+^{\mathrm{2w}}&\coloneqq\sqrt{\frac{\ii-J}{\ii+J}}, \label{eq4115a} \\
\tilde{\alpha}_-^{\mathrm{2w}}&\coloneqq\frac{1}{3}\qty[2\sqrt{\frac{\ii+2J}{\ii-2J}}+\sqrt{\frac{\ii-J}{\ii+J}}], \label{eq4115b} \\
\tilde{\gamma}^{\mathrm{2w}}&\coloneqq\frac{1}{6}\qty[+3+\sqrt{\frac{\ii+2J}{\ii-2J}}+2\sqrt{\frac{\ii-J}{\ii+J}}], \label{eq4115c} \\
\tilde{\delta}^{\mathrm{2w}}&\coloneqq\frac{1}{6}\qty[-3+\sqrt{\frac{\ii+2J}{\ii-2J}}+2\sqrt{\frac{\ii-J}{\ii+J}}], \label{eq4115d} \\
\tilde{\epsilon}_+^{\mathrm{2w}}&\coloneqq\frac{1}{3}\qty[-\sqrt{\frac{\ii+2J}{\ii-2J}}+\sqrt{\frac{\ii-J}{\ii+J}}],
\label{eq4115e}
\end{align}
\end{subequations}
in the $\textrm{SU}(2)$ case of the coupling parameter $J_x=J_y=J_z=J$.
We set $\epsilon=1$ to obtain Eq.~\eqref{eq4115}.
We can indeed confirm that squaring Eq.~\eqref{eq4113} yields Eq.~\eqref{eq4105} for both cases with Eqs.~\eqref{eq4114} and~\eqref{eq4115}.

This completes the algorithm for the unitary dynamics of two quantum walkers interacting via a magnetic impurity at the origin.
In Sec.~\ref{sec4}, we numerically simulate the dynamics, following the algorithm above.

\section{Analytical solution of the bound states of one quantum walker with a magnetic impurity}\label{sec3}

In this section, using the transfer-matrix method, we analytically find the eigenvalues and the eigenvectors of the bound states, in which one quantum walker is bound to the magnetic impurity.
We use the bound state obtained here for the initial state of the simulation of the dynamics with two quantum walkers in Sec.~\ref{sec4}.
We start our analysis by restricting ourselves to $J_z=0$, for which the Hamiltonian preserves chiral symmetry, as discussed in Sec.~\ref{sec2.3}.
To proceed further in the analysis, we will later impose the XX case $J_x=J_y=J$ in Eq.~\eqref{eq-xx} to reduce the order of the eigenvalue equation. 
As a result, we obtain a quadratic characteristic polynomial, 
which is solved explicitly and yields the eigenvalue formula in a closed form.

\subsection{Strategy of the analyses}

Let us first summarize how we obtain the eigenvalues and eigenvectors of the bound states~\cite{Kiumi21, Kiumi22, Kiumi22-2}.
We first assume that we know an eigenvalue $\lambda_\mu$ and the corresponding eigenstate $\hat{\Psi}_\mu(x)$ of the time-evolution operator.
From the eigenvalue equation for the time-evolution operator, 
\begin{align}
\hat{U}_{\mathrm{K}}^{\mathrm{1w}}\hat{\Psi}_\mu=\ee^{\ii\lambda_\mu}\hat{\Psi}_\mu,
\end{align}
we construct a transfer matrix $T_\mu$ as a function of $\lambda_\mu$ that relates the wave-function amplitude $\hat{\Psi}_\mu(x)$ to $\hat{\Psi}_\mu(x+1)$.
The transfer matrix $T_\mu$ is the same for every $x$ except for the origin, where the special transfer matrix $T_\mu^{(0)}$ relates $\hat{\Psi}_\mu(0)$ to $\hat{\Psi}_\mu(1)$, taking account of the magnetic impurity.
We will show that the transfer matrix $T_\mu$ has two degenerate eigenvalues $\zeta_\mu^\pm$.

\begin{figure}[h]
\centering
\includegraphics[width=0.5\textwidth]{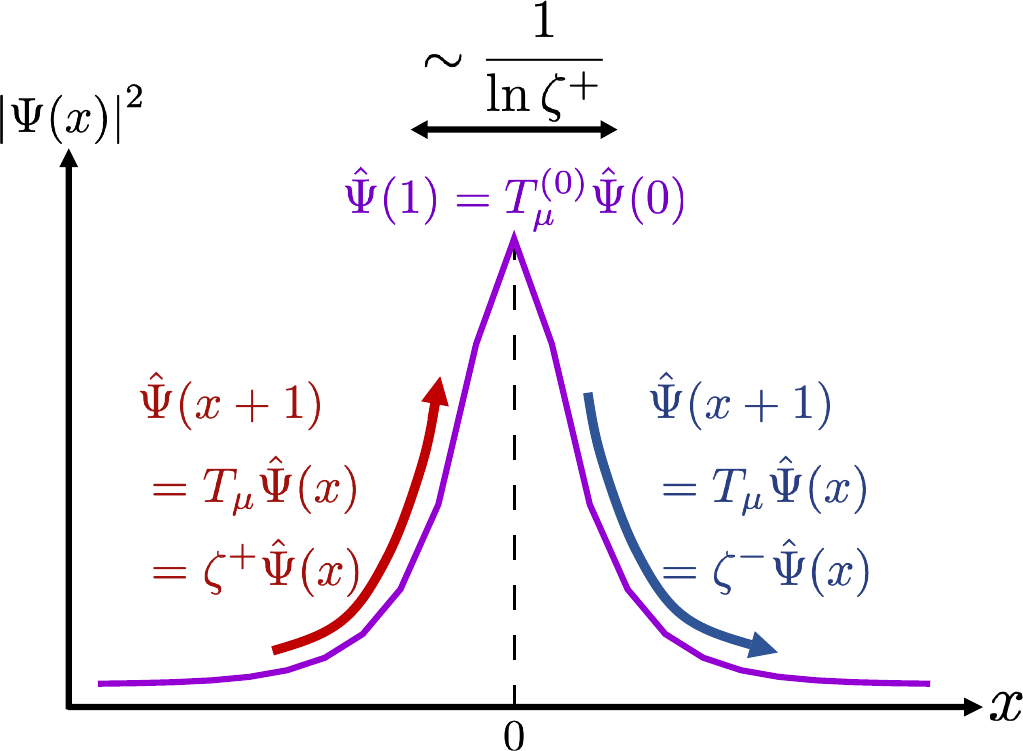}
\caption{Schematic figure of the eigenstate.} 
\label{fig06}
\end{figure}

We then require the eigenstate to take the form of a bound state, as shown in Fig.~\ref{fig06}.
In other words, the transfer matrix $T_\mu$ must increase the wave-function amplitude geometrically for $x<0$, but the special transfer matrix $T_\mu^{(0)}$ must switch the channel
so that the same transfer matrix $T_\mu$ may decrease the wave-function amplitude geometrically for 
$x>0$.

These requirements produce a solvable equation for the eigenvalue $\lambda_\mu$ of the time-evolution operator.
Once we obtain the eigenvalue of the time-evolution operator, we can next find the eigenvalues $\zeta_\mu^\pm$ of the transfer matrix $T_\mu$, which fix the eigenstate $\hat{\Psi}(x)$ of the time-evolution operator.

\subsection{Transfer-matrix analysis of eigenvalues}\label{sec3.2}

Before using the transfer-matrix method given above, we first reformulate the time-evolution operator~\eqref{eq2301} in the symmetric form to one in a form more convenient for the transfer-matrix method.
Note that reordering the elements of the operator~\eqref{eq2301} keeps the eigenvalue the same after we adjust the eigenvector.
For a pedagogical example, consider the eigenvalue equation $(AB)\vec{v}=\lambda\vec{v}$;
multiplying both sides by $B$ from the left and defining a new eigenvector $\vec{w}:=B\vec{v}$, we obtain $(BA)\vec{w}=\lambda\vec{w}$, and thus we conclude that the matrices $AB$ and $BA$ share the eigenvalues.

We similarly reorder the time-evolution operator~\eqref{eq2301},
\begin{align}
\hat{U}_{\mathrm{K}}^{\mathrm{1w}}\coloneqq
\sqrt{\hat{C}_0^{\mathrm{1w}}}\sqrt{\hat{C}^{\mathrm{1w}}}\hat{S}^{\mathrm{1w}}%
\sqrt{\hat{C}^{\mathrm{1w}}}\sqrt{\hat{C}_0^{\mathrm{1w}}}
=\qty(\sqrt{\hat{C}_0^{\mathrm{1w}}}\sqrt{\hat{C}^{\mathrm{1w}}})%
\qty(\hat{S}^{\mathrm{1w}}%
\sqrt{\hat{C}^{\mathrm{1w}}}\sqrt{\hat{C}_0^{\mathrm{1w}}}),
\end{align}
to the new one
\begin{align}
\tilde{\hat{U}}_{\mathrm{K}}^{\mathrm{1w}}&\coloneqq
\qty(\hat{S}^{\mathrm{1w}}%
\sqrt{\hat{C}^{\mathrm{1w}}}\sqrt{\hat{C}_0^{\mathrm{1w}}})
\qty(\sqrt{\hat{C}_0^{\mathrm{1w}}}\sqrt{\hat{C}^{\mathrm{1w}}})%
=
\hat{S}^{\mathrm{1w}}\tilde{\hat{C}}^{\mathrm{1w}}, \label{eq3002}
\end{align}
where the new coin operator is given by
\begin{align}
\tilde{\hat{C}}^{\mathrm{1w}}&\coloneqq\sqrt{\hat{C}^{\mathrm{1w}}}\hat{C}_0^{\mathrm{1w}}\sqrt{\hat{C}^{\mathrm{1w}}} 
=\sum_xC_x^{\mathrm{1w}}\otimes\dyad{x}, 
\end{align}
with
\begin{align}
 C_x^{\mathrm{1w}}=%
\begin{cases}
\sqrt{\displaystyle\sum_{S_0}C_{\varphi}\otimes\dyad{S_0}}S_{\mathrm{imp}}^{\mathrm{1w}}%
\sqrt{\displaystyle\sum_{S_0}C_{\varphi}\otimes\dyad{S_0}} &\mbox{for $x=0$},\\
\displaystyle\sum_{S_0}C_{\varphi}\otimes\dyad{S_0} & \mbox{for $x\ne0$}.
\end{cases}\label{eq3003}
\end{align}

The two time-evolution operators~\eqref{eq2301} and~\eqref{eq3002} share the same eigenvalues.
Multiplying the eigenvector of the new time-evolution operator~\eqref{eq3002} by  $\sqrt{\hat{C}^{\mathrm{1w}}_0}\sqrt{\hat{C}^{\mathrm{1w}}}$ recovers the eigenvector of the operator~\eqref{eq2301}.
The new time-evolution operator~\eqref{eq3002} is more convenient for obtaining analytical solutions using the transfer-matrix method.

We are now in a position to analytically find the bound-state eigenvalues of the time-evolution operator, which are exemplified numerically in Fig.~\ref{fig04}, using the transfer-matrix method~\cite{Kiumi21,Kiumi22,Kiumi23}.
Since the time-evolution operator $\tilde{\hat{U}}_{\mathrm{K}}^{\mathrm{1w}}$ is unitary, its eigenvalues take the form of $\tilde{U}^{\mathrm{1w}}_\mu=\ee^{\ii\lambda_\mu}$ in the eigenvalue equation
\begin{align}
\tilde{\hat{U}}_{\mathrm{K}}^{\mathrm{1w}}\Psi_\mu=\ee^{\ii\lambda_\mu}\Psi_\mu
\end{align}
with a real parameter $\lambda_\mu\in[0,2\pi)$  for $\mu=1,2,3,4$.
Note that the eigenstate $\Psi_\mu(x,\sigma;S_0)$ is a time-independent eigenfunction, different from the time-evolving wave function $\Psi(t; x, \sigma; S_0)$.

Hereafter, we find the transfer matrix $T_\mu$ for $x\ne0$ and the one $T_\mu^{(0)}$ for $x=0$ from the eigenvalue equation.
For $x\ne0$, $\tilde{\hat{C}}^{\mathrm{1w}}$ in Eq.~\eqref{eq3003} takes a simple form, and therefore the eigenvalue equation 
$\hat{S}^{\mathrm{1w}}\tilde{\hat{C}}^{\mathrm{1w}}\Psi_\mu(x,\sigma;S_0)=\ee^{\ii\lambda_\mu}\Psi_\mu(x,\sigma;S_0)$ 
reads
\begin{equation}\label{eq3008}
\hat{S}^{\mathrm{1w}}
\begin{pmatrix}
\cos\varphi\ \psi_{\mathrm{L}\uparrow}(x)-\sin\varphi\ \psi_{\mathrm{R}\uparrow}(x) \\
\sin\varphi\ \psi_{\mathrm{L}\uparrow}(x)+\cos\varphi\ \psi_{\mathrm{R}\uparrow}(x) \\
\cos\varphi\ \psi_{\mathrm{L}\downarrow}(x)-\sin\varphi\ \psi_{\mathrm{R}\downarrow}(x) \\
\sin\varphi\ \psi_{\mathrm{L}\downarrow}(x)+\cos\varphi\ \psi_{\mathrm{R}\downarrow}(x)
\end{pmatrix}
=
\begin{pmatrix}
\ee^{\ii\lambda_\mu}\psi_{\mathrm{L}\uparrow}(x) \\
\ee^{\ii\lambda_\mu}\psi_{\mathrm{R}\uparrow}(x) \\
\ee^{\ii\lambda_\mu}\psi_{\mathrm{L}\downarrow}(x) \\
\ee^{\ii\lambda_\mu}\psi_{\mathrm{R}\downarrow}(x)
\end{pmatrix},
\end{equation}
where $\psi_{\sigma S_0}$ with $\sigma=\LL,\RR$ and $S_0=\up,\down$ is a component of the eigenfunction $\Psi_\mu$, and
the subscript $\mu$ is suppressed in the notation of $\psi$ for brevity.
Equation~\eqref{eq3008} is followed by
\begin{equation}\label{eq3009}
\begin{pmatrix}
\cos\varphi\ \psi_{\mathrm{L}\uparrow}(x-1)-\sin\varphi\ \psi_{\mathrm{R}\uparrow}(x+1) \\
\sin\varphi\ \psi_{\mathrm{L}\uparrow}(x-1)+\cos\varphi\ \psi_{\mathrm{R}\uparrow}(x+1) \\
\cos\varphi\ \psi_{\mathrm{L}\downarrow}(x-1)-\sin\varphi\ \psi_{\mathrm{R}\downarrow}(x+1) \\
\sin\varphi\ \psi_{\mathrm{L}\downarrow}(x-1)+\cos\varphi\ \psi_{\mathrm{R}\downarrow}(x+1)
\end{pmatrix}
=
\begin{pmatrix}
\ee^{\ii\lambda_\mu}\psi_{\mathrm{L}\uparrow}(x) \\
\ee^{\ii\lambda_\mu}\psi_{\mathrm{R}\uparrow}(x) \\
\ee^{\ii\lambda_\mu}\psi_{\mathrm{L}\downarrow}(x) \\
\ee^{\ii\lambda_\mu}\psi_{\mathrm{R}\downarrow}(x)
\end{pmatrix}.
\end{equation}
Solving the first and second rows of Eq.~\eqref{eq3009}, we obtain
\begin{subequations}\label{eq3010}
\begin{align}
&\psi_{\mathrm{L}\uparrow}(x)%
=\frac{1}{\cos\varphi}\qty[\ee^{-\ii\lambda_\mu}\ \psi_{\mathrm{L}\uparrow}(x-1)%
                                      -\sin\varphi\ {\psi_{\mathrm{R}\uparrow}(x)}],\label{eq3010a} \\
&\psi_{\mathrm{R}\uparrow}(x+1)%
=\frac{1}{\cos\varphi}\qty[-\sin\varphi\ \psi_{\mathrm{L}\uparrow}(x-1)%
                                      +\ee^{\ii\lambda_\mu}\ {\psi_{\mathrm{R}\uparrow}(x)}].\label{eq3010b}
\end{align}
\end{subequations}
The same applies to the spin-down case given in the third and fourth rows of Eq.~\eqref{eq3009}.
Thus, for a new vector $\hatPsi_\mu(x)=(\psi_{\mathrm{L}\uparrow}(x-1), \psi_{\mathrm{R}\uparrow}(x), \psi_{\mathrm{L}\downarrow}(x-1), \psi_{\mathrm{R}\downarrow}(x))^\top$,
we obtain 
$\hatPsi_\mu(x+1)=T_\mu\hatPsi_\mu(x)$
for $x\neq0$, where
\begin{equation}\label{eq3011}
T_\mu=\frac{1}{\cos\varphi}
\begin{pmatrix}
\ee^{-\ii\lambda_\mu} & -\sin\varphi & 0 & 0 \\
-\sin\varphi & \ee^{\ii\lambda_\mu} & 0 & 0 \\
0 & 0 & \ee^{-\ii\lambda_\mu} & -\sin\varphi \\
0 & 0 & -\sin\varphi & \ee^{\ii\lambda_\mu}
\end{pmatrix}
\end{equation}
is the transfer matrix that describes for any eigenfunctions $\Psi_\mu(x,\sigma;S_0)$ with $x\neq0$ how the components  $\psi_{\LL S_0}(x-1)$ and $\psi_{\RR S_0}(x)$ are related to the neighboring components $\psi_{\LL S_0}(x)$ and $\psi_{\RR S_0}(x+1)$, as in Fig.~\ref{fig07}.
Note that the transfer matrix here is for the components of an eigenfunction $\Psi_\mu(x,\sigma;S_0)$, and is different from that of the T matrix, which appears in \ref{appB}.
\begin{figure}[h!]
\centering
\begin{tikzpicture}
\draw[thick] (0,0)--(0,4);
\draw[thick] (6,0)--(6,4);
\draw[thick] (12,0)--(12,4);
\draw[line width = 2, arrows = {-{Stealth[length=5mm]}}] (2,3) -- (0,3);
\draw (1.1,2.95)node[below]{$\psi_{\LL\up,\down}(x-1)$};
\draw[line width = 2, arrows = {-{Stealth[length=5mm]}}] (8,3) -- (6,3);
\draw (7,2.95)node[below]{$\psi_{\LL\up,\down}(x)$};
\draw[line width = 2, arrows = {-{Stealth[length=5mm]}}] (4,1) -- (6,1);
\draw (5,0.95)node[below]{$\psi_{\RR\up,\down}(x)$};
\draw[line width = 2, arrows = {-{Stealth[length=5mm]}}] (10,1) -- (12,1);
\draw (10.9,0.95)node[below]{$\psi_{\RR\up,\down}(x+1)$};
\draw[dashed, line width = 2, arrows = {-{Stealth[length=5mm]}}] (6,3) -- (4,3);
\draw[dashed, line width = 2, arrows = {-{Stealth[length=5mm]}}] (6,1) -- (8,1);
\draw (0,0)node[below,font=\large]{$x-1$};
\draw (6,0)node[below,font=\large]{$x$};
\draw (12,0)node[below,font=\large]{$x+1$};
\end{tikzpicture}
\caption{Transfer matrix $T_\mu$ that relates $\{\psi_{\mathrm{L}\uparrow,\downarrow}(x-1), \psi_{\mathrm{R}\uparrow,\downarrow}(x)\}$ and $\{\psi_{\mathrm{L}\uparrow,\downarrow}(x), \psi_{\mathrm{R}\uparrow,\downarrow}(x+1)\}$ for $x\neq0$.}
\label{fig07}
\end{figure}

In the same manner, we obtain the transfer matrix $T_\mu^{(0)}$, which takes account of the impurity effect, from the eigenvalue equation at $x=0$:
\begin{align}\label{eq3012}
\hat{S}^{\mathrm{1w}}\tilde{\hat{C}}^{\mathrm{1w}}\Psi_\mu(x=0,\sigma;S_0) 
=&\hat{S}^{\mathrm{1w}}\sqrt{\displaystyle\sum_{S_0}C_{\varphi}\otimes\dyad{S_0}}S_{\mathrm{imp}}^{\mathrm{1w}}%
\sqrt{\displaystyle\sum_{S_0}C_{\varphi}\otimes\dyad{S_0}}\Psi_\mu(x=0,\sigma;S_0)\notag \\
=&\ee^{\ii\lambda_\mu}\Psi_\mu(x=0,\sigma;S_0).
\end{align}
Each row of the matrix equation is given by
\begin{subequations}\label{eq3013}
\begin{align}
&\qty(A_-+A_+\cos{\varphi})\psi_{\LL\up}(x-1)-A_+\sin{\varphi} \psi_{\RR\up}(x+1) \notag \\
&\qquad+B_-\sin{\varphi}\psi_{\LL\down}(x-1)+\qty(B_++B_-\cos{\varphi})\psi_{\RR\down}(x+1)%
=\ee^{\ii\lambda_\mu}\psi_{\LL\up}(x), \label{eq3013a}\\
&A_+\sin{\varphi}\psi_{\LL\up}(x-1)+\qty(-A_-+A_+\cos{\varphi})\psi_{\RR\up}(x+1) \notag \\
&\qquad+\qty(B_+-B_-\cos{\varphi})\psi_{\LL\down}(x-1)+B_-\sin{\varphi}\psi_{\RR\down}(x+1)%
=\ee^{\ii\lambda_\mu}\psi_{\RR\up}(x), \label{eq3013b}\\
&-B_-\sin{\varphi}\psi_{\LL\up}(x-1)+\qty(B_+-B_-\cos{\varphi})\psi_{\RR\up}(x+1) \notag \\
&\qquad+\qty(-A_-+A_+\cos{\varphi})\psi_{\LL\down}(x-1)-A_+\sin{\varphi}\psi_{\RR\down}(x+1)%
=\ee^{\ii\lambda_\mu}\psi_{\LL\down}(x), \label{eq3013c}\\
&\qty(B_++B_-\cos{\varphi})\psi_{\LL\up}(x-1)-B_-\sin{\varphi}\psi_{\RR\up}(x+1) \notag \\
&\qquad+A_+\sin{\varphi}\psi_{\LL\down}(x-1)+\qty(A_-+A_+\cos{\varphi})\psi_{\RR\down}(x+1)%
=\ee^{\ii\lambda_\mu}\psi_{\RR\down}(x)\label{eq3013d}
\end{align}
\end{subequations}
with
\begin{equation}\label{eq3014}
A_\pm=\frac{\alpha_0^{\mathrm{1w}}\pm\gamma_0^{\mathrm{1w}}}{2},\quad
B_\pm=\frac{\beta_0^{\mathrm{1w}}\pm\delta_0^{\mathrm{1w}}}{2},
\end{equation}
where we used Eqs.~\eqref{eq2206} and~\eqref{eq2305}.
We thereby define the transfer matrix for $x=0$ in the form 
$\hatPsi_\mu(1)=T_\mu^{(0)}\hatPsi_\mu(0)$.

To further simplify the transfer matrix $T_\mu^{(0)}$ for $x=0$ and the resulting equations, we impose an additional condition of the XX interaction $J_x=J_y=J$.
By setting the XX case to use in Eq.~\eqref{eq3014}, we obtain the transfer matrix for $x=0$ as
\begin{align}\label{eq3015}
&T_\mu^{(0)}\times\qty(1-2J^2+\cos{2\varphi}) \notag \\
=&
\begin{pmatrix}
2\ee^{-\ii\lambda_\mu}(-{J}^2+\cos{\varphi}) & -\sin{2\varphi} & 
2\ii\ee^{-\ii\lambda_\mu}J\sin{\varphi} & 2\ii J(-1+\cos{\varphi}) \\
-\sin{2\varphi} & 2\ee^{\ii\lambda_\mu}({J}^2+\cos{\varphi}) &
-2\ii J(1+\cos{\varphi}) & 2\ii\ee^{\ii\lambda_\mu}J\sin{\varphi} \\
-2\ii\ee^{-\ii\lambda_\mu}J\sin{\varphi} & 2\ii J(1+\cos{\varphi}) &
2\ee^{-\ii\lambda_\mu}({J}^2+\cos{\varphi}) & -\sin{2\varphi} \\
-2\ii J(-1+\cos{\varphi}) & -2\ii\ee^{\ii\lambda_\mu} J\sin{\varphi} &
-\sin{2\varphi} & 2\ee^{\ii\lambda_\mu}(-{J}^2+\cos{\varphi})
\end{pmatrix}.
\end{align}
Using the transfer matrices~\eqref{eq3011} and~\eqref{eq3015}, an eigenfunction component can be written as follows:
\begin{equation}\label{eq3016}
\hatPsi_\mu(x)=\begin{cases}
{T_\mu}^{x-1}T_\mu^{(0)}\hatPsi_\mu(0) & \mbox{for $x>0$}, \\
{T_\mu}^x\hatPsi_\mu(0) & \mbox{for $x\leq0$},
\end{cases}
\end{equation}
where $\hatPsi_\mu(x)$ denotes the eigenfunction component at $x$ of the $\mu$th eigenvalue.
The eigenvalue $\ee^{\ii\lambda_\mu}$ of a bound state of the time-evolution operator $\tilde{\hat{U}}_{\mathrm{K}}^{\mathrm{1w}}=\hat{S}^{\mathrm{1w}}\tilde{\hat{C}}^{\mathrm{1w}}$ exists if and only if there exists  $\hatPsi_\mu(0)\in\mathbb{C}^4$ such that $\sum_{x\in\mathbb{Z}}\|\hatPsi_\mu(x)\|^2<\infty$, \textit{i.e.}, if the eigenfunction is normalizable.

We hereafter seek a bound state. The strategy is roughly as follows.
We first take $\hatPsi_\mu(x_-)$ for a negatively large value of $x_-$.
As we apply the transfer matrix $T_\mu$ to it again and again as ${T_\mu}^n$, the amplitude of $\hatPsi_\mu(x_-+n)$ should geometrically increase; it is multiplied by the factor ${\zeta_\mu^+}^n$, where $\zeta_\mu^+>1$ is a real eigenvalue of $T_\mu$. 
We eventually arrive at the impurity site $x=0$ when $n=\abs{x_-}$.
The multiplication of $T_\mu^{(0)}$ to $\hatPsi_\mu(0)$ should then switch the tendency. 
After it, as we apply the transfer matrix $T_\mu$ to it again and again as ${T_\mu}^n$, the amplitude should geometrically decrease 
by the factor ${\zeta_\mu^-}^n$,
where $\zeta_\mu^-<1$  is another real eigenvalue of $T_\mu$. For the bound-state eigenfunction to be symmetric with respect to $x=0$, we should have $\zeta_\mu^{-}=1/\zeta_\mu^{+}$

For a more detailed analysis, we diagonalize the $4\times4$ transfer matrix $T_\mu$ and find its four eigenvalues.
Since the transfer matrix $T_\mu$ in Eq.~\eqref{eq3011} is block-diagonalized, the characteristic equation is given in the form of
\begin{equation}\label{eq3017}
\frac{1}{\cos^4\varphi}\left|
\begin{matrix}
\ee^{-\ii\lambda_\mu}-\zeta_1^\pm \cos\varphi & -\sin\varphi \\
-\sin\varphi & \ee^{\ii\lambda_\mu}-\zeta_1^\pm\cos\varphi
\end{matrix}\right|\ 
\left|
\begin{matrix}
\ee^{-\ii\lambda_\mu}-\zeta_2^\pm \cos\varphi & -\sin\varphi \\
-\sin\varphi & \ee^{\ii\lambda_\mu}-\zeta_2^\pm \cos\varphi\ 
\end{matrix}
\right|=0
\end{equation}
with its four eigenvalues $\zeta_1^\pm$ and $\zeta_2^\pm$, where we suppressed the subscript $\mu$ from the notation for brevity.
We straightforwardly obtain the eigenvalues as
\begin{equation}\label{eq3018}
\zeta_1^\pm=\zeta_2^\pm=\frac{\cos\lambda_\mu\pm\sqrt{\cos^2\lambda_\mu-\cos^2\varphi}}{\cos\varphi}.
\end{equation}
We can indeed confirm $\zeta_\mu^-=1/\zeta_\mu^+$ for $\nu=1,2$.
Note that two of the eigenvalues with the same sign are degenerate.

The corresponding eigenvectors, $\bm{v}_1^\pm$ and $\bm{v}_2^\pm$, are given as follows:
\begin{subequations}\label{eq3019}
\begin{align}\label{eq3019a}
&\bm{v}_1^\pm=
\begin{pmatrix}
\sin\varphi \\
-\ii\sin\lambda_\mu\mp\sqrt{\cos^2\lambda_\mu-\cos^2\varphi} \\
0 \\
0
\end{pmatrix}
\propto
\begin{pmatrix}
\ii\sin\lambda_\mu\mp\sqrt{\cos^2\lambda_\mu-\cos^2\varphi} \\
\sin\varphi \\
0 \\
0
\end{pmatrix}, \\
\label{eq3019b}
&\bm{v}_2^\pm=
\begin{pmatrix}
0 \\
0 \\
\sin\varphi \\
-\ii\sin\lambda_\mu\mp\sqrt{\cos^2\lambda_\mu-\cos^2\varphi}
\end{pmatrix}
\propto
\begin{pmatrix}
0 \\
0 \\
\ii\sin\lambda_\mu\mp\sqrt{\cos^2\lambda_\mu-\cos^2\varphi}\\
\sin\varphi 
\end{pmatrix}.
\end{align}
\end{subequations}
Because of the degeneracy of the eigenvalues $\zeta_\mu^\pm$, any linear combination of $\bm{v}_1^+$ and $\bm{v}_2^+$ produces the eigenvalue $\zeta^+$, whlie any one of $\bm{v}_1^-$ and $\bm{v}_2^-$ produces the eigenvalue $\zeta^-$.
Note the following orthogonality relations:
\begin{subequations}\label{eq3020}
\begin{align}\label{eq3020a}
\bm{v}_1^+\cdot\bm{v}_1^-=\bm{v}_2^+\cdot\bm{v}_2^-&=0,\\
\label{eq3020b}
\bm{v}_1^+\cdot\bm{v}_2^\pm=\bm{v}_1^-\cdot\bm{v}_2^\pm&=0.
\end{align}
\end{subequations}

For $\cos^2\lambda_\mu-\cos^2\varphi\le0$, the eigenvalues~\eqref{eq3018} of the transfer matrix $T_\mu$ are complex and we have $|\zeta_1^\pm|=|\zeta_2^\pm|=1$.
This means that the eigenfunction component $\hatPsi_\mu(x)$ neither grows nor decays as we multiply $T_\mu$ to it, and hence
the eigenfunction $\Psi_\mu(x,\sigma;S_0)$ does not form a bound state.
Therefore, the eigenvalue $\lambda_\mu$ under the condition $\cos^2\lambda_\mu\le\cos^2\varphi$ should be in the two bulks of the eigenvalues. 
Indeed, in Fig.~\ref{fig04} for $\varphi=\pi/10$, the eigenvalues are densely distributed in the ranges $\pi/10\leq \lambda_\mu\leq 9\pi/10$ and $11\pi/10 \leq \lambda_\mu\leq 19\pi/10$ in all panels of Fig.~\ref{fig04}, which is consistent with the condition above. 

For  $\cos^2\lambda_\mu-\cos^2\varphi>0$, on the other hand, the eigenvalues \eqref{eq3018} of the transfer matrix $T_\mu$ are real.
We have the following four cases.
\begin{enumerate}
\item  $\cos\varphi>0$:
\label{casei}
 \begin{enumerate}
\item  $\cos\lambda_\mu>\cos\varphi>0$:
 We then have $\zeta_1^+=\zeta_2^+=1/\zeta_1^-=1/\zeta_2^->1$.
\label{casei-1}
\item  $\cos\lambda_\mu<-\cos\varphi<0$:
 We then have $\zeta_1^-=\zeta_2^-=1/\zeta_1^+=1/\zeta_2^+<-1$.
\label{casei-2}
\end{enumerate}
\item  $\cos\varphi<0$:
\label{caseii}
 \begin{enumerate}
\item  $\cos\lambda_\mu>-\cos\varphi>0$:
 We then have $\zeta_1^+=\zeta_2^+=1/\zeta_1^-=1/\zeta_2^-<-1$.
\label{caseii-1}
\item  $\cos\lambda_\mu<\cos\varphi<0$:
 We then have $\zeta_1^-=\zeta_2^-=1/\zeta_1^+=1/\zeta_2^+>1$.
\label{caseii-2}
\end{enumerate}
\end{enumerate}
In each of the two cases \ref{casei} and \ref{caseii}, we have four bound-state solutions. 
In the case~\ref{casei}, for example, we have two solutions in the case~\ref{casei-1} with the growth rate $\zeta_\mu^+>1$ and two more solutions in the case~\ref{casei-2} with the growth rate $\zeta_1^-<-1$.
Hereafter, let us focus on a solution in the case~\ref{casei-1} for simplicity.
\begin{thm}
There exists a bound-state eigenvalue $\ee^{\ii\lambda_\mu}$ of the time-evolution operator $\tilde{\hat{U}}_{\mathrm{K}}^{\mathrm{1w}}=\hat{S}^{\mathrm{1w}}\tilde{\hat{C}}^{\mathrm{1w}}$ if and only if
\begin{enumerate}
\item $\cos^2\lambda_\mu>\cos^2\varphi$;
\item $\exists\,\bm{v}\in\mathbb{C}^4$ s.t.\ $\bm{v}\in\mathrm{span}(\bm{v}_1^+,\bm{v}_2^+)$ and $T_\mu^{(0)}\bm{v}\in\mathrm{span}(\bm{v}_1^-, \bm{v}_2^-)$.
\end{enumerate}
\end{thm}
To put it in other words, we assume the eigenvector component in the form
\begin{align}
\hatPsi_\mu(0)&=a\bm{v}_1^++b\bm{v}_2^+,\label{eq3021}\\
T_\mu^{(0)}\hatPsi_\mu(0)&=c\bm{v}_1^-+d\bm{v}_2^-,\label{eq3022}
\end{align}
so that
\begin{equation}
\hatPsi_\mu(x)=\begin{cases}
T_\mu^x\hatPsi_\mu(0)=a (\zeta_1^+)^x\bm{v}_1^++b (\zeta_2^+)^x\bm{v}_2^+ & \mbox{for $x\le0$},\\
T_\mu^{x-1}T_\mu^{(0)}\hatPsi_\mu(0)=c (\zeta_1^-)^{x-1}\bm{v}_1^-+d (\zeta_2^-)^{x-1}\bm{v}_2^- & \mbox{for $x>0$}, 
\end{cases}\label{eq3023}
\end{equation}
with the coefficients $a$, $b$, $c$, and $d$ satisfying
\begin{equation}
T_\mu^{(0)}\qty(a\bm{v}_1^++b\bm{v}_2^+)=c\bm{v}_1^-+d\bm{v}_2^-. \label{eq3024}
\end{equation}
This assumption produces the situation shown in Fig.~\ref{fig06} as follows:
\begin{enumerate}
\item The eigenfunction component $\hatPsi_\mu(x)$ for $x<0$ is a linear combination $a\bm{v}_1^++b\bm{v}_2^+$, and hence the application of the transfer matrix $T_\mu$ multiplies the component by $\zeta_1^+=\zeta_2^+>1 $, increasing the amplitude geometrically as we approach the origin from the negative side of $x$.
\item The eigenfunction component  $\hatPsi_\mu(x)$ for $x>0$ is a linear combination $c\bm{v}_1^-+d\bm{v}_2^-$, and hence the application of the transfer matrix $T_\mu$ multiplies the component by $\zeta_1^-=\zeta_2^-=1/\zeta_1^+=1/\zeta_2^+<1$, decreasing the amplitude geometrically as we leave the origin to the positive side of $x$.
\item The application of $T_\mu^{(0)}$ at the origin $x=0$ switches the linear combination from $a\bm{v}_1^++b\bm{v}_2^+$ to $c\bm{v}_1^-+d\bm{v}_2^-$, as in Eq.~\eqref{eq3024}.
\end{enumerate}
%
In other words, starting from the origin $x=0$, the amplitude $\hatPsi_\mu(x)$ decays geometrically in both the positive and negative $x$ directions symmetrically, forming a cusp at the origin.

Because of the orthogonality relations~\eqref{eq3020a} and \eqref{eq3020b}, we have
\begin{subequations}\label{eq3025-0}
\begin{align}
\bm{v}_1^+\cdot\qty(c\bm{v}_1^-+d\bm{v}_2^-)&=0,\label{eq3025}\\
\bm{v}_2^+\cdot\qty(c\bm{v}_1^-+d\bm{v}_2^-)&=0.\label{eq3026}
\end{align}
\end{subequations}
Multiplying Eq.~\eqref{eq3024} by each of the vectors $\bm{v}_1^+$ and $\bm{v}_2^+$
from the left and using the relations in Eqs.~\eqref{eq3025} and~\eqref{eq3026}, we have the matrix equation
\begin{equation}
\begin{pmatrix}
\bm{v}_1^+\cdot T_\mu^{(0)}\bm{v}_1^+ & \bm{v}_1^+\cdot T_\mu^{(0)}\bm{v}_2^+ \\
\bm{v}_2^+\cdot T_\mu^{(0)}\bm{v}_1^+ & \bm{v}_2^+\cdot T_\mu^{(0)}\bm{v}_2^+
\end{pmatrix}
\begin{pmatrix}
a \\
b
\end{pmatrix}
=0.\label{eq3027}
\end{equation}
Note that 
$\bm{v}_1^+\cdot T_\mu^{(0)}\bm{v}_2^+=-\bm{v}_2^+\cdot T_\mu^{(0)}\bm{v}_1^+$.
For the coefficients $a$ and $b$ of the linear combination~\eqref{eq3021} to have a nontrivial solution, the determinant of the matrix in Eq.~\eqref{eq3027} must vanish:
\begin{equation}\label{eq3028}
\qty(\bm{v}_1^+\cdot T_\mu^{(0)}\bm{v}_1^+) \qty( \bm{v}_2^+\cdot T_\mu^{(0)}\bm{v}_2^+)=
\qty(\bm{v}_1^+\cdot T_\mu^{(0)}\bm{v}_2^+) \qty( \bm{v}_2^+\cdot T_\mu^{(0)}\bm{v}_1^+).
\end{equation}

For each of the four bound-state eigenvalues $\lambda_\mu$ of the time-evolution operator $\tilde{\hat{U}}_{\mathrm{K}}^{\mathrm{1w}}=\hat{S}^{\mathrm{1w}}\tilde{\hat{C}}^{\mathrm{1w}}$, we have the eigenvectors~\eqref{eq3019a}--\eqref{eq3019b} of the corresponding transfer matrix $T_\mu$ as well as the transfer matrix $T_\mu^{(0)}$ at the origin in Eq.~\eqref{eq3015}.
This makes Eq.~\eqref{eq3028} an equation of the bound-state solution $\lambda_\mu$ in turn.
We then explicitly have the eigenvalues $\zeta_\mu^\pm$ in Eq.~\eqref{eq3018} of the transfer matrix $T_\mu$ and the ratio of the coefficients $a/b$ from Eq.~\eqref{eq3027}, and then $c/d$ from Eq.~\eqref{eq3024}, which gives the eigenfunction $\hatPsi_\mu(x)$. This completes the procedure of finding bound-state eigenvalues and eigenfunctions in the case~\ref{casei-1}.

Equation~\eqref{eq3028} is reduced to a second-order polynomial of $\qty(\tilde{U}_\mu^{\mathrm{1w}})^2=\ee^{2\ii\lambda_\mu}$, which has the following two solutions:
\begin{equation}\label{eq3029}
\ee^{2\ii\lambda_\mu}=\frac{J^2+\cos{2\varphi}\mp\sqrt{2}\ii\sqrt{(1+2J^2+\cos{2\varphi})\sin^2{\varphi}}}{1+J^2}.
\end{equation}
Thus, the eigenvalues $\tilde{U}_\mu^{\mathrm{1w}}$ of the time-evolution operator $\tilde{\hat{U}}_{\mathrm{K}}^{\mathrm{1w}}$ are given as
\begin{align}\label{eq3030}
\ee^{\ii\lambda_\mu}%
&=
\sqrt{\frac{J^2+\cos{2\varphi}\mp\sqrt{2}\ii\sqrt{(1+2J^2+\cos{2\varphi})\sin^2{\varphi}}}{1+J^2}}\notag \\
&=
\frac{1}{\sqrt{1+J^2}}\qty[\sqrt{\frac{1+2J^2+\cos{2\varphi}}{2}}\mp\ii\sqrt{\sin^2{\varphi}}].
\end{align}
For the case~\ref{casei-1}, we chose only the two solutions with positive real parts, because they produce the eigenvectors with $\zeta_\mu^\pm>0$, as we will see below.

We have checked the consistency of the eigenvalues $\lambda_\mu$ of the time-evolution operator $\tilde{\hat{U}}_{\mathrm{K}}^{\mathrm{1w}}=\hat{S}^{\mathrm{1w}}\tilde{\hat{C}}^{\mathrm{1w}}$ obtained (i) based on Eq.~\eqref{eq3027} and (ii) from numerical diagonalization shown in Fig.~\ref{fig04}.
The solutions \eqref{eq3030}, which correspond to the plots in Fig.~\ref{fig04}, are listed in Table~\ref{tab01}.
\begin{table}[h]
\caption{Eigenvalues given in Eq.~\eqref{eq3030}, particularly in the range of $\cos\lambda_\mu>0$. 
We present the results of the calculation in double precision, which are accurate only to about 13 digits.
The parameter values are chosen to coincide with the plots in Figs.~\ref{fig04}(a)--(d) with $\varphi=\pi/10$, for which the continuum spectra range in $\pi/10\leq \lambda_\mu\leq9\pi/10$ and  $11\pi/10\leq \lambda_\mu\leq19\pi/10$.
We set the coupling parameter to the XX case as $J_x=J_y=J$ and $J_z=0$.
Note that in the case of $J=0$, four solutions are degenerate into two solutions, and that there are corresponding solutions in the range of $\cos\lambda_\mu<0$ for all cases.}

\label{tab01}
\centering
\begin{tabular}{rcl}
\hline
$J$ & $\sign\Im(\ee^{\ii\lambda_\mu})$ & Eigenvalues \\
\hline\hline
$0$ & $+$ & $0.951056516295 + 0.309016994375\,\ii$ \\
$0$ & $-$ & $0.951056516295 - 0.309016994375\,\ii$ \\
\hline
$1$ & $+$ & $0.975835154416 + 0.218508012224\,\ii$ \\
$1$ & $-$ & $0.975835154416 - 0.218508012224\,\ii$ \\
\hline
$3$ & $+$ & $0.995213971827 + 0.0977197537924\,\ii$ \\
$3$ & $-$ & $0.995213971827 - 0.0977197537924\,\ii$ \\
\hline
$20$ & $+$ & $0.999880926199 + 0.0154315722942\,\ii$ \\
$20$ & $-$ & $0.999880926199 - 0.015431572294\,\ii$ \\
\hline
\end{tabular}
\end{table}
In the case with $(J_x,J_y,J_z)=(0,0,0)$, the solutions are at the band edges $\ee^{\pm\ii\pi/10}\simeq0.9510565\pm0.3090170\,\ii$.
For the other cases, we find two solutions in the range of $\cos\lambda_\mu>\cos(\pi/10)$.

In the case~\ref{casei-2}, the following two points differ from the above.
First, the wave function grows from the left to the origin with $\zeta_1^-=\zeta_2^-<-1$.
Therefore, the eigenvectors $\bm{v}_\nu^+$ and $\bm{v}_\nu^-$ switch their roles in the argument in Theorem 1 and thereafter.
In Eq.~\eqref{eq3028} for bound-state eigenvalues, all $\bm{v}_\nu^+$ should be replaced with $\bm{v}_\nu^-$.
Second, since $\zeta_\mu^\pm$ are all negative, the amplitude of the bound-state eigenfunction alternates in sign as in $\sign \Psi_\mu(x)=(-1)^x$.
Finally, 
replacing $\bm{v}_\nu^+$ with $\bm{v}_\nu^-$ in Eq.~\eqref{eq3028} is  equivalent to replacing $\lambda_\mu$ with $-\lambda_\mu$ in Eq.~\eqref{eq3030}.
We indeed observe in Fig.~\ref{fig04} that to each of the two isolated eigenvalues with $\cos\lambda_\mu>0$, we have another isolated eigenvalue on the completely other side of the unit circle, which means $-\lambda_\mu$.

\subsection{Bound-state eigenfunctions}\label{sec3.3}
Let us now obtain the eigenvector corresponding to each eigenvalue of the transfer matrices $T_\mu$ and $T_\mu^{(0)}$.
We first find $\zeta_{1,2}^\pm$ explicitly.
Inserting a solution in Eq.~\eqref{eq3030}, which has a positive real part, to Eq.~\eqref{eq3018} yields
\begin{subequations}\label{eq3031}
\begin{align}
&\zeta_\mu^+\coloneqq\zeta_1^+=\zeta_2^+=%
\frac{1}{\cos{\varphi}\sqrt{1+J^2}}\qty[\sqrt{1+J^2-\sin^2{\varphi}}+\sqrt{J^2\sin^2{\varphi}}],\label{eq3031a} \\
&\zeta_\mu^-\coloneqq\zeta_1^-=\zeta_2^-=%
\frac{1}{\cos{\varphi}\sqrt{1+J^2}}\qty[\sqrt{1+J^2-\sin^2{\varphi}}-\sqrt{J^2\sin^2{\varphi}}],\label{eq3031b}
\end{align}
\end{subequations}
which are both positive.
Here, note that
\begin{align}
\frac{1}{\zeta_\mu^+}%
&=\cos{\varphi}\sqrt{1+J^2}\qty[\frac{1}{\sqrt{1+J^2-\sin^2{\varphi}}+\sqrt{J^2\sin^2{\varphi}}}] \notag \\
&=\cos{\varphi}\sqrt{1+J^2}%
\qty[\frac{\sqrt{1+J^2-\sin^2{\varphi}}-\sqrt{J^2\sin^2{\varphi}}}%
              {\cos^2{\varphi}(1+J^2)}] \notag \\
&=\zeta_\mu^-.\label{eq3032}
\end{align}
Note also that in the case of~\ref{casei-2}, we have two negative solutions $-\zeta_\mu^\pm<0$.

Continuing the case~\ref{casei-1}, we thereby obtain the eigenvector for $\zeta_\mu^+$ as
\begin{equation}\label{eq3033}
\bm{v}_1^+\propto
\begin{pmatrix}
\frac{-\ii\sqrt{\sin^2{\varphi}}-\sqrt{J^2\sin^2{\varphi}}}{\sin{\varphi}\sqrt{1+J^2}} \\
1 \\
0 \\
0
\end{pmatrix}
=
\begin{pmatrix}
\frac{-\ii-J}{\sqrt{1+J^2}} \\
1 \\
0 \\
0
\end{pmatrix}
\propto
\begin{pmatrix}
\sqrt{J+\ii} \\
-\sqrt{J-\ii} \\
0 \\
0
\end{pmatrix},
\end{equation}
where we used $J>0$ and $\sin{\varphi}>0$ for the equality.
We then normalize it as
\begin{equation}\label{eq3034}
\bm{v}_1^+=\frac{1}{\sqrt{2}\qty(J^2+1)^{1/4}}
\begin{pmatrix}
\sqrt{J+\ii} \\
-\sqrt{J-\ii} \\
0 \\
0
\end{pmatrix}.
\end{equation}
In the same manner, we obtain
\begin{align}
&\bm{v}_1^-=\frac{1}{\sqrt{2}\qty(J^2+1)^{1/4}}
\begin{pmatrix}
\sqrt{J-\ii} \\
\sqrt{J+\ii} \\
0 \\
0
\end{pmatrix},
\label{eq3035}\\
&\bm{v}_2^+=\frac{1}{\sqrt{2}\qty(J^2+1)^{1/4}}
\begin{pmatrix}
0 \\
0 \\
\sqrt{J+\ii} \\
-\sqrt{J-\ii}
\end{pmatrix},
\label{eq3036}\\
&\bm{v}_2^-=\frac{1}{\sqrt{2}\qty(J^2+1)^{1/4}}
\begin{pmatrix}
0 \\
0 \\
\sqrt{J-\ii} \\
\sqrt{J+\ii}
\end{pmatrix}.
\label{eq3037}
\end{align}

Now that we have the eigenvalues and the corresponding eigenvectors of the transfer matrix $T_\mu$, we write down $\hat{\Psi}(x)$ with linear combinations of them as
\begin{equation}\label{eq3038}
\hat{\Psi}(x)=
\begin{pmatrix}
\psi_{\mathrm{L}\uparrow}(x-1) \\
\psi_{\mathrm{R}\uparrow}(x) \\
\psi_{\mathrm{L}\downarrow}(x-1) \\
\psi_{\mathrm{R}\downarrow}(x)
\end{pmatrix}
=\begin{cases}
(\zeta_\mu^+)^xa_0\bm{v}_1^+ + (\zeta_\mu^+)^xb_0\bm{v}_2^+ & \mbox{for $x<0$}, \\
(\zeta_\mu^+)^{-x}c_0\bm{v}_1^- + (\zeta_\mu^+)^{-x}d_0\bm{v}_2^- & \mbox{for $x\ge0$},
\end{cases}
\end{equation}
with coefficients $a_0, b_0, c_0, d_0\in\mathbb{C}$ determined by the normalization condition, where we utilized the condition $\zeta_\mu^-=1/{\zeta_\mu^+}$.
Introducing an additional function as
\begin{equation}\label{eq3039}
\begin{cases}
a(x)=a_0(\zeta_\mu^+)^x,\quad b(x)=b_0(\zeta_\mu^+)^x & \mbox{for $x<0$}, \\
c(x)=c_0(\zeta_\mu^-)^x,\quad d(x)=d_0(\zeta_\mu^-)^x & \mbox{for $x>0$},
\end{cases}
\end{equation}
we rewrite Eq.~\eqref{eq3038} as
\begin{equation}\label{eq3040}
\hat{\Psi}(x)=
\begin{pmatrix}
\psi_{\mathrm{L}\uparrow}(x-1) \\
\psi_{\mathrm{R}\uparrow}(x) \\
\psi_{\mathrm{L}\downarrow}(x-1) \\
\psi_{\mathrm{R}\downarrow}(x)
\end{pmatrix}
=a(x)\bm{v}_1^+ + b(x)\bm{v}_2^+ + c(x)\bm{v}_1^- + d(x)\bm{v}_2^-.
\end{equation}
Thus, we have
\begin{align}
T_\mu\hat{\Psi}(x)=\hat{\Psi}(x+1)%
&=a(x+1)\bm{v}_1^+ + b(x+1)\bm{v}_2^+ + c(x+1)\bm{v}_1^- + d(x+1)\bm{v}_2^- \notag \\
&=a(x)\zeta_\mu^+\bm{v}_1^+ + b(x)\zeta_\mu^+\bm{v}_2^+ + c(x)\zeta_\mu^-\bm{v}_1^- + d(x)\zeta_\mu^-\bm{v}_2^-.\label{eq3041}
\end{align}
Therefore, we have the eigenstates of the time-evolution operator as
\begin{align}
&\begin{pmatrix}\psi_{\mathrm{L}\uparrow}(x) \\ \psi_{\mathrm{R}\uparrow}(x)\end{pmatrix}=%
\begin{cases}
a_0(\zeta_\mu^+)^x\begin{pmatrix}\zeta_\mu^+(\bm{v}_1^+)_1 \\ (\bm{v}_1^+)_2\end{pmatrix} & \mbox{for $x<0$}, \vspace{1ex}\\
c_0(\zeta_\mu^-)^x\begin{pmatrix}\zeta_\mu^-(\bm{v}_1^-)_1 \\ (\bm{v}_1^-)_2\end{pmatrix} & \mbox{for $x\ge0$},
\end{cases}\label{eq3042}\vspace{1ex}\\
&\begin{pmatrix}\psi_{\mathrm{L}\downarrow}(x) \\ \psi_{\mathrm{R}\downarrow}(x)\end{pmatrix}=%
\begin{cases}
b_0(\zeta_\mu^+)^x\begin{pmatrix}\zeta_\mu^+(\bm{v}_2^+)_1 \\ (\bm{v}_2^+)_2\end{pmatrix} & \mbox{for $x<0$}, \vspace{1ex}\\
d_0(\zeta_\mu^-)^x\begin{pmatrix}\zeta_\mu^-(\bm{v}_2^-)_1 \\ (\bm{v}_2^-)_2\end{pmatrix} & \mbox{for $x\ge0$}.
\end{cases}\label{eq3043}
\end{align}

\begin{figure}[h!]
\centering
\includegraphics[width=\textwidth]{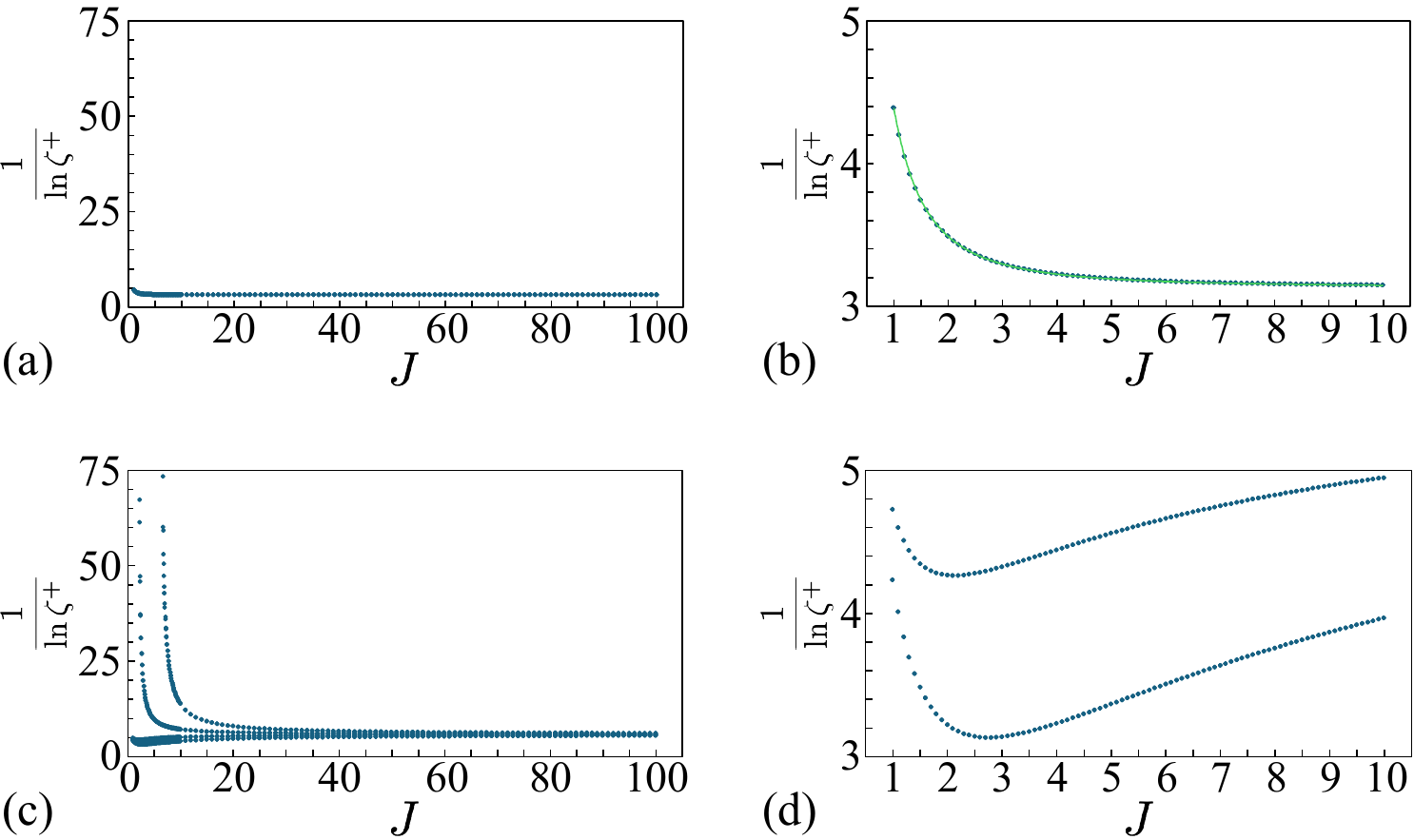}
\caption{Dependence of the localization length of bound states on the coupling strength in the XX case [(a), (b)] and the Heisenberg case [(c), (d)] obtained from numerical diagonalization. 
For panel~(b), we also show values obtained analytically using Eqs.~\eqref{eq3030} and~\eqref{eq3044} with the green curve.
For (a) and (b), we set the XX case $J_x=J_y=J$ and $J_z=0$. For (c) and (d), we set the $\textrm{SU}(2)$ case $J_x=J_y=J_z=J$. We focus only on sharp peaks on the right panels [(b), (d)].}
\label{fig08}
\end{figure}

From the above discussion, we know that the localization length of the bound states is given by
\begin{equation}\label{eq3044}
\frac{1}{\ln{\zeta_\mu^+}}=%
\frac{1}{\ln\qty[\qty(|\cos\lambda_\mu|+\sqrt{\cos^2\lambda_\mu-\cos^2\varphi})/\cos\varphi]}.
\end{equation}
In Fig.~\ref{fig08}, we plot the localization length of bound states obtained from numerical diagonalization of the time-evolution operator for different values of the coupling constant $J$. 
We first obtain the eigenvalues $\ee^{\ii\lambda_\mu}$ of the time-evolution operator for different $J$ from numerical diagonalization and then insert the values into Eq.~\eqref{eq3044}.
In the XX case [Fig.~\ref{fig08}(a) and (b)], we also show analytical values obtained by inserting Eq.~\eqref{eq3030} into Eq.~\eqref{eq3044} and confirm that they match the values obtained from numerical diagonalization.
We notice from Eq.~\eqref{eq3044} that the localization length becomes shorter as the eigenvalue approaches $\ee^{\ii\lambda_\mu}=\pm1$, because $|\cos\lambda_\mu|$ and $\cos^2\lambda_\mu$ become larger.
However, in the SU(2) Heisenberg case for which we cannot obtain eigenvalues analytically, the eigenvalues themselves do not behave monotonically under the variation of $J$, as shown in Sec.~\ref{sec2.3} and Fig.~\ref{fig04}.
This leads to non-monotonic behavior of the localization length in the Heisenberg case [Fig.~\ref{fig04}(c) and (d)]. 

\section{Numerical simulations of two walkers interacting with a magnetic impurity}\label{sec4}

In the present section, we finally
investigate the dynamics of two quantum walkers with an impurity spin at the origin, following the algorithm that we presented in Subsec.~\ref{sec4.1}.
We first simulate in Secs.~\ref{sec4.2.1} and~\ref{sec4.2.2} the collision dynamics starting from two types of initial conditions with the XX interaction~\eqref{eq-xx}.
We then simulate the collision dynamics in the case of the SU(2) Heisenberg interaction~\eqref{eq-su2}.
We find a slight precursor of the Kondo screening in our two-body model.

The following numerics are limited both by the memory space and the CPU time.
To calculate only the time evolution of the state, we only multiply the coin and shift operators, which is not memory- and time-consuming. 
To calculate the entanglement as in Figs.~\ref{fig10} and~\ref{fig14} below, however, we need to handle the density matrix.
To store the density of states of the two particles with an additional spin-$1/2$ impurity, we need a matrix of size $2(L_x)^2\times 2(L_x)^2$.
In most of the cases below, we set $L_x=201$, and hence we need more than $400$ GB of memory for a single matrix.
We then repeatedly apply the time-evolution operator to both sides of the density matrix. 
This also limits us to simulating the dynamics of the system of size $L_x=201$ mostly up to around $100$ steps,
even using a supercomputer of the Institute for Solid State Physics, the University of Tokyo.
Trying to go beyond the $100$ time steps is not fruitful, because the edges of the scattered wave functions wrap around the system under the periodic boundary conditions and interfere with each other, rendering the data useless.
In other words, making a longer simulation would be meaningful only if we could simulate a larger system.
To compute the entanglement negativity, we need additional memory, which largely limits us to a system size of $L_x=80$.

To put it short, the collision dynamics does not have any finite-size effects until the corresponding finite-time window. 
To run a longer simulation correctly, we would need a larger system, which is difficult due to the limitations of the current computer's memory space.

\subsection{Collision of delta functions in the XX case}\label{sec4.2.1}
We first study the dynamics starting with two delta peaks and numerically simulate their collision in the XX case.
We compare the dynamics with the different statistics of (a)~fermions, (b)~bosons, and (c)~distinguishable particles with the following initial states:
\begin{subequations}\label{eq4202}
\begin{align}
&\ket{\Psi_\mathrm{f0}}=%
\frac{1}{2}\left[\ket{x_1=0,x_2=x_0}\otimes\qty(\ket{\LL,\LL,\down}-\ket{\RR,\LL,\up}) 
-\ket{x_1=x_0,x_2=0}\otimes\qty(\ket{\LL,\LL,\down}-\ket{\LL,\RR,\up})\right], \label{eq4202a} \\
&\ket{\Psi_\mathrm{b0}}=%
\frac{1}{2}\left[\ket{x_1=0,x_2=x_0}\otimes\qty(\ket{\LL,\LL,\down}-\ket{\RR,\LL,\up}) 
+\ket{x_1=x_0,x_2=0}\qty(\ket{\LL,\LL,\down}-\ket{\LL,\RR,\up})\right],  \label{eq4202b} \\
&\ket{\Psi_\mathrm{d0}}=%
\frac{1}{\sqrt{2}}\ket{x_1=0,x_2=x_0}\otimes\qty(\ket{\LL,\LL,\down}-\ket{\RR,\LL,\up}), \label{eq4202c}
\end{align}
\end{subequations}
respectively, where we dropped the argument $(t=0; x_1, \sigma_1; x_2, \sigma_2; S_0)$ from the left-hand sides.
We will see that the time evolution of the probability distribution is identical for the fermions and bosons.
By contrast, we will see in the next subsection~\ref{sec4.2.2} that for the time evolution from a different type of initial condition, the difference between the fermions and the bosons emerges.

All the computation in this section was done with the parameter values
\begin{equation}\label{eq4201}
\varphi=\frac{\pi}{10},\quad J_x=J_y=J=3,\quad J_z=0,\quad x_0=21,
\end{equation}
and the system size $L_x=201$ with $-100\le x\le100$. We did not see any essential difference when we changed the value of $J$.

\begin{figure}[h!]
\centering
\includegraphics[width=\textwidth]{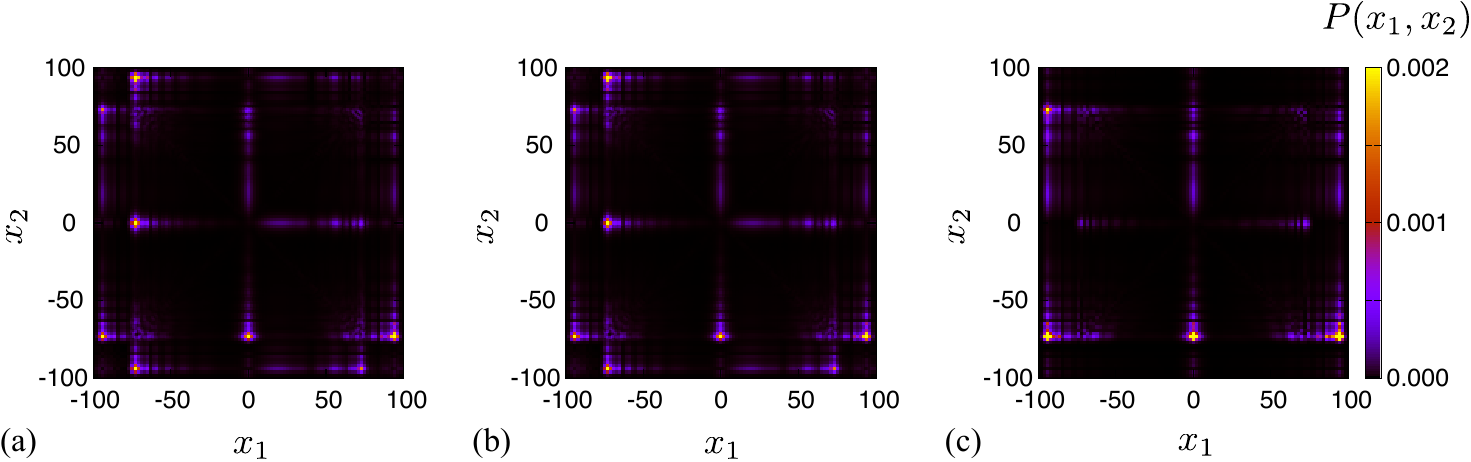}
\caption{
Color plots of the probability distribution $P(x_1,x_2)$ after $100$ time steps of evolution for different initial states of (a) identical fermions $\ket{\Psi_\mathrm{f0}}$, (b) identical bosons $\ket{\Psi_\mathrm{b0}}$, and (c) distinguishable particles $\ket{\Psi_\mathrm{d0}}$ with $J=3$. 
}
\label{fig09}
\end{figure}

Figure~\ref{fig09} exemplifies the probability distribution $P(x_1,x_2)$ after $100$ time steps of evolution from the initial condition of two delta peaks prepared at $(x_1, x_2)=(0, 21)$ and $(x_1, x_2)=(21, 0)$. 
\begin{table}[h]
\caption{Interpretation of the bright spots in Fig.~\ref{fig09}(a) and (b). We list only the spots in the region $x_1>x_2$. The spots in the region $x_1<x_2$ have the same interpretation under exchanging $x_1$ and $x_2$.}
\label{tab02}
\centering
\begin{tabular}{cll}
\hline
Peak position $(x_1,x_2)$ & Peak of $x_1$ & Peak of $x_2$ \\
\hline\hline
around $(0,-75)$ & bound to the impurity at the origin & transmitted wave from $x_0=21$ \\
\hline
around $(+75, 0)$ & reflected wave from $x_0=21$ & bound to the impurity at the origin \\
\hline
around $(-75,-95)$ & transmitted wave from $x_0=21$ & dissipative wave from the origin \\
\hline
around $(+95,-75)$ & dissipative wave from the origin & transmitted wave from the $x_0=21$ \\
\hline
around $(+75,-95)$ & reflected wave from the $x_0=21$ & dissipative wave from the origin \\
\hline
\end{tabular}
\end{table}
We find that the probability distributions of the cases of fermions and bosons [Fig.~\ref{fig09}(a), (b)] 
are equal to each other exactly, while we have checked that the phases of the wave functions are different by $\pi$.
This result is 
plausible because we have indirect interaction via the localized impurity described by the same scattering matrix~\eqref{eq2206} regardless of the statistics, only at the origin.
We list up in Table~\ref{tab02} the components that constitutes each peak in the fermionic and bosonic cases in Fig.~\ref{fig09}(a) and (b).
We can similarly understand the bright spots in Fig.~\ref{fig09}(c).


We next evaluate the entanglement between two quantum walkers before and after the collision.
It is not 
straightforward
to compute the entanglement between the two quantum walkers, because this is a three-body problem.
Tracing out the impurity-spin degree of freedom, we find the mixed state
\begin{equation}\label{eq4203}
\rho_{12}(x_1,x_2,\sigma_1,\sigma_2;y_1,y_2,\tau_1,\tau_2)%
=\sum_{S_0=\up,\down}\dyad{\Psi(x_1,\sigma_1;x_2,\sigma_2;S_0)}{\Psi(y_1,\tau_1;y_2,\tau_2;S_0)},
\end{equation}
where we dropped time $t$ from the argument for brevity.

We here use the entanglement negativity $\mathcal{N}(\rho_{12})$~\cite{Zyczkowski98,Zyczkowski99,Lee00,Eisert06,Vidal02,Horodecki09}, defined as 
\begin{equation}\label{eq4204}
\mathcal{N}(\rho_{12})\coloneqq\sum_{\lambda_i<0}\qty|\lambda_i|,
\end{equation}
where $\{\lambda_i\}$ are all of the eigenvalues of the partially transposed density matrix ${\rho_{12}}^{\Gamma_1}$ defined by
\begin{equation}\label{eq4205}
{\rho_{12}}^{\Gamma_1}(x_1,x_2,\sigma_1,\sigma_2;y_1,y_2,\tau_1,\tau_2)\coloneqq%
\rho_{12}(y_1,x_2,\tau_1,\sigma_2;x_1,y_2,\sigma_1,\tau_2).
\end{equation}
This is an entanglement monotone, and the two walkers are entangled whenever $\mathcal{N}(\rho_{12})>0$.
%
Due to limited computational resources, we employ
a smaller system size $L_x=81$ than in $L_x=201$ in Fig.~\ref{fig09} with $-40\le x\le40$ and $x_0=20$.

\begin{figure}[h!]
\centering
\includegraphics[width=0.6\textwidth]{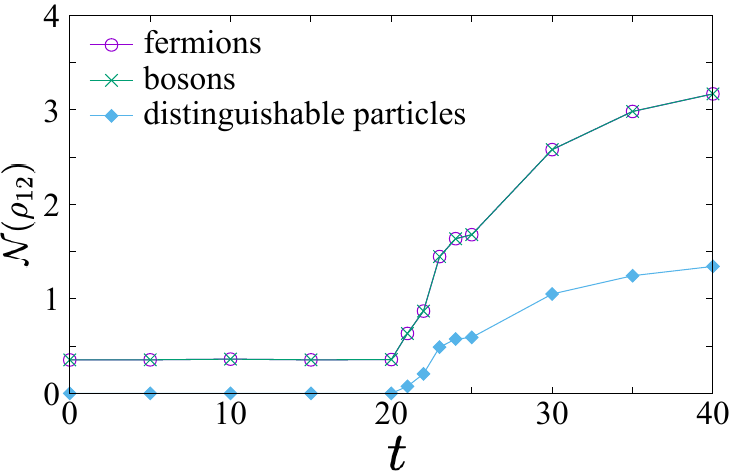}
\caption{
Entanglement negativity $\mathcal{N}$ versus time step with fermions (purple circles), bosons (green crosses), and distinguishable particles (blue diamonds). We set $J=3$.}
\label{fig10}
\end{figure}

Figure~\ref{fig10} shows the time evolution of the entanglement negativity $\mathcal{N}$ with $J=3$.
We observe a sudden increase in the entanglement negativity when one of the quantum walkers passes through the origin, from $t=22$ to $t=23$ for all three cases.
This indicates that the two initial delta functions are strongly entangled upon collision and confirms that the two walkers indirectly interact via the magnetic impurity under the Kondo-type interaction in Eq.~\eqref{eq4101}.
Note that the small amount of entanglement negativity at $t=0$ in the cases of fermions and bosons is due to the anti-symmetrization and the symmetrization of the initial states~\eqref{eq4202a} and~\eqref{eq4202b}, respectively.
Furthermore, we notice that the entanglement negativity of fermions and bosons at each time step of consideration takes exactly the same values, which is consistent with the exact matching of the probability distribution shown in Fig.~\ref{fig09}(a) and~(b).


\subsection{Collision of a bound state and a delta function in the XX case}\label{sec4.2.2}
\begin{figure}
\centering
\begin{subfigure}[c]{0.31\textwidth}
\includegraphics[width=\textwidth]{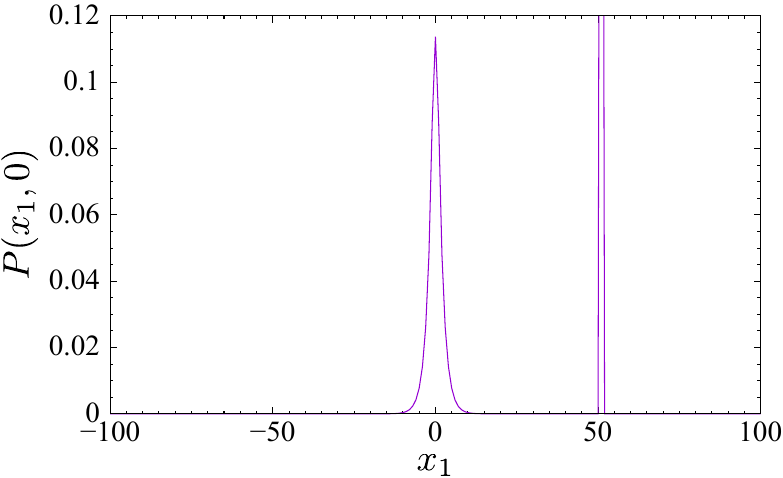}
\vspace{-1.5\baselineskip}
\subcaption{$t=0$: initial state}
\end{subfigure}
\\
\begin{subfigure}[c]{0.31\textwidth}
\includegraphics[width=\textwidth]{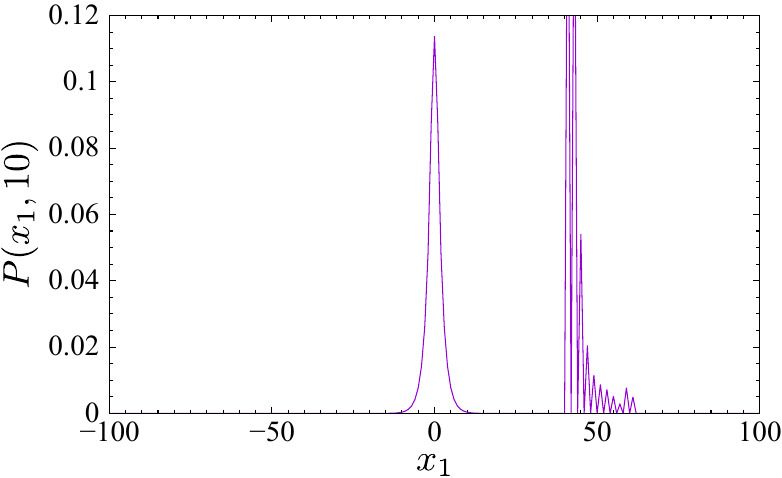}
\vspace{-1.5\baselineskip}
\subcaption{$t=10$}
\end{subfigure}
\hspace{0.1\textwidth}
\begin{subfigure}[c]{0.31\textwidth}
\includegraphics[width=\textwidth]{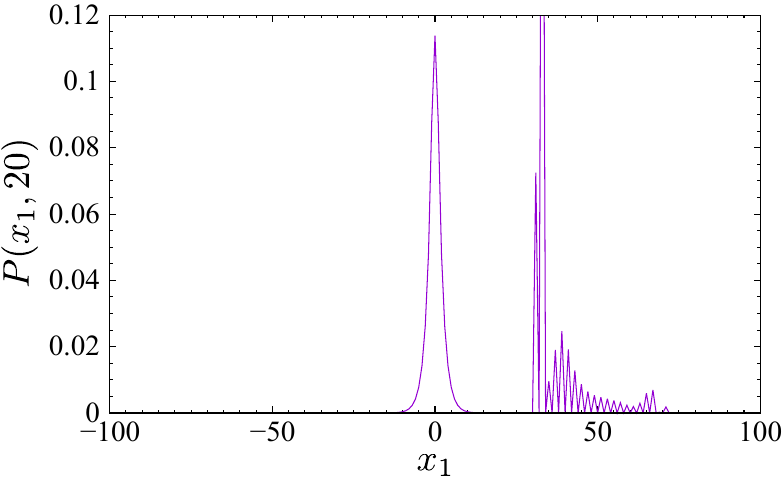}
\vspace{-1.5\baselineskip}
\subcaption{$t=20$}
\end{subfigure}
\\
\vspace{0.5\baselineskip}
\begin{subfigure}[c]{0.31\textwidth}
\includegraphics[width=\textwidth]{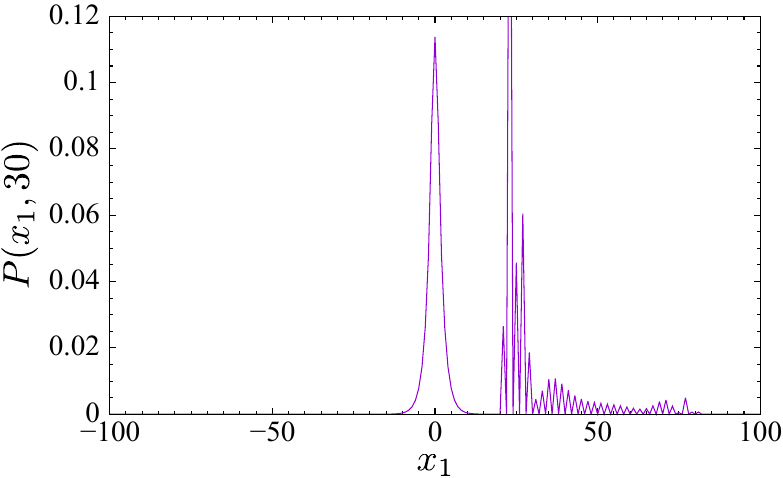}
\vspace{-1.5\baselineskip}
\subcaption{$t=30$}
\end{subfigure}
\hspace{0.1\textwidth}
\begin{subfigure}[c]{0.31\textwidth}
\includegraphics[width=\textwidth]{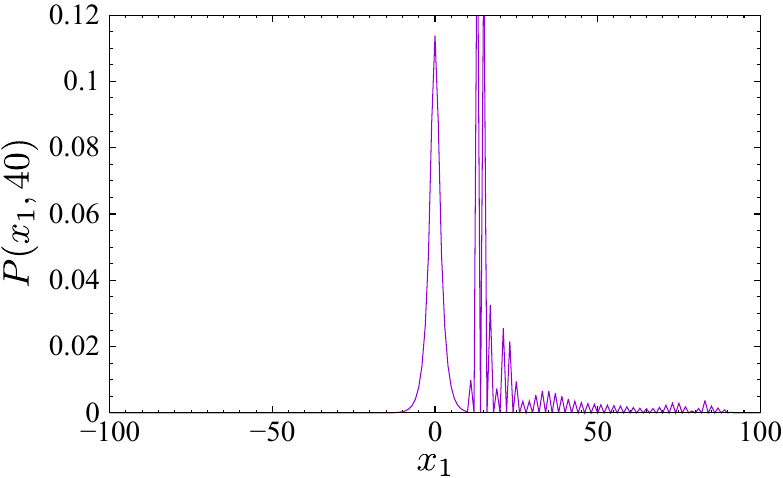}
\vspace{-1.5\baselineskip}
\subcaption{$t=40$}
\end{subfigure}
\\
\vspace{0.5\baselineskip}
\begin{subfigure}[c]{0.31\textwidth}
\includegraphics[width=\textwidth]{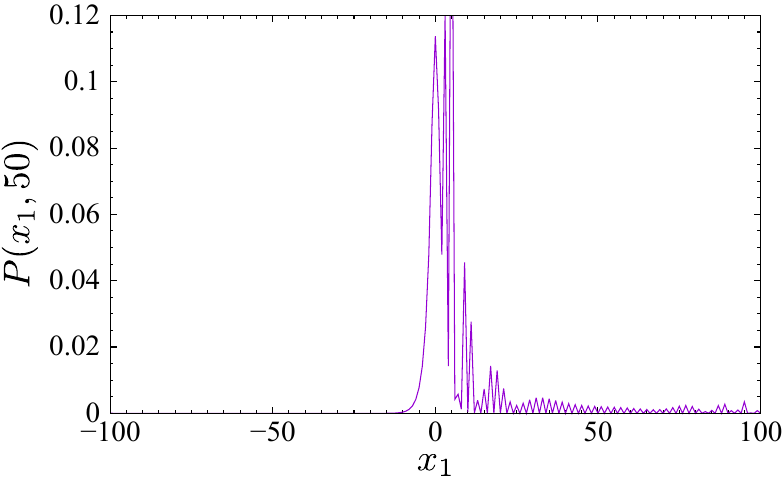}
\vspace{-1.5\baselineskip}
\subcaption{$t=50$}
\end{subfigure}
\hspace{0.1\textwidth}
\begin{subfigure}[c]{0.31\textwidth}
\includegraphics[width=\textwidth]{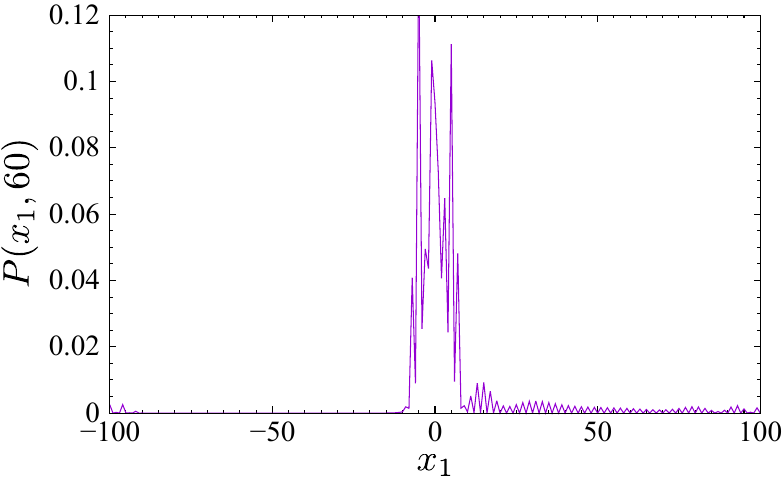}
\vspace{-1.5\baselineskip}
\subcaption{$t=60$}
\end{subfigure}
\\
\vspace{0.5\baselineskip}
\begin{subfigure}[c]{0.31\textwidth}
\includegraphics[width=\textwidth]{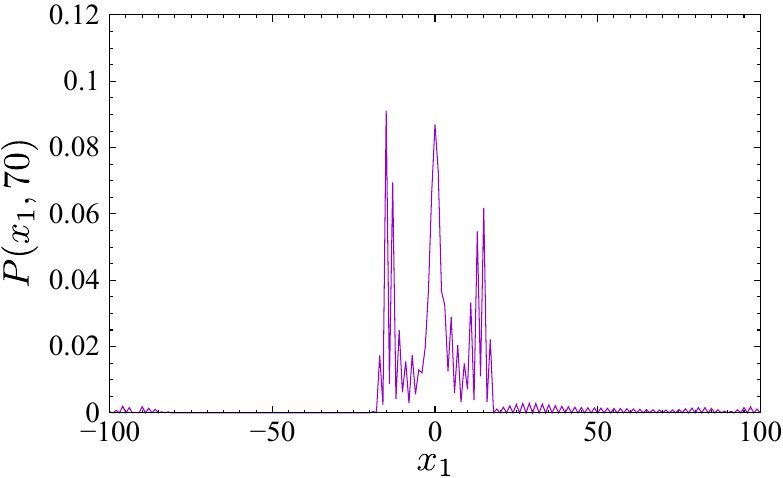}
\vspace{-1.5\baselineskip}
\subcaption{$t=70$}
\end{subfigure}
\hspace{0.1\textwidth}
\begin{subfigure}[c]{0.31\textwidth}
\includegraphics[width=\textwidth]{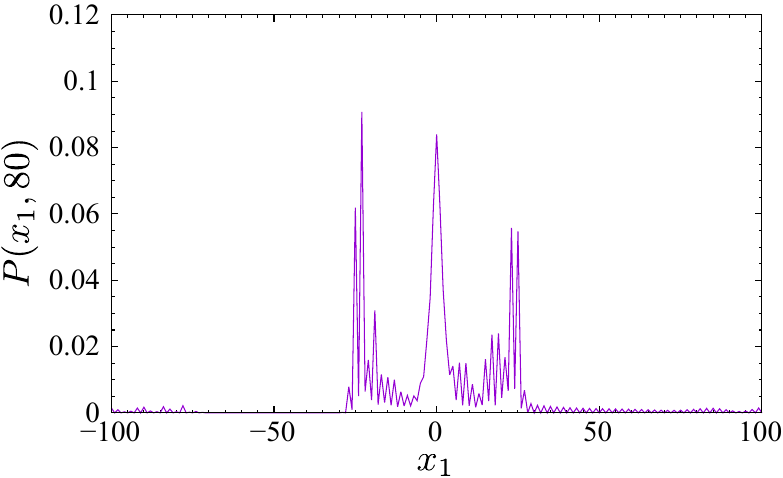}
\vspace{-\baselineskip}
\subcaption{$t=80$}
\end{subfigure}
\\
\vspace{0.5\baselineskip}
\begin{subfigure}[c]{0.31\textwidth}
\includegraphics[width=\textwidth]{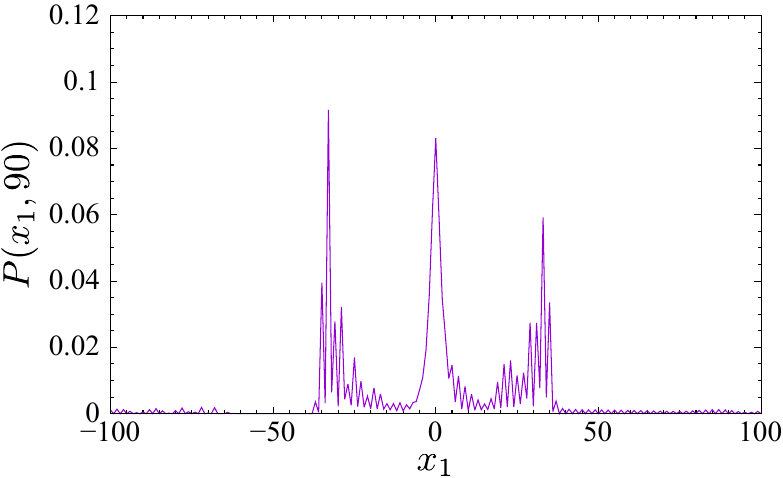}
\vspace{-1.5\baselineskip}
\subcaption{$t=90$}
\end{subfigure}
\hspace{0.1\textwidth}
\begin{subfigure}[c]{0.31\textwidth}
\includegraphics[width=\textwidth]{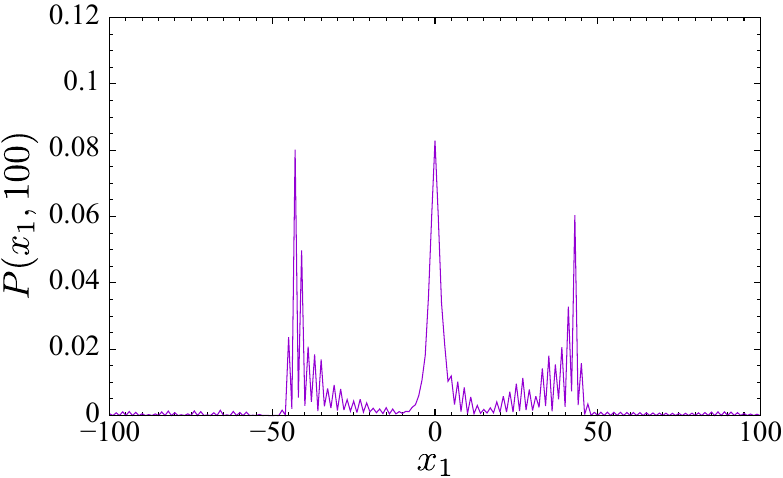}
\vspace{-1.5\baselineskip}
\subcaption{$t=100$}
\end{subfigure}
\caption{Time evolution of the probability distribution $P(x_1,t)$ for the two fermions from $t=0$ to $t=100$ every 10 time steps. We set the XX case $J_x=J_y=3, J_z=0$ with $x_0=51$.}
\label{fig11}
\end{figure}

We next simulate the time evolution of the initial state, as shown in Fig.~\ref{fig11}(a).
One of the quantum walkers is in a bound eigenstate around the magnetic impurity spin at the origin, which we analytically obtained in Sec.~\ref{sec3}, whereas the other quantum walker is in the left-going state of a delta peak.
As shown in Fig.~\ref{fig04}, there appear four isolated eigenvalues in the case with chiral symmetry for any values of the coupling strength $J$.
We use the bound eigenstate of the eigenvalue in the first quadrant for the initial state.
We have confirmed that the dynamics is completely the same for all four bound eigenstates, although they have different amplitudes of singlet and triplet components at the origin.
The computation here was done with the parameter values
\begin{equation}\label{eq4208}
\varphi=\frac{\pi}{10},\quad x_0=51,
\end{equation}
and the system size $L_x=201$ with $-100\le x\le100$.

We first consider the case of fermions:
\begin{align}
&\ket{\Psi_\mathrm{f0}}=%
\frac{1}{\sqrt{2}}\left[\ket{x_1=0,x_2=x_0}\otimes\ket{\mathrm{b.s.}}_{1,S}\ket{\sigma_2=\LL} 
-\ket{x_1=x_0,x_2=0}\otimes\ket{\mathrm{b.s.}}_{2,S}\ket{\sigma_1=\LL}\right], \label{eq4209}
\end{align}
where $\ket{\mathrm{b.s.}}_{\mu,S}$ with $\mu=1,2$ denotes the bound state of the $\mu$th walker.
Figure~\ref{fig11} exemplifies the subsequent time evolution of the probability distribution.
As we see here, a component that goes to the right from the delta peak on the right comes out from the left edge because of the periodic boundary condition, and almost interferes with the transmission wave to the left.
We therefore stop the time evolution here.

\begin{figure}
\centering
\includegraphics[width=\textwidth]{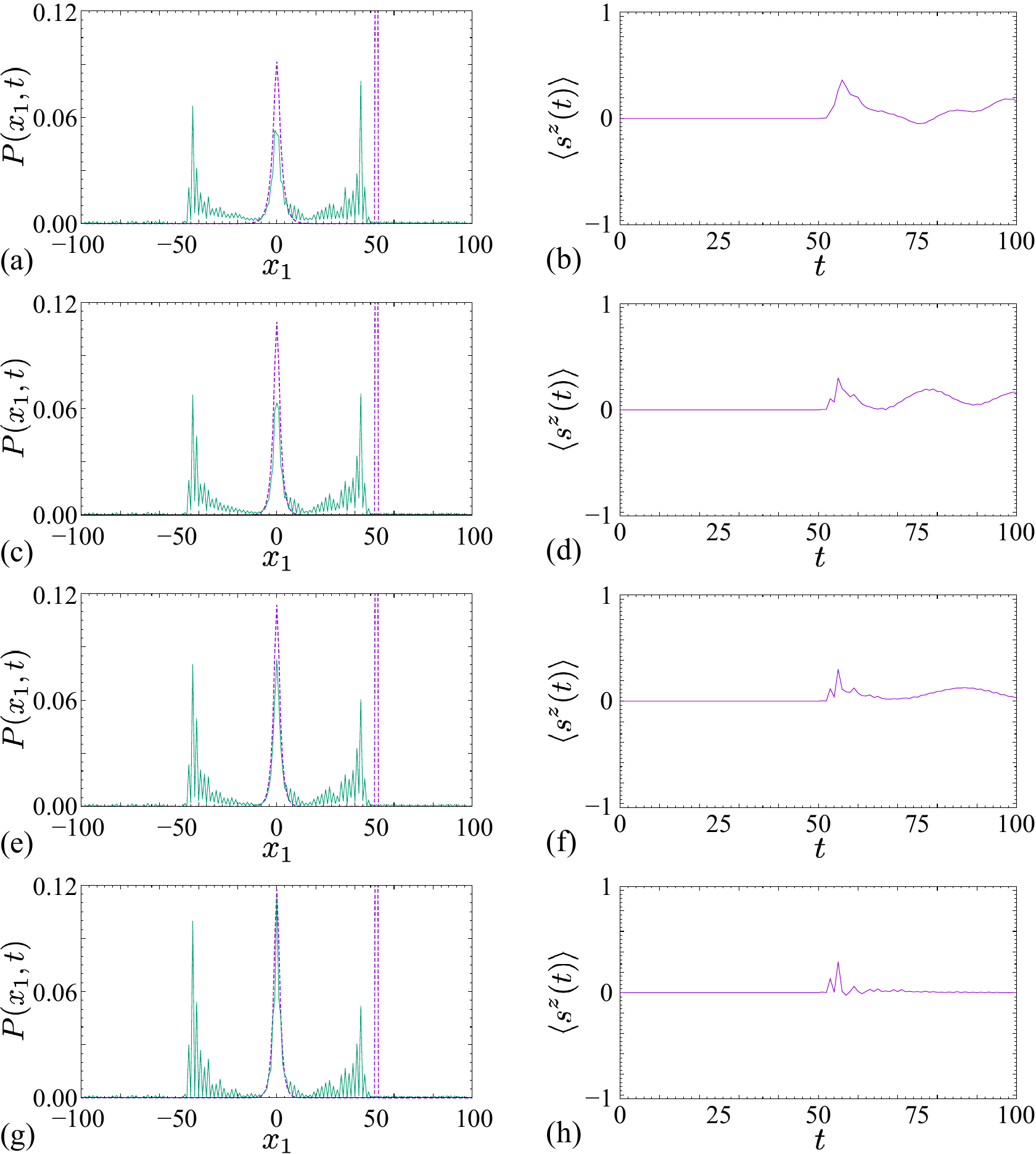}
\caption{
Probability distribution [(a), (c), (e), (g)] after $100$ time steps of evolution (solid green curve) with the fermionic initial state defined by Eq.~\eqref{eq4209} (dashed purple curve) and expectation value of $s_z$ versus time step [(b), (d), (f), (h)] in the XX case $J_x=J_y=J$ and $J_z=0$. We set $J=1$ [(a), (b)], $J=2$ [(c), (d)], $J=3$ [(e), (f)], and $J=10$ [(g), (h)].}
\label{fig12}
\end{figure}

Figure~\ref{fig12} shows the probability distribution after $100$ time steps of evolution together with the initial state and the time dependence of the expectation values of $s_z$ for four values of the coupling strength $J$.
We see in Fig.~\ref{fig12} that, as we naturally expect, the initial bound state is less perturbed by the other walker coming from the right as we increase the coupling strength $J$, and as the particle at the origin is bound to the impurity more strongly.
See also Fig.~\ref{fig08} for the dependence of the localization length of bound states on the coupling strength. 
The bound state becomes sharper as we increase $J$.
In other words, a more strongly localized state lets the other particle pass it more easily.

\begin{figure}[h!]
\centering
\includegraphics[width=0.8\textwidth]{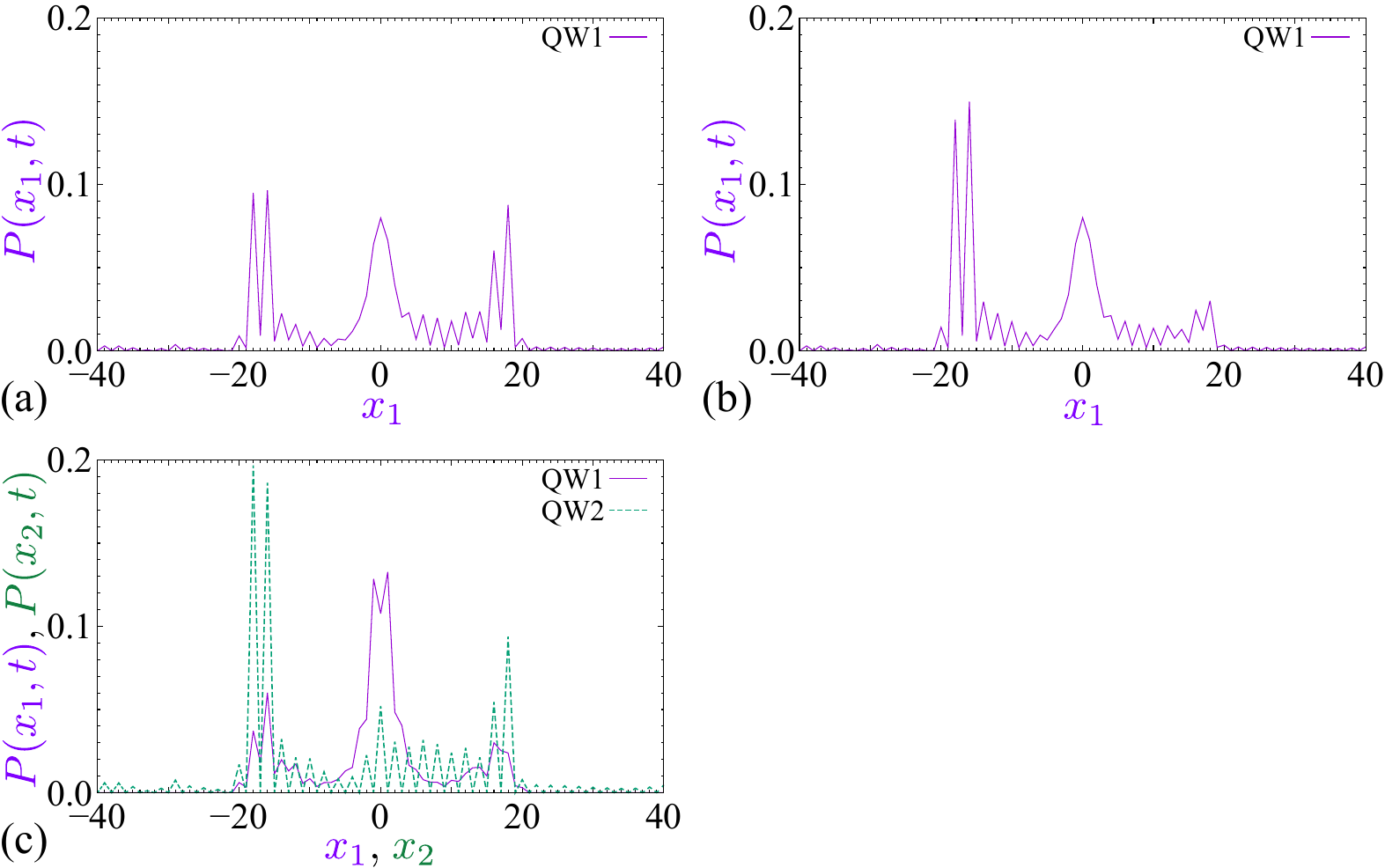}
\caption{
Probability distribution after $40$ time steps of evolution with (a)~fermions, (b)~bosons, and (c)~distinguishable particles. We set $J=3$, $L_x=81$, and $x_0=20$. In panel (c), the probability distributions of both quantum walkers, $P(x_1)$ (solid purple line) and $P(x_2)$ (dashed green line), are plotted, and hence the total probability sums up to $2$.}
\label{fig13}
\end{figure}

We then compare the dynamics with the different statistics of (a)~fermions, (b)~bosons, and (c)~distinguishable particles with the following initial states:
\begin{subequations}\label{eq4210}
\begin{align}
&\ket{\Psi_\mathrm{f0}}=%
\frac{1}{\sqrt{2}}\left[\ket{x_1=0,x_2=x_0}\otimes\ket{\mathrm{b.s.}}_{1,S}\ket{\sigma_2=\LL} 
-\ket{x_1=x_0,x_2=0}\otimes\ket{\mathrm{b.s.}}_{2,S}\ket{\sigma_1=\LL}\right],\label{eq4210a} \\
&\ket{\Psi_\mathrm{b0}}=%
\frac{1}{\sqrt{2}}\left[\ket{x_1=0,x_2=x_0}\otimes\ket{\mathrm{b.s.}}_{1,S}\ket{\sigma_2=\LL} 
+\ket{x_1=x_0,x_2=0}\otimes\ket{\mathrm{b.s.}}_{2,S}\ket{\sigma_1=\LL}\right],\label{eq4210b} \\
&\ket{\Psi_\mathrm{d0}}=%
\left[\ket{x_1=0,x_2=x_0}\otimes\ket{\mathrm{b.s.}}_{1,S}\ket{\sigma_2=\LL}\right],\label{eq4210c}
\end{align}
\end{subequations}
respectively.
Note that Eq.~\eqref{eq4210a} is identical to Eq.~\eqref{eq4209}.
Figure~\ref{fig13} shows the probability distribution after $40$ time steps of the evolution in each case for the coupling strength $J=3$.

\begin{figure}[h!]
\centering
\includegraphics[width=0.6\textwidth]{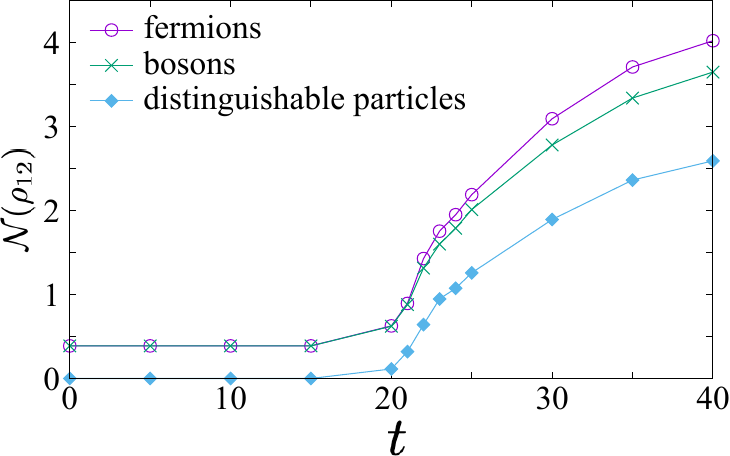}
\caption{
Entanglement negativity $\mathcal{N}$ versus time step with fermions (purple circles), bosons (green crosses), and distinguishable particles (blue diamonds). We set $J=3$.}
\label{fig14}
\end{figure}

A difference emerges between the cases of fermions and bosons.
We observe a markedly smaller transmission and greater reflection in the case of fermions [Fig.~\ref{fig13}(a)] than in the case of bosons [Fig.~\ref{fig13}(b)].
Note that there was no difference between the two statistics when we let the two delta peaks collide with each other; see Fig.~\ref{fig09} in Subsec.~\ref{sec4.2.1}.
This difference
may be related to the difference in the statistics in the entanglement negativity. 
Figure~\ref{fig14} shows the time development of the entanglement negativity. 
While the delta function coming from the right goes through the bound state ($15\lesssim t \lesssim 30$), especially after the delta function goes through the origin ($t\simeq 22$), we observe a drastic enhancement of the entanglement negativity, which indicates that the two quantum walkers are entangled with each other after the collision.
The nonzero entanglement negativity before the collision in the cases of fermions and bosons comes from antisymmetrization and symmetrization of the initial states~\eqref{eq4210a} and~\eqref{eq4210b}, respectively. 
In the case of distinguishable particles,
the entanglement negativity remains zero before the collision.

We saw in Fig.~\ref{fig09} that the probability distributions for fermions and bosons are equal to each other when we let two delta functions collide with each other. We also saw in Fig.~\ref{fig10} that the entanglement of negativities of fermions and bosons are equal to each other. In the present case of the collision between a bound state and a delta function, in contrast, the difference emerges between the cases of fermions and bosons both in the time evolution of the probability, as in Fig.~\ref{fig13}, and in the time evolution of the entanglement negativity, as in Fig.~\ref{fig14}. The entanglement negativity for fermions is larger than that for bosons.
%
The comparison between Figs.~\ref{fig13} and~\ref{fig14} may imply that the smaller transmission in the case of fermions is caused by a higher entanglement negativity of fermions. 
Pursuing the reason 
is beyond the scope of the present paper.

\subsection{Collision of a bound state and a delta function in the Heisenberg case}\label{sec4.2.3}

\begin{figure}
\centering
\includegraphics[width=\textwidth]{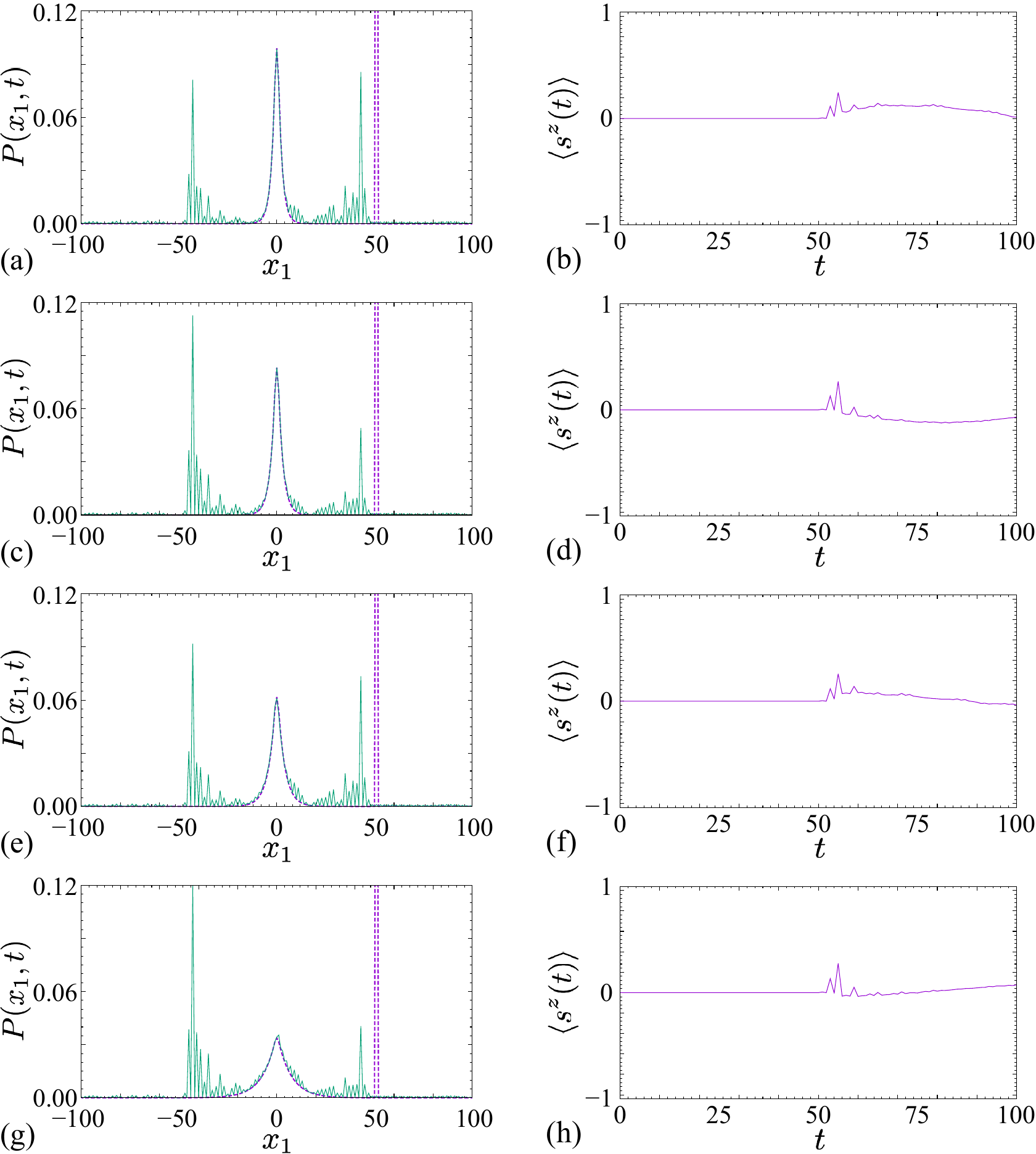}
\caption{
Probability distributions [(a), (c), (e), (g)] after $100$ time steps of evolution (solid green curve) with the fermionic initial state defined by Eq.~\eqref{eq4209} (dashed purple curve) and expectation value of $s_z$ versus time step [(b), (d), (f), (h)] for the $\textrm{SU}(2)$ case $J_x=J_y=J_z=J$ and $J=10$. For each row, we start from a different eigenstate. The second row [(c), (d)] corresponds to the eigenstate with more of a singlet-state component than a triplet-state component, and the other three rows are opposite.}
\label{fig15}
\end{figure}

Lastly, we 
study the collision dynamics with the SU(2) Heisenberg interaction $J_x=J_y=J_z=J$.
Figure~\ref{fig15} shows the probability distributions after $100$ time steps of evolution together with the initial state and the time dependence of the expectation values of $s_z$, for different initial bound states.
We here assume the fermion statistics~\eqref{eq4209}.
In contrast to the XX case in Fig.~\ref{fig12}, the dynamics after the collision looks different depending on which eigenstate we start with.
In the XX case, the four bound states are connected to symmetry, and hence we did not have any options of starting from essentially different initial states, while it is possible in the present Heisenberg case. 
We here compare the dynamics from different initial bound states for the specific value of $J=10$ in Fig.~\ref{fig15}.

We have confirmed that the dynamics is identical for each pair of eigenstates with symmetric eigenvalues with respect to the origin, and hence we consider only eigenstates with positive imaginary parts.
We characterize each bound state, focusing on how much the quantum walker is likely to 
form a singlet state with the localized magnetic impurity at the origin; 
see Table~\ref{tab03}.
Here, we define the singlet state of the bound quantum walker and the impurity spin as
\begin{equation}\label{eq4206}
\frac{1}{\sqrt{2}}\ket{x=0}\otimes\qty(\ket{\LL,\down}-\ket{\RR,\up})
\end{equation}
and their triplet states as
\begin{subequations}\label{eq4207}
\begin{align}
&\ket{x=0}\otimes\ket{\LL,\up} \label{eq4207a} \\
&\frac{1}{\sqrt{2}}\ket{x=0}\otimes\qty(\ket{\LL,\down}+\ket{\RR,\up}) \label{eq4207b} \\
&\ket{x=0}\otimes\ket{\RR,\down}.
\end{align}
\end{subequations}
Since we define the singlet and triplet states of the bound quantum walker and the magnetic impurity as in Eqs.~\eqref{eq4206} and~\eqref{eq4207}, we claim that the second bound state in Table~\ref{tab03} is the closest to the singlet state.

\begin{table}[h!]
\begin{center}
\caption{Amplitudes of the wave-function components at $x=0$ of bound states used in Fig.~\ref{fig15}. We list values significantly greater than the machine precision.}
\label{tab03}
\begin{tabular}{c|cc|c}
\hline
row in Fig.~\ref{fig15} & QW & spin & wave function component at $x=0$ \\
\hline\hline
1 & $\LL$ & $\up$ & $+\,0.2225226688\hspace{0.185cm}$\\
 & $\RR$ & $\up$ & $+\,0.2225226688\,\ii$\\
 & $\LL$ & $\down$ & $+\,0.2225226688\,\ii$\\
 & $\RR$ & $\down$ & $-\,0.2225226688\hspace{0.185cm}$\\
\hline
2 & $\LL$ & $\up$ & $-\,0.1950927928\,\ii$\\
 & $\RR$ & $\up$ & $-\,0.2123314659\hspace{0.185cm}$ \\
 & $\LL$ & $\down$ & $+\,0.2123314659\hspace{0.185cm}$\\
 & $\RR$ & $\down$ & $-\,0.1950927928\,\ii$ \\
\hline
3 & $\LL$ & $\up$ & $+\,0.1830123562\hspace{0.185cm}$\\
 & $\RR$ & $\up$ & $+\,0.1681540299\,\ii$\\
 & $\LL$ & $\down$ & $-\,0.1681540299\,\ii$ \\
 & $\RR$ & $\down$ & $+\,0.1830123562\hspace{0.185cm}$\\
\hline
4 & $\LL$ & $\up$ & $+\,0.1299180682\,\ii$\\
 & $\RR$ & $\up$ & $+\,0.1299180682\hspace{0.185cm}$\\
 & $\LL$ & $\down$ & $+\,0.1299180682\hspace{0.185cm}$\\
 & $\RR$ & $\down$ & $-\,0.1299180682\,\ii$\\
\hline
\end{tabular}
\end{center}
\end{table}

Figure~\ref{fig15} shows the collision dynamics in the order of peak width from the top row to the bottom one;
in the top row, we used the sharpest bound state for the initial state, while in the bottom row we used the widest one.
We notice that the bound state in the second row is least perturbed by the other walker coming from the right, among the top three cases.
Table~\ref{tab03} then shows the amplitudes of the wave-function components of the corresponding four bound states at the origin in the same order as in Fig.~\ref{fig15}, giving more details than Fig.~\ref{fig15}, which shows the square modulus of the wave function.
We see in Table~\ref{tab03} that the bound state on the second row, which seems most protected and screened from the collision of the other walker in Fig.~\ref{fig15}, has a more singlet-state amplitude than a triplet-state amplitude at the origin.
In short, the state closest to the singlet state is least perturbed, seemingly transparent to the other walker.

This is
in contrast to our observation in Sec.~\ref{sec4.2.2}, where the dynamics did not depend on the amplitude of the singlet component~\eqref{eq4207}.
This is the essential difference between the XX and Heisenberg cases.
We have confirmed up to $J=10$ this tendency in which the second sharpest bound state is the closest to the singlet and least perturbed.
Note that this tendency does not apply to the fourth state with the widest peaks, which appears only for $J\ge7$. 
This may be because the initial bound state is too wide and easy to penetrate.

Consequently, we speculate that we might have seen Kondo screening-like behavior in its lowest level of two-body case.
While the original Kondo effect is a truly many-body effect, it is known to have an ``onion-skin" structure that consists of many levels of real-space renormalization-group procedure. 
As in Refs.~\cite{Mitchell11,Matsueda12}, the Kondo screening of the magnetic impurity may be described by a step-by-step real-space renormalization of a nearby electronic state into the singlet state that screens the central impurity at the origin.
In this sense, our numerical result of a two-body case may imply Kondo screening, as the very first step of the renormalization-group procedure toward many-body screening.

Note, however, that the pure singlet state cannot be the eigenstate of the Hamiltonian~\eqref{eq2201}, as we mentioned at the end of Sec.~\ref{sec3}.
This is because the model~\eqref{eq4101} 
includes an additional interaction beyond the two types.
As we have emphasized, the interaction between each quantum walker and the impurity, $\Hm$, induces a Kondo-type interaction.
In addition, 
we also have the spin-orbit coupling $\qty(p_1\sigma_1^z\otimes\sigma_2^0+p_2\sigma_1^0\otimes\sigma_2^z)\otimes s^0$ for the walkers.
In this sense, our model~\eqref{eq4101} has a new feature of the Kondo-type interaction with a spin-orbit coupling.
While it may exhibit 
a
more drastic feature in two spatial dimensions, it is beyond the scope of the present paper.

\section{Summary}\label{secsummary}

In the present paper, we introduced a Kondo-like interaction between quantum walkers.
Quantum walkers indirectly interact with each other via a localized magnetic impurity at the origin.
We first interpreted the dynamics of a quantum walker as those of a massless Dirac particle propagating between a series of delta potentials, and then added the interaction with the localized magnetic impurity based on this interpretation.
We then analytically obtained scattering matrices of two Dirac particles with a magnetic impurity.

To demonstrate the utility of our model with interactions, we analytically studied the bound state of a single quantum walker and numerically studied the collision dynamics of two quantum walkers.
For a single quantum walker and a single magnetic impurity, we analytically obtained the bound-state eigenvalues and eigenfunctions. 
To analytically solve the model, we set the interaction to the XX case, $J_x=J_y=J$ and $J_z=0$, to simplify the eigenvalue equation. 

In the three-body case of two quantum walkers and one magnetic impurity, we studied the collision dynamics of the two walkers.
As the initial state, we used three cases, two with the XX interaction and one with the Heisenberg interaction.
For cases with the XX interaction, we first tried an initial state consisting of two delta functions: one at the origin with the magnetic impurity and the other on its right.
Upon collision, we observed a dramatic increase in the entanglement negativity between the two walkers, underscoring the indirect interaction between the walkers via the magnetic impurity under the Kondo-like interaction in our model.

In the case of the collision of two delta functions, we did not observe any difference between fermionic and bosonic walkers, but observed the difference 
for the initial condition that consists of a bound eigenstate 
and a delta function. 
One of the quantum walkers is bound by the magnetic impurity spin at the origin, whereas the other quantum walker, starting in the left-going state of a delta peak, collides with the bound state from the right.
For this initial condition, the second walker transmits less in the case of fermions than in the case of bosons.
This difference in the transmission is consistent with the difference in the entanglement.
The fermionic walkers are more strongly entangled with each other than the bosonic walkers after the collision.

We finally analyzed the dynamics in the $\textrm{SU}(2)$ Heisenberg case $J_x=J_y=J_z=J$.
We used the same initial state as in the second case of the XX case, but observed markedly different behavior.
In contrast to the XX case, 
the bound state is hardly perturbed by the other walker coming from the right when the bound state is more likely to form a singlet state with the localized magnetic impurity.
We speculate that this is the revelation of the Kondo screening in its lowest level.
While the full Kondo screening is truly a many-body phenomenon, it emerges in infinitely many levels of renormalization, and 
what we observed 
can be
the starting point of the renormalization-group procedure.
The Kondo physics emerges not in the XX case but in the $\textrm{SU}(2)$ Heisenberg case, as in the original Kondo model~\eqref{eq1004}.

We finally mention a basic difference between the original Kondo model~\eqref{eq1004} and our model~\eqref{eq4101}.
The latter not only has the interaction with each quantum walker and the impurity, $H_m$, which induces the spin-spin interaction between quantum walkers and the localized magnetic impurity, but also spin-orbit coupling 
for the walkers.
Thus, it will be an interesting problem to extend 
our model to two dimensions, comparing with Rashba spin-orbit coupling and Dresselhaus spin-orbit coupling.

\section*{Acknowledgments}\label{Acknowledgments}
The authors thank Maximilian H\"{u}nenberger, Ken-Ichiro Imura, Kensuke Kobayashi, Hiroyasu Tajima, Kazuaki Takasan, and Shion Yamashika for discussions.
The computations in this work were performed partly using the facilities of the Supercomputer Center, the Institute for Solid State Physics, the University of Tokyo.
This work was also supported by JSPS KAKENHI [Grant Numbers JP24KJ0655 (M.Y.), JP24K00545 (N.H.~and H.O.), JP23K22411 (H.O.~and N.H.), JP24H00945, JP26H02015, JP26K06970, JP26K17046 (K.K.), and JP22KJ1408 (C.K.)].
M.Y.~was supported by the RIKEN Junior Research Associate Program.
K.K.~was supported in part by JST [via the FOREST program (Grant Number JPMJFR256P)].
C.K.~was supported in part by JST [via the PRESTO program (Grant Number JPMJPR25F1) and the ASPIRE program (Grant Number JPMJAP2319)].
F.N.~was supported in part by JST [via the CREST Quantum Frontiers program (Grant Number JPMJCR24I2), the Quantum Leap Flagship Program (Q-LEAP), the Moonshot R\&D (Grant Number JPMJMS256E), and the ASPIRE program (Grant Number JPMJAP2513)], and the Office of Naval Research (ONR) Global (via Grant Number N62909-23-1-2074).

\appendix
\renewcommand{\theequation}{\Alph{section}\arabic{equation}}
\renewcommand{\thefigure}{\Alph{section}\arabic{figure}}

\setcounter{equation}{0}
\setcounter{figure}{0}

\section{Scattering matrix of a massless Dirac particle due to a delta potential}
\label{appA}

In the present Appendix, we show details of Subsec.~\ref{sec2.1}, 
the identification of a quantum walker with a massless Dirac particle scattered by a series of delta potentials.
We start by analyzing the scattering of a Dirac particle by Dirac's delta potential.
We will show that it is equivalent to the coin operator $C_{\varphi}$ in Eq.~\eqref{eq1001-1}.

We consider the Hamiltonian of the form
\begin{equation}\label{eq2101}
H_{{\mathrm{D}}_\delta}\coloneqq\epsilon p\sigma^z+m\delta(x)\sigma^y=
\begin{pmatrix}
\epsilon p & -\ii m \delta(x) \\
\ii m \delta(x) & -\epsilon p
\end{pmatrix},
\end{equation}
where $\epsilon$ and $m$ are positive parameters, and $\delta(x)$ is Dirac's delta function.
The momentum $p$ is defined as $p \coloneqq -\ii
\dd/\dd{x}$.
For $x\neq0$, we do not have any potentials, and hence
the eigenfunction is a plane wave.
For $k>0$, we have two positive-energy solutions
\begin{subequations}\label{eq2102}
\begin{align}
\begin{pmatrix}\epsilon p & \\ & -\epsilon p\end{pmatrix}\begin{pmatrix}\ee^{\ii kx} \\ 0\end{pmatrix} &= \epsilon k\begin{pmatrix}\ee^{\ii kx} \\ 0\end{pmatrix}, \label{eq2102a}
\\
\begin{pmatrix}\epsilon p & \\ & -\epsilon p\end{pmatrix}\begin{pmatrix}0 \\ \ee^{-\ii kx}\end{pmatrix} &= \epsilon k\begin{pmatrix}0 \\ \ee^{-\ii kx}\end{pmatrix}, \label{eq2102b}
\end{align}
\end{subequations}
and two negative-energy solutions
\begin{subequations}\label{eq2103}
\begin{align}
\begin{pmatrix}
\epsilon p & \\ & -\epsilon p\end{pmatrix}\begin{pmatrix}\ee^{-\ii kx} \\ 0\end{pmatrix} &= -\epsilon k\begin{pmatrix}\ee^{-\ii kx} \\ 0
\end{pmatrix},
 \label{eq2103a}
\\
\begin{pmatrix}
\epsilon p & \\ & -\epsilon p\end{pmatrix}\begin{pmatrix}0 \\ \ee^{\ii kx}\end{pmatrix} &= -\epsilon k\begin{pmatrix}0 \\ \ee^{\ii kx}
\end{pmatrix}. 
\label{eq2103b}
\end{align}
\end{subequations}
Since we identify the first element of the state vector as a left-going walker and the second element as a right-going walker for quantum walk, we focus on the latter negative-energy solutions hereafter.

We analyze the situation in Fig.~\ref{figappA1}(a).
\begin{figure}
\centering
\begin{subfigure}[c]{0.5\textwidth}
\centering
\begin{tikzpicture}
\draw[line width = 0.5, arrows = {-{Stealth[length=3mm]}}] (0,3) -- (6,3);
\draw[thick] (3,1)--(3,5);
\draw[line width = 2, arrows = {-{Stealth[length=5mm]}}] (0.5,3.75) -- (2.5,3.75);
\draw (1.5,4)node[above]{$A\ee^{\ii kx}\mqty(0 \\ 1)$};
\draw[line width = 2, arrows = {-{Stealth[length=5mm]}}] (3.5,3.75) -- (5.5,3.75);
\draw (4.5,4)node[above]{$C\ee^{\ii kx}\mqty(0 \\ 1)$};
\draw[line width = 2, arrows = {-{Stealth[length=5mm]}}] (2.5,2.25) -- (0.5,2.25);
\draw (1.5,2)node[below]{$B\ee^{-\ii kx}\mqty(1 \\ 0)$};
\draw[line width = 2, arrows = {-{Stealth[length=5mm]}}] (5.5,2.25) -- (3.5,2.25);
\draw (4.5,2)node[below]{$D\ee^{-\ii kx}\mqty(1 \\ 0)$};
\draw (3,0.3)node{$m_0\delta(x)\mqty( & -\ii \\ \ii & )$};
\draw (3.25,3.25)node{$0$};
\draw (6,3)node[right]{$x$};
\end{tikzpicture}
\vspace{0.5\baselineskip}
\caption{}
\end{subfigure}
\hspace*{0.08\textwidth}
\begin{subfigure}[c]{0.38\textwidth}
\centering
\vspace{0.8cm}
\begin{tikzpicture}
\draw[line width = 0.5, arrows = {-{Stealth[length=3mm]}}] (3,3) -- (7,3);
\draw[thick] (5,2.5)--(5,3.5);
\draw[line width = 2, arrows = {-{Stealth[length=5mm]}}] (4,3.75) -- (6,3.75);
\draw (4,3.75)node[left]{$C$};
\draw[line width = 2, arrows = {-{Stealth[length=5mm]}}] (6,4.25) -- (4,4.25);
\draw (4,4.25)node[left]{$B$};
\draw[line width = 2, arrows = {-{Stealth[length=5mm]}}] (6,2.25) -- (4,2.25);
\draw (4,2.25)node[left]{$D$};
\draw[line width = 2, arrows = {-{Stealth[length=5mm]}}] (4,1.75) -- (6,1.75);
\draw (4,1.75)node[left]{$A$};
\draw[line width = 0.5, arrows = {-{Stealth[length=5mm]}}] (3.2,2) arc (270:90:1cm);
\draw (1.4,3)node[above]{coin};
\draw (1.4,3)node[below]{operation};
\draw (5.25,3.25)node{$0$};
\draw (7,3)node[right]{$x$};
\end{tikzpicture}
\vspace{3.5\baselineskip}
\caption{}
\end{subfigure}
\caption{Scattering of a Dirac particle due to a delta potential at $x=0$.}
\label{figappA1}
\end{figure}
More specifically, we have
\begin{equation}\label{eq2104}
\vec{\psi}(x)=\begin{cases}
\begin{pmatrix}B\ee^{-\ii k x} \\ A\ee^{\ii kx}\end{pmatrix} &\quad\mbox{for $x<0$},
 \vspace{1ex} \\
\begin{pmatrix}D\ee^{-\ii k x} \\ C\ee^{\ii kx}\end{pmatrix} &\quad\mbox{for $x>0$},
\end{cases}
\end{equation}
whose eigenvalue is $E=-\epsilon k<0$.

Because the Hamiltonian~\eqref{eq2101} has a first-order derivative and Dirac's delta function, the wave function is discontinuous at the origin.
Let us find the amplitude of the jump.
For a wave function $\vec{\psi}= \left(\psi_1~\psi_2 \right)^\top$, the eigenvalue equation $H_{{\mathrm{D}}_\delta}\vec{\psi}=E\vec{\psi}$ reads
\begin{subequations}\label{eq2105}
\begin{align}
-\ii \epsilon \dv{x}\psi_1(x) -\ii m \delta(x)\psi_2(x)&=E\psi_1(x),\label{eq2105a}
\\
+\ii \epsilon \dv{x}\psi_2(x) +\ii m \delta(x)\psi_1(x)&=E\psi_2(x).\label{eq2105b}
\end{align}
\end{subequations}
Integrating each of them over an infinitesimal segment $[-\eta,\eta]$ and taking the limit $\eta\to0$, we eliminate the right-hand side in both equations, and find
\begin{subequations}\label{eq2106}
\begin{align}
-\ii\epsilon\qty(\psi_1(0+)-\psi_1(0-)) -\ii m \psi_2(0)&=0,\label{eq2106a}
\\
+\ii\epsilon\qty(\psi_2(0+)-\psi_2(0-)) +\ii m \psi_1(0)&=0.\label{eq2106b}
\end{align}
\end{subequations}
Since the values of $\psi_1(0)$ and $\psi_2(0)$ at $x=0$ are not determined by the differential equations~\eqref{eq2105},
we simply assume the following~\cite{Nishino11,Nishino15}:
\begin{subequations}\label{eq2107}
\begin{align}
\psi_1(0)&=\frac{1}{2}\qty(\psi_1(0+)+\psi_1(0-)),\label{eq2107a}
\\
\psi_2(0)&=\frac{1}{2}\qty(\psi_2(0+)+\psi_2(0-)).\label{eq2107b}
\end{align}
\end{subequations}
We note that the difference in the choice of these values is known to be absorbed by a renormalization of the system parameters for the Kondo model; see p.~138 of Ref.~\cite{Hewson93}.

We thereby obtain the equations for the jump at the origin in the form
\begin{equation}\label{eq2108}
\qty(\epsilon\sigma^0+\frac{m}{2}\sigma^x)\vec{\psi}(0+)=\qty(\epsilon\sigma^0-\frac{m}{2}\sigma^x)\vec{\psi}(0-).
\end{equation}
Note that the amplitude of the jump does not depend on the energy.
Let us now use Eq.~\eqref{eq2104} for Eq.~\eqref{eq2108}.
We then find
\begin{equation}\label{eq2109}
\qty(\epsilon\sigma^0+\frac{m}{2}\sigma^x)\begin{pmatrix}D \\ C\end{pmatrix} = \qty(\epsilon\sigma^0-\frac{m}{2}\sigma^x)\begin{pmatrix}B \\ A\end{pmatrix},
\end{equation}
which describes the scattering of a massless Dirac particle due to a single delta potential.

Equation~\eqref{eq2109} gives the transfer matrix, or the $T$ matrix, which relates the wave amplitudes on one side of the scattering potential to those on the other side.
We are rather interested in the scattering matrix, or the $S$ matrix, which relates the incoming wave amplitudes to the outgoing wave amplitudes in the form
\begin{equation}\label{eq2110}
\begin{pmatrix} B \\ C\end{pmatrix} = S_{\mathrm{D}} \begin{pmatrix}D \\ A\end{pmatrix}.
\end{equation}
We find the first and second rows of Eq.~\eqref{eq2110} by eliminating $B$ and $C$, respectively, from Eq.~\eqref{eq2109}.
We thereby obtain
\begin{equation}\label{eq2111}
S_{\mathrm{D}}\coloneqq\frac{1}{\epsilon^2+m^2/4}
\begin{pmatrix} \epsilon^2-m^2/4 & \epsilon m \\
-\epsilon m &\epsilon^2-m^2/4\end{pmatrix}.
\end{equation}
We indeed confirm that this is an orthogonal matrix, and therefore a unitary matrix, because we have
\begin{equation}\label{eq2112}
\qty(\epsilon^2-\frac{m^2}{4})^2+\qty(\epsilon m)^2=\qty(\epsilon^2+\frac{m^2}{4})^2.
\end{equation}

We then identify the amplitudes $D$ and $A$ of the incoming waves in Fig.~\ref{figappA1}(a) as the amplitudes of the left-going and right-going states, respectively, of a quantum walk before the coin operation, whereas we identify the amplitudes $B$ and $C$ of the outgoing waves as those after the coin operation.
See Fig.~\ref{figappA1}(b).
This lets us identify the scattering matrix $S_{\mathrm{D}}$ in Eq.~\eqref{eq2110} as the coin operator $C_{\varphi}$ in Eq.~\eqref{eq1001-1} of the quantum walk.
Parameterizing the matrix elements as in Eq.~\eqref{eq2113}, 
we can indeed reduce the $S$ matrix in Eq.~\eqref{eq2111} into the form $\ee^{-\ii\sigma^y\varphi}$. 
We thus know that the angle $\varphi$ of the coin operator is determined by $\epsilon$ and $m$ of the Dirac particle in the form of Eqs.~\eqref{eq2113}.

\section{Scattering matrix of one walker with a magnetic impurity}
\label{appB}

In this Appendix, we show the derivation of the scattering matrix~\eqref{eq2206} for a single quantum walker with a magnetic impurity.
We use two methods, one similar to the formulation in \ref{appA} and the other using a Green's function.
They produce the same result, but we develop the second method because it is more convenient to analyze the case of two walkers in \ref{appC}.

\subsection{Scattering due to a magnetic impurity}\label{appB.1}

Instead of the Dirac Hamiltonian~\eqref{eq2101} in \ref{appA}, in the present Appendix, we consider scattering due to the Hamiltonian of the form
\begin{equation}\label{eqJ101}
H\coloneqq\epsilon p \sigma^z\otimes s^0 +\Hm\delta(x)
\end{equation}
with
\begin{equation}\label{eqJ102}
\Hm\coloneqq J_x\sigma^x \otimes s^x+J_y\sigma^y \otimes s^y+J_z\sigma^z \otimes s^z,
\end{equation}
where $J_x$, $J_y$, and $J_z$ are real parameters.
The new degree of freedom in Eq.~\eqref{eqJ102} represented by a set of Pauli matrices $\{s^x,s^y,s^z\}$ is a magnetic impurity localized at $x=0$, with $s^0$ being the identity operator. 

The magnetic impurity Hamiltonian~\eqref{eqJ102} is transformed to
\begin{equation}\label{eqJ103}
\Hm=J_x\qty(\sigma^++\sigma^-)\otimes \qty(s^++s^-)-J_y\qty(\sigma^+-\sigma^-)\otimes \qty(s^+-s^-)+J_z\sigma^z \otimes s^z,
\end{equation}
with $\sigma^\pm\coloneqq(\sigma^x\pm \ii \sigma^y)/2$ and $s^\pm\coloneqq(s^x\pm\ii s^y)/2$,
and hence is represented by the matrix
\begin{equation}\label{eqJ104}
\Hm=
\qty(\begin{array}{c|cc|c}
J_z & & & J_x-J_y \\ \hline
& -J_z & J_x+J_y & \\
& J_x+J_y & -J_z & \\ \hline
J_x-J_y & & & J_z
\end{array}
)
\end{equation}
under the representation bases $\{\ket{\LL\up},\ket{\RR\up},\ket{\LL\down},\ket{\RR\down}\}$ in this order.
Here, $\ket{\LL}$ and $\ket{\RR}$ are the vectors that span the space of $\{\sigma^x,\sigma^y,\sigma^z\}$ with the eigenvalues $+1$ and $-1$ of $\sigma^z$, respectively, while $\ket{\up}$ and $\ket{\down}$ are the vectors that span the space of $\{s^x,s^y,s^z\}$ with the eigenvalues $+1$ and $-1$ of $s^z$, respectively.
We used the abbreviation $\ket{\LL\up}\coloneqq\ket{\LL} \otimes\ket{\up}$ and so on.
The eigenvectors of the magnetic interaction Hamiltonian~\eqref{eqJ104} are given by diagonalizing the unitary transformation
\begin{equation}\label{eqJ105}
\begin{pmatrix}
\ket{\psi^{(1)}} \\
\ket{\psi^{(2)}} \\
\ket{\psi^{(3)}} \\
\ket{\psi^{(4)}} 
\end{pmatrix}
=U\begin{pmatrix}
\ket{\LL\up}\\
\ket{\RR\up}\\
\ket{\LL\down}\\
\ket{\RR\down}
\end{pmatrix}
\end{equation}
with the eigenvalues
\begin{equation}\label{eqJ106}
+J_x-J_y+J_z,\quad
+J_x+J_y-J_z,\quad
-J_x-J_y-J_z,\quad
-J_x+J_y+J_z,
\end{equation}
and the unitary operator
\begin{equation}\label{eqJ107}
U=\frac{1}{\sqrt{2}}
\begin{pmatrix}
+1 &&& +1\\
& +1 & +1 & \\
& +1 & -1 & \\
+1 &&& -1
\end{pmatrix}.
\end{equation}

Instead of Eqs.~\eqref{eq2105}, the eigenvalue equation $H\vec{\psi}=E\vec{\psi}$
now reads
\begin{subequations}\label{eqJ108}
\begin{align}
-\ii\epsilon\dv{x} \psi_{\LL\up}+\delta(x)\qty[+J_z\psi_{\LL\up}+\qty(J_x-J_y)\psi_{\RR\down}]&=E\psi_{\LL\up},\label{eqJ108a}
\\
+\ii\epsilon\dv{x} \psi_{\RR\up}+\delta(x)\qty[-J_z\psi_{\RR\up}+\qty(J_x+J_y)\psi_{\LL\down}]&=E\psi_{\RR\up},\label{eqJ108b}
\\
-\ii\epsilon\dv{x} \psi_{\LL\down}+\delta(x)\qty[+\qty(J_x+J_y)\psi_{\RR\up}-J_z\psi_{\LL\down}]&=E\psi_{\LL\down},\label{eqJ108c}
\\
+\ii\epsilon\dv{x} \psi_{\RR\down}+\delta(x)\qty[+\qty(J_x-J_y)\psi_{\LL\up}+J_z\psi_{\RR\down}]&=E\psi_{\RR\down}\label{eqJ108d}
\end{align}
\end{subequations}
for the wave function $\vec{\psi}=(\psi_{\LL\up}\ \ \psi_{\RR\up}\ \ \psi_{\LL\down}\ \ \psi_{\RR\down})^\top$.
Integrating each of Eqs.~\eqref{eqJ108} over an infinitesimally narrow region $x\in[-\eta,\eta]$ and taking the limit $\eta\to0$, we obtain
\begin{subequations}\label{eqJ109}
\begin{align}
&-\ii\epsilon\qty[\psi_{\LL\up}(0+)-\psi_{\LL\up}(0-)]
+\frac{1}{2}\qty[+J_z\qty(\psi_{\LL\up}(0+)+\psi_{\LL\up}(0-))-\qty(J_x-J_y)\qty(\psi_{\RR\down}(0+)-\psi_{\RR\down}(0-))]=0,\label{eqJ109a}
\\
&+\ii\epsilon\qty[\psi_{\RR\up}(0+)-\psi_{\RR\up}(0-)]
+\frac{1}{2}\qty[-J_z\qty(\psi_{\RR\up}(0+)-\psi_{\RR\up}(0-))+\qty(J_x+J_y)\qty(\psi_{\LL\down}(0+)+\psi_{\LL\down}(0-))]=0,\label{eqJ109b}
\\
&-\ii\epsilon\qty[\psi_{\LL\down}(0+)-\psi_{\LL\down}(0-)]
+\frac{1}{2}\qty[+\qty(J_x+J_y)\qty(\psi_{\RR\up}(0+)+\psi_{\RR\up}(0-))-J_z\qty(\psi_{\LL\down}(0+)-\psi_{\LL\down}(0-))]=0,\label{eqJ109c}
\\
&+\ii\epsilon\qty[\psi_{\RR\down}(0+)-\psi_{\RR\down}(0-)]
+\frac{1}{2}\qty[+\qty(J_x-J_y)\qty(\psi_{\LL\up}(0+)-\psi_{\LL\up}(0-))+J_z\qty(\psi_{\RR\down}(0+)+\psi_{\RR\down}(0-))]=0,\label{eqJ109d}
\end{align}
\end{subequations}
where we assumed 
\begin{subequations}\label{eqJ110}
\begin{align}
\psi_{\LL\up}(0)&=\frac{1}{2}\qty[\psi_{\LL\up}(0+)+\psi_{\LL\up}(0-)],\label{eqJ110a}
\\
\psi_{\RR\up}(0)&=\frac{1}{2}\qty[\psi_{\RR\up}(0+)+\psi_{\RR\up}(0-)],\label{eqJ110b}
\\
\psi_{\LL\down}(0)&=\frac{1}{2}\qty[\psi_{\LL\down}(0+)+\psi_{\LL\down}(0-)],\label{eqJ110c}
\\
\psi_{\RR\down}(0)&=\frac{1}{2}\qty[\psi_{\RR\down}(0+)+\psi_{\RR\down}(0-)],\label{eqJ110d}
\end{align}
\end{subequations}
similarly to Eqs.~\eqref{eq2107}.
Equations~\eqref{eqJ109} are summarized in the form
\begin{align}\label{eqJ111}
&\qty[
\ii\epsilon
\begin{pmatrix}
-1 &&&\\
&+1&&\\
&&-1&\\
&&&+1
\end{pmatrix}
+\frac{1}{2}
\begin{pmatrix}
+J_z && &J_x-J_y \\
& -J_z&J_x+J_y&\\
&J_x+J_y&-J_z&\\
J_x-J_y&&&+J_z
\end{pmatrix}]
\begin{pmatrix}
\psi_{\LL\up}(0+)\\
\psi_{\RR\up}(0+)\\
\psi_{\LL\down}(0+)\\
\psi_{\RR\down}(0+)
\end{pmatrix}=
\notag\\
&\qquad
\qty[
\ii\epsilon
\begin{pmatrix}
-1 &&&\\
&+1&&\\
&&-1&\\
&&&+1
\end{pmatrix}
-\frac{1}{2}
\begin{pmatrix}
+J_z && &J_x-J_y \\
& -J_z&J_x+J_y&\\
&J_x+J_y&-J_z&\\
J_x-J_y&&&+J_z
\end{pmatrix}]
\begin{pmatrix}
\psi_{\LL\up}(0-)\\
\psi_{\RR\up}(0-)\\
\psi_{\LL\down}(0-)\\
\psi_{\RR\down}(0-)
\end{pmatrix}
\end{align}
or more compactly
\begin{equation}\label{eqJ112}
\qty(-\ii\epsilon\sigma^z\otimes s^0+\frac{\Hm}{2})\vec{\psi}(0+)
=\qty(-\ii\epsilon\sigma^z\otimes s^0-\frac{\Hm}{2})\vec{\psi}(0-).
\end{equation}

Equation~\eqref{eqJ111} provides the transfer matrix, or the $T$ matrix, which relates the wave functions on the left and on the right.
We now transform the $T$ matrix to the scattering matrix, or the $S$ matrix, which relates the incoming wave functions to the outgoing wave functions.
The incoming wave functions are $\psi_{\RR\up}(0-)$ and $\psi_{\RR\down}(0-)$ from the left and
$\psi_{\LL\up}(0+)$ and $\psi_{\LL\down}(0+)$ from the right.
The outgoing wave functions are $\psi_{\LL\up}(0-)$ and $\psi_{\LL\down}(0-)$ to the left and 
$\psi_{\RR\up}(0+)$ and $\psi_{\RR\down}(0+)$ to the right.
Solving Eq.~\eqref{eqJ111} with respect to the outgoing waves, 
we find
\begin{equation}\label{eqJ113}
\begin{pmatrix}
\psi_{\LL\up}(0-)\\
\psi_{\RR\up}(0+)\\
\psi_{\LL\down}(0-)\\
\psi_{\RR\down}(0+)
\end{pmatrix}
=
S_{\mathrm{imp}}^{\mathrm{1w}}
\begin{pmatrix}
\psi_{\LL\up}(0+)\\
\psi_{\RR\up}(0-)\\
\psi_{\LL\down}(0+)\\
\psi_{\RR\down}(0-)
\end{pmatrix}
\end{equation}
with the $S$ matrix
\begin{equation}\label{eqJ114}
S_{\mathrm{imp}}^{\mathrm{1w}}\coloneqq
\begin{pmatrix}
\alpha^{\mathrm{1w}} & 0 & 0 & \beta^{\mathrm{1w}} \\
 0 & \gamma^{\mathrm{1w}} & \delta^{\mathrm{1w}} & 0 \\
 0 & \delta^{\mathrm{1w}} & \gamma^{\mathrm{1w}} & 0 \\
\beta^{\mathrm{1w}} & 0 & 0 & \alpha^{\mathrm{1w}} \\
\end{pmatrix},
\end{equation}
where
\begin{subequations}\label{eqJ115}
\begin{align}
\alpha^{\mathrm{1w}}
&\coloneqq-\frac{4\epsilon^2-\qty(J_x-J_y)^2+{J_z}^2}%
{(2\ii\epsilon+J_z)^2-\qty(J_x-J_y)^2}, \label{eqJ115a}\\
\beta^{\mathrm{1w}}
&\coloneqq-\frac{4\ii\epsilon\qty(J_x-J_y)}%
{(2\ii\epsilon+J_z)^2-\qty(J_x-J_y)^2}, \label{eqJ115b}\\
\gamma^{\mathrm{1w}}&\coloneqq-\frac{4\epsilon^2-\qty(J_x+J_y)^2+{J_z}^2}%
{(2\ii\epsilon-J_z)^2-\qty(J_x+J_y)^2}, \label{eqJ115c}\\
\delta^{\mathrm{1w}}&\coloneqq-\frac{4\ii\epsilon\qty(J_x+J_y)}%
{(2\ii\epsilon-J_z)^2-\qty(J_x+J_y)^2}. \label{eqJ115d}
\end{align}
\end{subequations}
Owing to $\abs{\alpha^{\mathrm{1w}}}^2+\abs{\beta^{\mathrm{1w}}}^2=\abs{\gamma^{\mathrm{1w}}}^2+\abs{\delta^{\mathrm{1w}}}^2=1$, this $S$ matrix is indeed a unitary matrix, and hence the flux is conserved.
This is used in Eq.~\eqref{eq2206} in Subsec.~\ref{sec2.2}.

\subsection{Identification of the $S$ matrix with a new coin operator}\label{appB.2}

\begin{figure}
\begin{subfigure}[t]{0.45\textwidth}
\includegraphics[width=\textwidth]{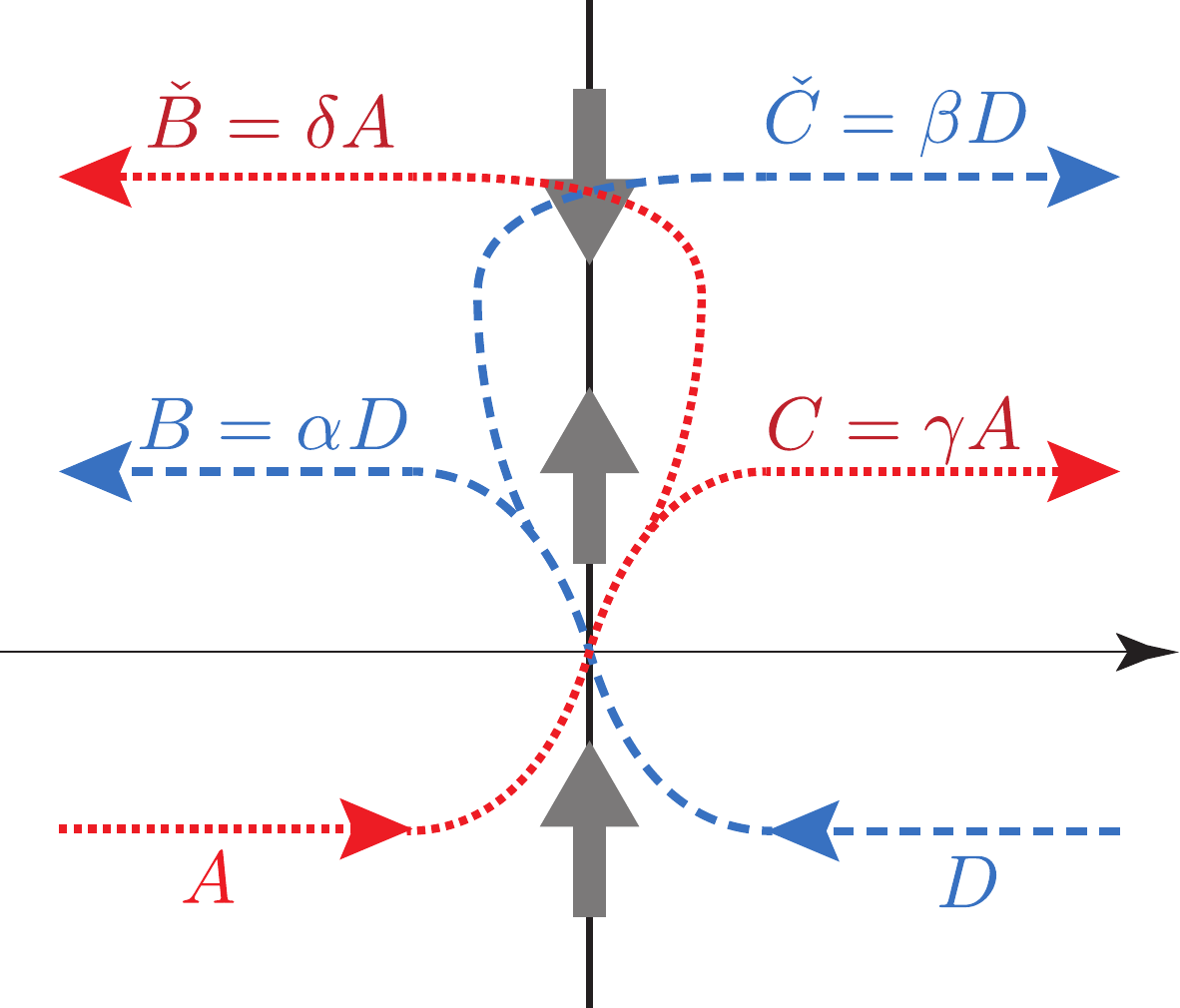}
\caption{}
\end{subfigure}
\hfill
\begin{subfigure}[t]{0.45\textwidth}
\includegraphics[width=\textwidth]{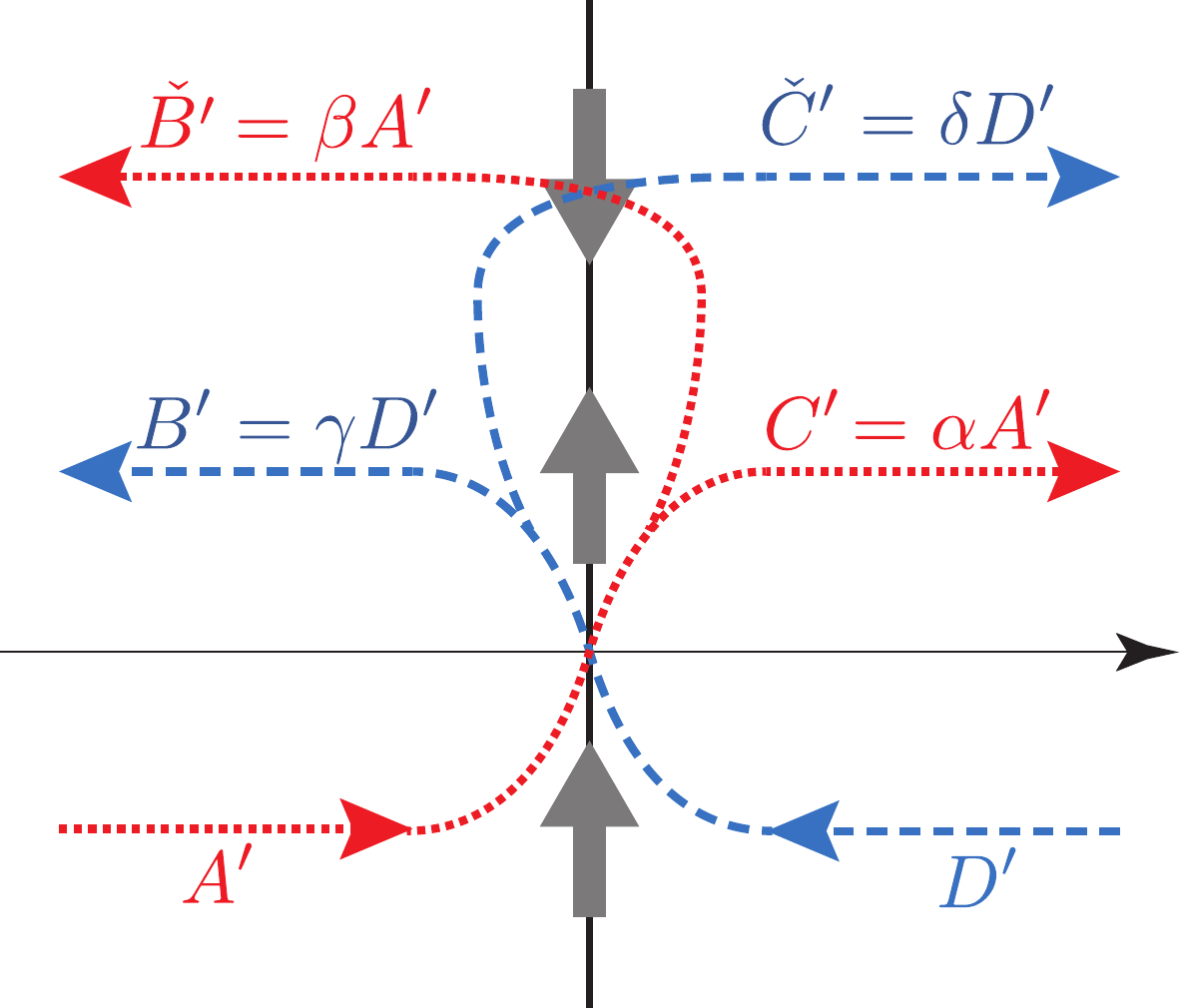}
\caption{}
\end{subfigure}
\caption{Schematic picture of scattering processes for the magnetic impurity in the states (a) $\ket{\up}$ and (b) $\ket{\down}$.}
\label{figappB2}
\end{figure}

In order to understand the $S$ matrix~\eqref{eqJ114}, let us first consider the case for which the magnetic impurity was in the state $\ket{\up}$. We thereby assume the incident wave of the form [see Fig.~\ref{figappB2}(a)] 
\begin{equation}\label{eqJ116}
\begin{pmatrix}
\psi_{\LL\up}(0+)\\
\psi_{\RR\up}(0-)\\
\psi_{\LL\down}(0+)\\
\psi_{\RR\down}(0-)
\end{pmatrix}
=\begin{pmatrix}
D \\
A \\
0 \\
0
\end{pmatrix}.
\end{equation}
We then have the scattered wave in the form
\begin{equation}\label{eqJ117}
\begin{pmatrix}
\psi_{\LL\up}(0-)\\
\psi_{\RR\up}(0+)\\
\psi_{\LL\down}(0-)\\
\psi_{\RR\down}(0+)
\end{pmatrix}
=\begin{pmatrix}
B \\
C \\
\check{B} \\
\check{C}
\end{pmatrix}
=\begin{pmatrix}
\alpha D \\
\delta A \\
\gamma A \\
\beta D
\end{pmatrix}.
\end{equation}
This means that the left-going wave $D$ that comes from the right is transmitted to the left as $B=\alpha D$ without spin flipping and reflected to the right as $C'=\beta D$ with spin flipping.
The right-going wave $A$ that comes from the left is transmitted to the right as $C= \delta A$ without spin flipping and reflected to the left as $B'=\gamma A$.
In other words, the transmission wave does not flip the impurity spin, acquiring the amplitude $\alpha$ or $\delta$, while the reflection wave flips the spin, acquiring the amplitude $\beta$ or $\gamma$.

Let us next consider the case in which the impurity spin was in the state $\ket{\down}$.
Assuming the incident wave of the form
\begin{equation}\label{eqJ118}
\begin{pmatrix}
\psi_{\LL\up}(0+)\\
\psi_{\RR\up}(0-)\\
\psi_{\LL\down}(0+)\\
\psi_{\RR\down}(0-)
\end{pmatrix}
=
\begin{pmatrix}
0 \\
0 \\
D' \\
A'
\end{pmatrix},
\end{equation}
we have the scattered wave of the form
\begin{equation}\label{eqJ119}
\begin{pmatrix}
\psi_{\LL\up}(0-)\\
\psi_{\RR\up}(0+)\\
\psi_{\LL\down}(0-)\\
\psi_{\RR\down}(0+)
\end{pmatrix}
=\begin{pmatrix}
\check{B'} \\
\check{C'} \\
B' \\
C'
\end{pmatrix}
=\begin{pmatrix}
\beta A' \\
\gamma D' \\
\delta D' \\
\alpha A'
\end{pmatrix}.
\end{equation}
This again means that the transmitted wave does not flip the impurity spin, acquiring the amplitude $\alpha$ or $\delta$, while the reflected wave flips the spin, acquiring the amplitude $\beta$ or $\gamma$.

This spin flipping induces interaction between two Dirac particles.
Suppose that a Dirac particle named $1$ is scattered by an up spin first and another Dirac particle named $2$ is scattered next.
Since the first particle may flip the impurity spin, the second particle may be scattered by a down spin.
Suppose instead that the particle $2$ is scattered by an up spin first and the particle $1$ is scattered next.
In this case, the particle $2$ is scattered only by the up spin, not the down spin.
This difference in the processes means that there are interactions between the two particles via the impurity spin.

\subsection{Green's function formalism to obtain the $S$ matrix}\label{appB.3}
Here, we show that the $S$ matrix in Eq.~\eqref{eqJ114} is also obtained using the Green's function formalism~\cite{Datta-textbook}. 
The Green’s function formalism is more tractable in the case of two quantum walkers.
Let us consider scattering by the Hamiltonian $H=H_0+V$.
The Schr\"{o}dinger equation reads
\begin{equation}\label{eqJ120}
(H_0+V)\ket{\psi}=E\ket{\psi}
\end{equation}
with the eigenenergy $E$ and the corresponding eigenstate $\ket{\psi}$.
The eigenstate of the kinetic term $H_0$ of the Hamiltonian is given by
\begin{equation}\label{eqJ121}
H_0\ket{\psi_0}=E\ket{\psi_0}.
\end{equation}
From the above two equations, we have
\begin{equation}\label{eqJ122}
(E-H_0)\qty(\ket{\psi}-\ket{\psi_0})=V\ket{\psi}.
\end{equation}
Assuming the regularity, we obtain
\begin{equation}\label{eqJ123}
\ket{\psi}-\ket{\psi_0}=(E-H_0)^{-1}V\ket{\psi}.
\end{equation}

For a one-particle system, we obtain the position-representation equation as follows:
\begin{align}
\psi(x)-\psi_0(x)%
&=\int_{-\infty}^{+\infty}\dd y\int_{-\infty}^{+\infty}dz \mel{x}{(E-H_0)^{-1}}{y}\mel{y}{V}{z}\ip{z}{\psi} \notag \\
&=\int_{-\infty}^{+\infty}\dd y \mel{x}{(E-H_0)^{-1}}{y}V(y)\psi(y),\label{eqJ124}
\end{align}
where $\psi(x)\coloneqq\braket{x}{\psi}$ is an eigenfunction of the total Hamiltonian $H=H_0+V$, $\psi_0(x)\coloneqq\braket{x}{\psi_0}$ is an eigenfunction of the kinetic term $H_0$, and $V(y)\coloneqq\ev{V}{y}$ is the potential function.
Thus, we define the Green's function as
\begin{equation}\label{eqJ125}
G_0(x-y)\coloneqq\mel{x}{(E-H_0)^{-1}}{y},
\end{equation}
and obtain
\begin{equation}\label{eqJ126}
\psi(x)=\psi_0(x)+\int^\infty_{-\infty}\dd y\ G_0(x-y)V(y)\psi(y).
\end{equation}
When the scattering potential is in the form of delta function as $V(x)=V\delta(x)$, we have
\begin{equation}\label{eqJ127}
\psi(x)=\psi_0(x)+G_0(x)V\psi(0).
\end{equation}

We now consider scattering due to the Hamiltonian~\eqref{eqJ101} with~\eqref{eqJ102}, \textit{i.e.},
\begin{align}
H_0&=\epsilon p\sigma^zs^0,\label{eqJ128}\\
V&=J_x\qty(\sigma^++\sigma^-)\otimes \qty(s^++s^-)-J_y\qty(\sigma^+-\sigma^-)\otimes \qty(s^+-s^-)+J_z\sigma^z \otimes s^z\notag \\
&=
\qty(\begin{array}{c|cc|c}
J_z & & & J_x-J_y \\ \hline
& -J_z & J_x+J_y & \\
& J_x+J_y & -J_z & \\ \hline
J_x-J_y & & & J_z
\end{array}
).\label{eqJ129}
\end{align}
Then we straightforwardly obtain the Green's function in momentum space as
\begin{equation}\label{eqJ130}
G_0(p)=(E-\epsilon p\sigma^zs^0)^{-1}=%
\begin{pmatrix}
\qty(E-\epsilon p)^{-1} &  &  &  \\
 & \qty(E+\epsilon p)^{-1} &  &  \\
 &  & \qty(E-\epsilon p)^{-1} &  \\
 &  &  & \qty(E+\epsilon p)^{-1}
\end{pmatrix}.
\end{equation}
We perform the Fourier transform on each of the elements in $G_0(p)$ to obtain the Green's function in real space.
We also note that $\psi_0(x)$ in Eq.~\eqref{eqJ127} is an eigenfunction of the kinetic term $H_0$ in Eq.~\eqref{eqJ128}, and hence given by $\ee^{-\ii kx}$ for the left-going state and $\ee^{\ii kx}$ for the right-going state.

For the $(1,1)$ and $(3,3)$ elements, which are multiplied to left-going components of the wave function, we take the contour $C_1$ for the complex integral as shown in Fig.~\ref{figappB3}(a) so that the corresponding elements in the real-space Green's function $G_0(x)$ may have finite values for $x<0$.
Thus, we have
\begin{align}
\int_{-\infty}^\infty\frac{\ee^{\ii px}}{E-\epsilon p-\ii\eta}\frac{\dd p}{2\pi}%
&=\theta(-x)\qty[\oint_{C_1}\frac{\ee^{\ii zx}}{E-\epsilon z-\ii\eta}\frac{\dd z}{2\pi}%
-\lim_{R\rightarrow\infty}\qty(\int_{2\pi}^\pi\frac{\ee^{\ii R\ee^{\ii\phi}x}}{E-\epsilon R\ee^{\ii\phi}-\ii\eta}\frac{\dd\phi}{2\pi})]\notag \\
&=\theta(-x)\qty[-\frac{\ii}{2\pi\ii\epsilon}\oint_{C_1}\frac{\ee^{\ii zx}}{z-(E-\ii\eta)/\epsilon}\dd z-0]%
=\frac{\ii}{\epsilon}\ee^{\ii Ex/\epsilon}\theta(-x) \label{eqJ131}
\end{align}
with the Heaviside step function $\theta$.
Let us note here that $\ee^{\ii Ex/\epsilon}$ describes a left-going wave as we 
take negative-energy solutions $E=-\epsilon k$ as shown in Sec.~\ref{sec2.1}.
On the other hand, for the $(2,2)$ and $(4,4)$ elements, which are multiplied to right-going components of the wave function, we take the contour $C_2$ for the complex integral as shown in Fig.~\ref{figappB3}(b) so that the corresponding elements in the real-space Green's function $G_0(x)$ may have finite values for $x>0$.
Thus, we have
\begin{align}
\int_{-\infty}^\infty\frac{\ee^{\ii px}}{E+\epsilon p-\ii\eta}\frac{\dd p}{2\pi}%
&=\theta(+x)\qty[\oint_{C_2}\frac{\ee^{\ii zx}}{E+\epsilon z-\ii\eta}\frac{\dd z}{2\pi}%
-\lim_{R\rightarrow\infty}\qty(\int_0^\pi\frac{\ee^{\ii R\ee^{\ii\phi}x}}{E+\epsilon R\ee^{\ii\phi}-\ii\eta}\frac{\dd\phi}{2\pi})]\notag \\
&=\theta(+x)\qty[\frac{\ii}{2\pi\ii\epsilon}\oint_{C_2}\frac{\ee^{\ii zx}}{z+(E-\ii\eta)/\epsilon}\dd z-0]%
=\frac{\ii}{\epsilon}\ee^{-\ii Ex/\epsilon}\theta(+x).\label{eqJ132}
\end{align}
To summarize, we obtain
\begin{equation}\label{eqJ133}
G_0(x)=\frac{\ii}{\epsilon}
\begin{pmatrix}
\ee^{+\ii Ex/\epsilon}\theta(-x) & & & \\
 & \ee^{-\ii Ex/\epsilon}\theta(+x) & & \\
 & & \ee^{+\ii Ex/\epsilon}\theta(-x) & \\
 & & & \ee^{-\ii Ex/\epsilon}\theta(+x)
\end{pmatrix}.
\end{equation}

\begin{figure}
\centering
\begin{subfigure}[c]{0.48\textwidth}
\centering
\begin{tikzpicture}
\draw[line width = 0.5, arrows = {-{Stealth[length=3mm]}}] (0,3) -- (6,3);
\draw[line width = 0.5, arrows = {-{Stealth[length=3mm]}}] (3,0) -- (3,6);
\draw[line width = 2, arrows = {-{Stealth[length=5mm]}}] (4.415,1.585) arc (315:290:2);
\draw[line width = 2, arrows = {-{Stealth[length=5mm]}}] (3,1) arc (270:230:2);
\draw[line width = 2] (5,3) arc (360:180:2);
\draw[line width = 2, arrows = {-{Stealth[length=5mm]}}] (1,3)--(2.3,3);
\draw[line width = 2, arrows = {-{Stealth[length=5mm]}}] (3,3)--(4.3,3);
\draw[line width = 2] (0.965,3)--(5.035,3);
\draw (6,3)node[right]{$\Re z$};
\draw (3,6)node[above]{$\Im z$};
\draw (1.3,1.3)node{$C_1$};
\draw[line width = 2] (4.15,1.85)--(3.85,2.15);
\draw[line width = 2] (4.15,2.15)--(3.85,1.85);
\draw (4,2.1)node[above]{$\frac{E-\ii\eta}{\epsilon}$};
\end{tikzpicture}
\caption{}
\end{subfigure}
\hspace*{0.02\textwidth}
\begin{subfigure}[c]{0.48\textwidth}
\centering
\begin{tikzpicture}
\draw[line width = 0.5, arrows = {-{Stealth[length=3mm]}}] (0,3) -- (6,3);
\draw[line width = 0.5, arrows = {-{Stealth[length=3mm]}}] (3,0) -- (3,6);
\draw[line width = 2, arrows = {-{Stealth[length=5mm]}}] (4.415,4.415) arc (45:70:2);
\draw[line width = 2, arrows = {-{Stealth[length=5mm]}}] (3,5) arc (90:130:2);
\draw[line width = 2] (5,3) arc (0:180:2);
\draw[line width = 2, arrows = {-{Stealth[length=5mm]}}] (1,3)--(2.3,3);
\draw[line width = 2, arrows = {-{Stealth[length=5mm]}}] (3,3)--(4.3,3);
\draw[line width = 2] (0.965,3)--(5.035,3);
\draw (6,3)node[right]{$\Re z$};
\draw (3,6)node[above]{$\Im z$};
\draw (1.3,4.7)node{$C_2$};
\draw[line width = 2] (1.85,4.15)--(2.15,3.85);
\draw[line width = 2] (2.15,4.15)--(1.85,3.85);
\draw (2,3.9)node[below]{$\frac{-E+\ii\eta}{\epsilon}$};
\end{tikzpicture}
\caption{}
\end{subfigure}
\caption{Contours for complex integrals.}
\label{figappB3}
\end{figure}

Assuming $\psi(0)=[\psi(0+)+\psi(0-)]/2$ in Eq.~\eqref{eqJ127}, we have
\begin{subequations}\label{eqJ134}
\begin{align}
\psi_{\mathrm{L}\uparrow}(x)&=%
\begin{cases}
A\ee^{-\ii kx} & (x>0), \\
A\ee^{-\ii kx}+\displaystyle\frac{\ii}{2\epsilon}\,\ee^{+\ii Ex/\epsilon}%
\left[+J_z\qty(\psi_{\mathrm{L}\uparrow}(0+)+\psi_{\mathrm{L}\uparrow}(0-))\right. & \\
       \left.\hspace{3.3cm}+(J_x-J_y)\qty(\psi_{\mathrm{R}\downarrow}(0+)+\psi_{\mathrm{R}\downarrow}(0-))\right] & (x<0),
\end{cases}\label{eqJ134a}\\
\psi_{\mathrm{R}\uparrow}(x)&=%
\begin{cases}
B\ee^{\ii kx}+\displaystyle\frac{\ii}{2\epsilon}\,\ee^{-\ii Ex/\epsilon}%
\left[-J_z\qty(\psi_{\mathrm{R}\uparrow}(0+)+\psi_{\mathrm{R}\uparrow}(0-))\right. & \\
       \left.\hspace{3.3cm}+(J_x+J_y)\qty(\psi_{\mathrm{L}\downarrow}(0+)+\psi_{\mathrm{L}\downarrow}(0-))\right] & (x>0), \\
B\ee^{\ii kx} & (x<0),
\end{cases} \label{eqJ134b}\\
\psi_{\mathrm{L}\downarrow}(x)&=%
\begin{cases}
C\ee^{-\ii kx} & (x>0), \\
C\ee^{-\ii kx}+\displaystyle\frac{\ii}{2\epsilon}\,\ee^{+\ii Ex/\epsilon}%
\left[-J_z\qty(\psi_{\mathrm{L}\downarrow}(0+)+\psi_{\mathrm{L}\downarrow}(0-))\right. & \\
       \left.\hspace{3.3cm}+(J_x+J_y)\qty(\psi_{\mathrm{R}\uparrow}(0+)+\psi_{\mathrm{R}\uparrow}(0-))\right] & (x<0),
\end{cases}\label{eqJ134c}\\
\psi_{\mathrm{R}\downarrow}(x)&=%
\begin{cases}
D\ee^{\ii kx}+\displaystyle\frac{\ii}{2\epsilon}\,\ee^{-\ii Ex/\epsilon}%
\left[+J_z\qty(\psi_{\mathrm{R}\downarrow}(0+)+\psi_{\mathrm{R}\downarrow}(0-))\right. & \\
       \left.\hspace{3.3cm}+(J_x-J_y)\qty(\psi_{\mathrm{L}\uparrow}(0+)+\psi_{\mathrm{L}\uparrow}(0-))\right] & (x>0), \\
D\ee^{\ii kx} & (x<0),
\end{cases}\label{eqJ134d}
\end{align}
\end{subequations}
where $A, B, C, D\in\mathbb{C}$ are arbitrary coefficients of the wave function $\psi_0(x)$ in Eq.~\eqref{eqJ127}.
Inserting $x=0+$ and $0-$ in each equation and subtracting the latter from the former, we have
\begin{subequations}\label{eqJ135}
\begin{align}
&-\ii\epsilon\qty[\psi_{\mathrm{L}\uparrow}(0+)-\psi_{\mathrm{L}\uparrow}(0-)] 
+\frac{\ee^{-\ii Ex/\epsilon}}{2}%
\qty[+J_z\qty(\psi_{\mathrm{L}\uparrow}(0+)+\psi_{\mathrm{L}\uparrow}(0-))+%
       (J_x-J_y)\qty(\psi_{\mathrm{R}\downarrow}(0+)+\psi_{\mathrm{R}\downarrow}(0-))]=0, \label{eqJ135a}\\
&+\ii\epsilon\qty[\psi_{\mathrm{R}\uparrow}(0+)-\psi_{\mathrm{R}\uparrow}(0-)] 
+\frac{\ee^{+\ii Ex/\epsilon}}{2}%
\qty[-J_z\qty(\psi_{\mathrm{R}\uparrow}(0+)+\psi_{\mathrm{R}\uparrow}(0-))+%
       (J_x+J_y)\qty(\psi_{\mathrm{L}\downarrow}(0+)+\psi_{\mathrm{L}\downarrow}(0-))]=0, \label{eqJ135b}\\
&-\ii\epsilon\qty[\psi_{\mathrm{L}\downarrow}(0+)-\psi_{\mathrm{L}\downarrow}(0-)] 
+\frac{\ee^{-\ii Ex/\epsilon}}{2}%
\qty[-J_z\qty(\psi_{\mathrm{L}\downarrow}(0+)+\psi_{\mathrm{L}\downarrow}(0-))+%
       (J_x+J_y)\qty(\psi_{\mathrm{R}\uparrow}(0+)+\psi_{\mathrm{R}\uparrow}(0-))]=0, \label{eqJ135c}\\
&+\ii\epsilon\qty[\psi_{\mathrm{R}\downarrow}(0+)-\psi_{\mathrm{R}\downarrow}(0-)] 
+\frac{\ee^{+\ii Ex/\epsilon}}{2}%
\qty[+J_z\qty(\psi_{\mathrm{R}\downarrow}(0+)+\psi_{\mathrm{R}\downarrow}(0-))+%
       (J_x-J_y)\qty(\psi_{\mathrm{L}\uparrow}(0+)+\psi_{\mathrm{L}\uparrow}(0-))]=0,\label{eqJ135d}
\end{align}
\end{subequations}
which eliminates all of the states with the arbitrary coefficients.
Since the exponential function is regular at $x=0$ and $\lim_{x\rightarrow0}\ee^x\rightarrow1$, we have
\begin{subequations}\label{eqJ136}
\begin{align}
&-\ii\epsilon\qty[\psi_{\mathrm{L}\uparrow}(0+)-\psi_{\mathrm{L}\uparrow}(0-)] 
+\frac{1}{2}%
\qty[+J_z\qty(\psi_{\mathrm{L}\uparrow}(0+)+\psi_{\mathrm{L}\uparrow}(0-))+%
       (J_x-J_y)\qty(\psi_{\mathrm{R}\downarrow}(0+)+\psi_{\mathrm{R}\downarrow}(0-))]=0, \label{eqJ136a}\\
&+\ii\epsilon\qty[\psi_{\mathrm{R}\uparrow}(0+)-\psi_{\mathrm{R}\uparrow}(0-)]
+\frac{1}{2}%
\qty[-J_z\qty(\psi_{\mathrm{R}\uparrow}(0+)+\psi_{\mathrm{R}\uparrow}(0-))+%
       (J_x+J_y)\qty(\psi_{\mathrm{L}\downarrow}(0+)+\psi_{\mathrm{L}\downarrow}(0-))]=0, \label{eqJ136b}\\
&-\ii\epsilon\qty[\psi_{\mathrm{L}\downarrow}(0+)-\psi_{\mathrm{L}\downarrow}(0-)] 
+\frac{1}{2}%
\qty[-J_z\qty(\psi_{\mathrm{L}\downarrow}(0+)+\psi_{\mathrm{L}\downarrow}(0-))+%
       (J_x+J_y)\qty(\psi_{\mathrm{R}\uparrow}(0+)+\psi_{\mathrm{R}\uparrow}(0-))]=0, \label{eqJ136c}\\
&+\ii\epsilon\qty[\psi_{\mathrm{R}\downarrow}(0+)-\psi_{\mathrm{R}\downarrow}(0-)] 
+\frac{1}{2}%
\qty[+J_z\qty(\psi_{\mathrm{R}\downarrow}(0+)+\psi_{\mathrm{R}\downarrow}(0-))+%
       (J_x-J_y)\qty(\psi_{\mathrm{L}\uparrow}(0+)+\psi_{\mathrm{L}\uparrow}(0-))]=0.\label{eqJ136d}
\end{align}
\end{subequations}
This exactly matches what we have in Eq.~\eqref{eqJ109}, and hence we obtain the same scattering matrix $S_{\mathrm{imp}}^{\mathrm{1w}}$ using the Green's function formalism in the one-dimensional case.

\section{Scattering matrix of two walkers with a magnetic impurity}
\label{appC}
In this Appendix, we derive the scattering matrix~\eqref{eq4105} when two quantum walkers are simultaneously scattered by the localized magnetic impurity at the origin, extending the Green's function formalism in \ref{appB.2}.
From Eq.~\eqref{eqJ123}, we obtain a position-representation equation for a two-particle system as follows:
\begin{align}
&\psi(x_1,x_2)-\psi_0(x_1,x_2)\notag \\
=&\int_{-\infty}^{+\infty}\dd y_1\int_{-\infty}^{+\infty}\dd y_2\int_{-\infty}^{+\infty}\dd z_1\int_{-\infty}^{+\infty}\dd z_2%
\mel{x_1,x_2}{(E-H_0)^{-1}}{y_1,y_2}\mel{y_1,y_2}{V}{z_1,z_2}\ip{z_1,z_2}{\psi} \notag \\
=&\int_{-\infty}^{+\infty}\dd y_1\int_{-\infty}^{+\infty}\dd y_2%
\mel{x_1,x_2}{(E-H_0)^{-1}}{y_1,y_2}V(y_1,y_2)\psi(y_1,y_2),\label{eqJ201}
\end{align}
where $\psi(x_1,x_2)\coloneqq\braket{x_1,x_2}{\psi}$ is an eigenfunction of the total Hamiltonian, $\psi_0(x_1,x_2)\coloneqq\braket{x_1,x_2}{\psi_0}$ is an eigenfunction of the kinetic term $H_0$, and $V(y_1,y_2)\coloneqq\ev{V}{y_1,y_2}$ is the potential function.
Thus, we define the Green's function 
as
\begin{equation}\label{eqJ202}
G_0(x_1-y_1,x_2-y_2)\coloneqq\mel{x_1,x_2}{(E-H_0)^{-1}}{y_1,y_2},
\end{equation}
and obtain
\begin{equation}\label{eqJ203}
\psi(x_1,x_2)=\psi_0(x_1,x_2)%
+\int_{-\infty}^{+\infty}\dd y_1\int_{-\infty}^{+\infty}\dd y_2%
G_0(x_1-y_1,x_2-y_2)V(y_1,y_2)\psi(y_1,y_2).
\end{equation}
When the scattering potential is in the form of delta functions as $V(x_1,x_2)=V_1\delta(x_1)+V_2\delta(x_2)$, we have
\begin{align}
&\psi(x_1,x_2)-\psi_0(x_1,x_2)\notag \\
=&\int_{-\infty}^{+\infty}\dd y_1\int_{-\infty}^{+\infty}\dd y_2%
G_0(x_1-y_1,x_2-y_2)V(y_1,y_2)\psi(y_1,y_2) \notag \\
=&\int_{-\infty}^{+\infty}\dd y_2G_0(x_1,x_2-y_2)V_1\psi(0,y_2)%
+\int_{-\infty}^{+\infty}\dd y_1G_0(x_1-y_1,x_2)V_2\psi(y_1,0)\label{eqJ204}.
\end{align}

We now consider scattering due to the Hamiltonian $H_0+V$, where
\begin{align}
H_0(p_1,p_2)%
&=\epsilon p_1\sigma_1^z\sigma_2^0s^0+\epsilon p_2\sigma_1^0\sigma_2^zs_0 \notag \\
&=\epsilon p_1
\begin{pmatrix}
1 & & & & & & & \\
 & -1 & & & & & & \\
 & & 1 & & & & & \\
 & & & -1 & & & & \\
 & & & & 1 & & & \\
 & & & & & -1 & & \\
 & & & & & & 1 & \\
 & & & & & & & -1
\end{pmatrix}
+\epsilon p_2
\begin{pmatrix}
1 & & & & & & & \\
 & 1 & & & & & & \\
 & & -1 & & & & & \\
 & & & -1 & & & & \\
 & & & & 1 & & & \\
 & & & & & 1 & & \\
 & & & & & & -1 & \\
 & & & & & & & -1
\end{pmatrix},\label{eqJ205}
\end{align}
\begin{align}
V(x_1,x_2) 
&=V_1\delta(x_1)+V_2\delta(x_2) \notag \\
&=\sigma_2^0(J_x\sigma_1^xs^x+J_y\sigma_1^ys^y+J_z\sigma_1^zs^z)\delta(x_1)%
+\sigma_1^0(J_x\sigma_2^xs^x+J_y\sigma_2^ys^y+J_z\sigma_2^zs^z)\delta(x_2) \notag \\
&=\qty(\begin{array}{cccc|cccc}
J_z & & & & & J_x-J_y & & \\
& -J_z & & & J_x+J_y & & & \\
& & J_z & & & & & J_x-J_y \\
& & & -J_z & & & J_x+J_y & \\ \hline
& J_x+J_y & & & -J_z & & & \\
J_x-J_y & & & & & J_z & & \\
& & & J_x+J_y & & & -J_z & \\
& & J_x-J_y & & & & & J_z
\end{array}
)\delta(x_1)\notag \\
&+\qty(\begin{array}{cccc|cccc}
J_z & & & & & & J_x-J_y & \\
& J_z & & & & & & J_x-J_y \\
& & -J_z & & J_x+J_y & & & \\
& & & -J_z & & J_x+J_y & & \\ \hline
& & J_x+J_y & & -J_z & & & \\
& & & J_x+J_y & & -J_z & & \\
J_x-J_y & & & & & & J_z & \\
& J_x-J_y & & & & & & J_z
\end{array}
)\delta(x_2). \label{eqJ206}
\end{align}
Then we straightforwardly obtain the Green's function in momentum space as
\begin{equation}\label{eqJ207}
G_0(p_1, p_2)%
=\begin{pmatrix}
\frac{1}{E-\epsilon p_+} & & & & & & & \\
 & \frac{1}{E+\epsilon p_-} & & & & & & \\
 & & \frac{1}{E-\epsilon p_-} & & & & & \\
 & & & \frac{1}{E+\epsilon p_+} & & & & \\
 & & & & \frac{1}{E-\epsilon p_+} & & & \\
 & & & & & \frac{1}{E+\epsilon p_-} & & \\
 & & & & & & \frac{1}{E-\epsilon p_-} & \\
 & & & & & & & \frac{1}{E+\epsilon p_+}
\end{pmatrix}
\end{equation}
with
\begin{equation}\label{eqJ208}
p_+\coloneqq p_1+p_2,\quad p_-\coloneqq p_1-p_2.
\end{equation}
We perform the Fourier transform on each of the elements in $G_0(p_1,p_2)$ to obtain the Green's function in real space.
While we can also find $\psi_0(x_1,x_2)$ as an eigenfunction of the kinetic term $H_0$ in Eq.~\eqref{eqJ205}, we do not specify it here because it will not appear in the final equations below.

Let us first 
investigate the $(1,1)$ element of $G_0(p_1,p_2)$:
\begin{align}
\int_{-\infty}^\infty\int_{-\infty}^\infty%
\frac{\ee^{\ii p_1x_1}\ee^{\ii p_2x_2}}{E-\epsilon(p_1+p_2)-\ii\eta}\frac{\dd p_1}{2\pi}\frac{\dd p_2}{2\pi} 
=&\int_{-\infty}^\infty\int_{-\infty}^\infty%
\frac{\ee^{\ii p_+x_+/2}\ee^{\ii p_-x_-/2}}{E-\epsilon p_+-\ii\eta}\frac{\dd p_+}{2\pi}\frac{\dd p_-}{2\pi}\frac{1}{2}
\notag \\
=&\frac{1}{4\pi}\int_{-\infty}^\infty\ee^{\ii p_-x_-/2}\dd p_-%
\times\int_{-\infty}^\infty\frac{\ee^{\ii p_+x_+/2\epsilon}}{E-\epsilon p_+-\ii\eta}\frac{\dd p_+}{2\pi},\label{eqJ209}
\end{align}
where we made the variable transformation $x_\pm\coloneqq x_1\pm x_2$. 
For the first integral on the right-hand side of Eq.~\eqref{eqJ209}, using the formula
\begin{equation}\label{eqJ210}
\delta(x)=\frac{1}{2\pi}\int_{-\infty}^\infty\ee^{\ii px}\dd p,
\end{equation}
we obtain
\begin{equation}\label{eqJ211}
\frac{1}{4\pi}\int_{-\infty}^\infty\ee^{\ii p_-x_-/2}\dd p_-%
=\frac{1}{4\pi}\int_{-\infty}^\infty\ee^{\ii p_-x_-/2}\dd (p_-/2)\frac{\dd p_-}{\dd (p_-/2)}%
=\delta(x_-),
\end{equation}
which means that the $(1,1)$ component has an amplitude only for $x_1=x_2$.
For the second integral, we take the contour $C_1$ for the complex integral as shown in Fig.~\ref{figappB3}(a) for the following reason.
The $(1,1)$ element of the Green's function is multiplied by $\psi_{\mathrm{LL\uparrow}}$, which means that the internal states of both walkers are left-going.
The condition that we now have, $x_1=x_2$, reads $x_+=x_1+x_2=2x_1$.
Since we now want the real-space Green's function $G_0(x_1, x_2)$ to have a finite value at $x_1<0$ and $x_2<0$, we take the contour $C_1$ for the complex integral as shown in Fig.~\ref{figappB3}(a).
Thus, we have
\begin{align}
\int_{-\infty}^\infty\frac{\ee^{\ii p_+x_+/2\epsilon}}{E-\epsilon p_+-\ii\eta}\frac{\dd p_+}{2\pi} 
=&\theta(-x_+)\qty[\oint_{C_1}\frac{\ee^{\ii zx_+/2}}{E-\epsilon z%
-\ii\eta}\frac{\dd z}{2\pi}-\lim_{R\rightarrow\infty}\qty(\int_{2\pi}^\pi\frac{\ee^{\ii R\ee^{\ii\phi}x_+/2}}{E-\epsilon R\ee^{\ii\phi}-\ii\eta}\frac{\dd\phi}{2\pi})] \notag \\
=&\theta(-x_+)\qty[-\frac{\ii}{2\pi\ii\epsilon}\oint_{C_1}\frac{\ee^{\ii zx_+/2}}{z-(E-\ii\eta)/\epsilon}\dd z-0]%
=\frac{\ii}{\epsilon}\ee^{\ii Ex_+/(2\epsilon)}\theta(-x_+).
\label{eqJ212}
\end{align}
Inserting Eqs.~\eqref{eqJ210} and~\eqref{eqJ212} to Eq.~\eqref{eqJ209} yields
\begin{equation}\label{eqJ213}
\int_{-\infty}^\infty\int_{-\infty}^\infty\frac{\ee^{\ii p_1x_1}\ee^{\ii p_2x_2}}{E-\epsilon(p_1+p_2)-\ii\eta}\frac{\dd p_1}{2\pi}\frac{\dd p_2}{2\pi}
=\frac{\ii}{\epsilon}\ee^{\ii Ex_+/(2\epsilon)}\delta(x_-)\theta(-x_+).
\end{equation}

Similarly, we perform the Fourier transform to each element of Eq.~\eqref{eqJ207}, following Table~\ref{appA_tab1}, and obtain the real-space Green's function as

\begin{align}
\tilde{G}_0(x_1,x_2)
=&\frac{\ii}{\epsilon}\delta(x_-)\theta(-x_+)\qty(\begin{array}{cccc|cccc}
1 & & & & & & & \\
 & & & & & & & \\
 & & & & & & & \\
 & & & & & & & \\ \hline
 & & & & 1 & & & \\
 & & & & & & & \\
 & & & & & & & \\
 & & & & & & &
\end{array})
+\frac{\ii}{\epsilon}\delta(x_+)\theta(+x_-)\qty(\begin{array}{cccc|cccc}
 & & & & & & & \\
 & 1 & & & & & & \\
 & & & & & & & \\
 & & & & & & & \\ \hline
 & & & & & & & \\
 & & & & & 1 & & \\
 & & & & & & & \\
 & & & & & & &
\end{array}) \notag \\
&+\frac{\ii}{\epsilon}\delta(x_+)\theta(-x_-)\qty(\begin{array}{cccc|cccc}
 & & & & & & & \\
 & & & & & & & \\
 & & 1 & & & & & \\
 & & & & & & & \\ \hline
 & & & & & & & \\
 & & & & & & & \\
 & & & & & & 1 & \\
 & & & & & & &
\end{array})
+\frac{\ii}{\epsilon}\delta(x_-)\theta(+x_+)\qty(\begin{array}{cccc|cccc}
 & & & & & & & \\
 & & & & & & & \\
 & & & & & & & \\
 & & & 1 & & & & \\ \hline
 & & & & & & & \\
 & & & & & & & \\
 & & & & & & & \\
 & & & & & & & 1
\end{array}),\label{eqJ215}
\end{align}
where we set the exponential factors $\exp[\pm\ii Ex_\pm/(2\epsilon)]$ to unity because we only focus on the case with $x_1,x_2=0\pm$.
Note here that each pair of $(1,1)$ and $(5,5)$, $(2,2)$ and $(6,6)$, $(3,3)$ and $(7,7)$, and $(4,4)$ and $(8,8)$ elements has the same value because only the internal degree of freedom of the localized magnetic impurity differs.

\begin{table}[h!]
\begin{center}
\caption{Internal states of the wave function and how to take contours for each integral}
\label{appA_tab1}
\begin{tabular}{ccc|ccc}
\hline
element & QW1 & QW2 & the other integral & finite value in the Green's func. for & contour \\
\hline\hline
$(1,1)$ & $\LL$ & $\LL$ & $\delta(x_-)$ & $x_1<0$\hspace{0.5em}and\hspace{0.5em}$x_2<0$\hspace{0.7em}\textit{i.e.}\hspace{0.7em}$x_+<0$ & $C_1$ \\
$(2,2)$ & $\RR$ & $\LL$ & $\delta(x_+)$ & $x_1>0$\hspace{0.5em}and\hspace{0.5em}$x_2<0$\hspace{0.7em}\textit{i.e.}\hspace{0.7em}$x_->0$ & $C_2$ \\
$(3,3)$ & $\LL$ & $\RR$ & $\delta(x_+)$ & $x_1<0$\hspace{0.5em}and\hspace{0.5em}$x_2>0$\hspace{0.7em}\textit{i.e.}\hspace{0.7em}$x_-<0$ & $C_1$ \\
$(4,4)$ & $\RR$ & $\RR$ & $\delta(x_-)$ & $x_1>0$\hspace{0.5em}and\hspace{0.5em}$x_2>0$\hspace{0.7em}\textit{i.e.}\hspace{0.7em}$x_+>0$ & $C_2$ \\
\hline
\end{tabular}
\end{center}
\end{table}

From Eqs.~\eqref{eqJ206} and~\eqref{eqJ215}, we obtain
\begin{align}
&G_0(x_1,x_2-y_2)V_1 \notag \\
=&\frac{\ii}{\epsilon}\fontsize{8pt}{8pt}\selectfont\qty(\begin{array}{llll | llll}
+J_z & & & & & +J_- & & \\
\delta(x_-+y_2) & & & & & \delta(x_-+y_2)  & & \\
\theta(-x_++y_2) & & & & & \theta(-x_++y_2) & & \\
& -J_z & & & +J_+ & & & \\
& \delta(x_+-y_2) & & & \delta(x_+-y_2) & & & \\
& \theta(x_-+y_2) & & & \theta(x_-+y_2) & & & \\
& & +J_z & & & & & +J_- \\
& & \delta(x_+-y_2) & & & & & \delta(x_+-y_2) \\
& & \theta(-x_--y_2) & & & & & \theta(-x_--y_2) \\
& & & -J_z & & & +J_+ & \\
& & & \delta(x_-+y_2) & & & \delta(x_-+y_2) & \\
& & & \theta(x_+-y_2) & & & \theta(x_+-y_2) & \\ \hline
& +J_+ & & & -J_z & & & \\
& \delta(x_-+y_2) & & & \delta(x_-+y_2)  & & & \\
& \theta(-x_++y_2) & & & \theta(-x_++y_2) & & & \\
+J_- & & & & & +J_z & & \\
\delta(x_+-y_2) & & & & & \delta(x_+-y_2) & & \\
\theta(x_-+y_2) & & & & & \theta(x_-+y_2) & & \\
& & & +J_+ & & & -J_z & \\
& & & \delta(x_+-y_2) & & & \delta(x_+-y_2) & \\
& & & \theta(-x_--y_2) & & & \theta(-x_--y_2) & \\
& & +J_- & & & & & +J_z \\
& & \delta(x_-+y_2) & & & & & \delta(x_-+y_2) \\
& & \theta(x_+-y_2) & & & & & \theta(x_+-y_2) 
\end{array}), \notag \\ \label{eqJ216} \\
&G_0(x_1-y_1,x_2)V_2 \notag \\
=&\frac{\ii}{\epsilon}\fontsize{8pt}{8pt}\selectfont\qty(\begin{array}{llll | llll}
+J_z & & & & & & +J_- & \\
\delta(x_--y_1) & & & & & & \delta(x_--y_1) & \\
\theta(-x_++y_1) & & & & & & \theta(-x_++y_1) & \\
& +J_z & & & & & & +J_- \\
& \delta(x_+-y_1) & & & & & & \delta(x_+-y_1) \\
& \theta(x_--y_1) & & & & & & \theta(x_--y_1) \\
& & -J_z & & +J_+ & & & \\
& & \delta(x_+-y_1) & & \delta(x_+-y_1) & & & \\
& & \theta(-x_-+y_1) & & \theta(-x_-+y_1) & & & \\
& & & -J_z & & +J_+ & & \\
& & & \delta(x_--y_1) & & \delta(x_--y_1) & & \\
& & & \theta(x_+-y_1) & & \theta(x_+-y_1) & & \\ \hline
& & +J_+ & & -J_z & & & \\
& & \delta(x_--y_1) & & \delta(x_--y_1) & & & \\
& & \theta(-x_++y_1) & & \theta(-x_++y_1) & & & \\
& & & +J_+ & & -J_z & & \\
& & & \delta(x_+-y_1) & & \delta(x_+-y_1) & & \\
& & & \theta(x_--y_1) & & \theta(x_--y_1) & & \\
+J_- & & & & & & +J_z & \\
\delta(x_+-y_1) & & & & & & \delta(x_+-y_1) & \\
\theta(-x_-+y_1) & & & & & & \theta(-x_-+y_1) & \\
& +J_- & & & & & & +J_z \\
& \delta(x_--y_1) & & & & & & \delta(x_--y_1) \\
& \theta(x_+-y_1) & & & & & & \theta(x_+-y_1) 
\end{array}),\notag \\ \label{eqJ217}
\end{align}
with 
$J_{\pm}\coloneqq J_x\pm J_y$.
Inserting Eqs.~\eqref{eqJ216} and~\eqref{eqJ217} into Eq.~\eqref{eqJ204} yields
\begin{align}
\psi(x_1,x_2)-\psi_0(x_1,x_2) 
=&\int_{-\infty}^{+\infty}\dd y_2G_0(x_1,x_2-y_2)V_1\psi(0,y_2)%
+\int_{-\infty}^{+\infty}\dd y_1G_0(x_1-y_1,x_2)V_2\psi(y_1,0)\notag \\
=&
\begin{pmatrix}
+J_z\theta_{1,-}\psi_{\mathrm{LL}\uparrow}(0,-x_-)%
+ J_-\theta_{1,-}\psi_{\mathrm{RL}\downarrow}(0,-x_-)  \\
-J_z\theta_{1,+}\psi_{\mathrm{RL}\uparrow}(0,+x_+)%
+J_+\theta_{1,+}\psi_{\mathrm{LL}\downarrow}(0,+x_+) \\
+J_z\theta_{1,-}\psi_{\mathrm{LR}\uparrow}(0,+x_+)%
+J_-\theta_{1,-}\psi_{\mathrm{RR}\downarrow}(0,+x_+) \\
-J_z\theta_{1,+}\psi_{\mathrm{RR}\uparrow}(0,-x_-)%
+J_+\theta_{1,+}\psi_{\mathrm{LR}\downarrow}(0,-x_-) \\
-J_z\theta_{1,-}\psi_{\mathrm{LL}\downarrow}(0,-x_-)%
+ J_+\theta_{1,-}\psi_{\mathrm{RL}\uparrow}(0,-x_-)  \\
+J_z\theta_{1,+}\psi_{\mathrm{RL}\downarrow}(0,+x_+)%
+J_-\theta_{1,+}\psi_{\mathrm{LL}\uparrow}(0,+x_+) \\
-J_z\theta_{1,-}\psi_{\mathrm{LR}\downarrow}(0,+x_+)%
+J_+\theta_{1,-}\psi_{\mathrm{RR}\uparrow}(0,+x_+) \\
+J_z\theta_{1,+}\psi_{\mathrm{RR}\downarrow}(0,-x_-)%
+J_-\theta_{1,+}\psi_{\mathrm{LR}\uparrow}(0,-x_-)
\end{pmatrix} \notag \\
&+
\begin{pmatrix}
+J_z\theta_{2,-}\psi_{\mathrm{LL}\uparrow}(+x_-,0)%
+J_-\theta_{2,-}\psi_{\mathrm{LR}\downarrow}(+x_-,0) \\
+J_z\theta_{2,-}\psi_{\mathrm{RL}\uparrow}(+x_+,0)%
+J_-\theta_{2,-}\psi_{\mathrm{RR}\downarrow}(+x_+,0) \\
-J_z\theta_{2,+}\psi_{\mathrm{LR}\uparrow}(+x_+,0)%
+J_+\theta_{2,+}\psi_{\mathrm{LL}\downarrow}(+x_+,0) \\
-J_z\theta_{2,+}\psi_{\mathrm{RR}\uparrow}(+x_-,0)%
+J_+\theta_{2,+}\psi_{\mathrm{RL}\downarrow}(+x_-,0) \\
-J_z\theta_{2,-}\psi_{\mathrm{LL}\downarrow}(+x_-,0)%
+J_+\theta_{2,-}\psi_{\mathrm{LR}\uparrow}(+x_-,0) \\
-J_z\theta_{2,-}\psi_{\mathrm{RL}\downarrow}(+x_+,0)%
+J_+\theta_{2,-}\psi_{\mathrm{RR}\uparrow}(+x_+,0) \\
+J_z\theta_{2,+}\psi_{\mathrm{LR}\downarrow}(+x_+,0)%
+J_-\theta_{2,+}\psi_{\mathrm{LL}\uparrow}(+x_+,0) \\
+J_z\theta_{2,+}\psi_{\mathrm{RR}\downarrow}(+x_-,0)%
+J_-\theta_{2,+}\psi_{\mathrm{RL}\uparrow}(+x_-,0)
\end{pmatrix},\label{eqJ219}
\end{align}
with
\begin{equation}\label{eqJ220}
\theta_{1,\pm}\coloneqq\theta(\pm x_1),\quad\theta_{2,\pm}\coloneqq\theta(\pm x_2).
\end{equation}

We now consider which elements of the wave function are relevant to actual scattering to obtain the $S$ matrix, as in Table~\ref{appA_tab2}.
\begin{table}[h!]
\begin{center}
\caption{Contribution of each element of the wave function to scattering.}
\label{appA_tab2}
\begin{tabular}{cccc|c}
\hline
QW1 & QW2 & $x_1$ & $x_2$ & contribution \\
\hline\hline
$\mathrm{L}$ & $\mathrm{L}$ & $0+$ & $0+$ & $\mathrm{\textcolor{blue}{in}^{*F}}$  \\
$\mathrm{L}$ & $\mathrm{L}$ & $0+$ & $0-$ & N/A \\
$\mathrm{L}$ & $\mathrm{L}$ & $0-$ & $0+$ & N/A \\
$\mathrm{L}$ & $\mathrm{L}$ & $0-$ & $0-$ & $\mathrm{\textcolor{red}{out}^{*F}}$ \\
\hline
$\mathrm{R}$ & $\mathrm{L}$ & $0+$ & $0+$ & N/A  \\
$\mathrm{R}$ & $\mathrm{L}$ & $0+$ & $0-$ & $\mathrm{\textcolor{red}{out}}$ \\
$\mathrm{R}$ & $\mathrm{L}$ & $0-$ & $0+$ & $\mathrm{\textcolor{blue}{in}}$ \\
$\mathrm{R}$ & $\mathrm{L}$ & $0-$ & $0-$ & N/A \\
\hline
$\mathrm{L}$ & $\mathrm{R}$ & $0+$ & $0+$ & N/A  \\
$\mathrm{L}$ & $\mathrm{R}$ & $0+$ & $0-$ & $\mathrm{\textcolor{blue}{in}}$ \\
$\mathrm{L}$ & $\mathrm{R}$ & $0-$ & $0+$ & $\mathrm{\textcolor{red}{out}}$ \\
$\mathrm{L}$ & $\mathrm{R}$ & $0-$ & $0-$ & N/A \\
\hline
$\mathrm{R}$ & $\mathrm{R}$ & $0+$ & $0+$ & $\mathrm{\textcolor{red}{out}^{*F}}$  \\
$\mathrm{R}$ & $\mathrm{R}$ & $0+$ & $0-$ & N/A \\
$\mathrm{R}$ & $\mathrm{R}$ & $0-$ & $0+$ & N/A \\
$\mathrm{R}$ & $\mathrm{R}$ & $0-$ & $0-$ & $\mathrm{\textcolor{blue}{in}^{*F}}$ \\
\hline
\end{tabular}
\end{center}
\end{table}
Here, the superscript $\mathrm{*F}$ means that the applicable terms do not contribute to scattering if we consider fermions due to Pauli's exclusion principle.
Using Eq.~\eqref{eqJ219} and writing down the relevant terms to scattering, we obtain the following:
\begin{subequations}\label{eqJ221}
\begin{align}
&\psi_{\mathrm{LL\uparrow}}(0-,0-)-(\psi_0)_{\mathrm{LL\uparrow}}(0-,0-) \notag \\
=&\frac{\ii}{\epsilon}[+J_z\psi_{\mathrm{LL\uparrow}}(0,-(0-)+(0-))%
                                   +J_-\psi_{\mathrm{RL\downarrow}}(0,-(0-)+(0-))\notag \\
          &\hspace{2mm}+J_z\psi_{\mathrm{LL\uparrow}}((0-)-(0-),0)%
                                   +J_-\psi_{\mathrm{LR\downarrow}}((0-)-(0-),0)] \notag \\
=&\frac{\ii}{\epsilon}[+J_z\psi_{\mathrm{LL\uparrow}}(0,0)%
                                   +J_-\psi_{\mathrm{RL\downarrow}}(0,0)
                                   +J_z\psi_{\mathrm{LL\uparrow}}(0,0)%
                                   +J_-\psi_{\mathrm{LR\downarrow}}(0,0)] \notag \\
=&\frac{\ii}{2\epsilon}\{+2J_z[\psi_{\mathrm{LL\uparrow}}(0+,0+)+\psi_{\mathrm{LL\uparrow}}(0-,0-)]\notag \\
          &\hspace{4mm}+J_-[\psi_{\mathrm{RL\downarrow}}(0+,0-)+\psi_{\mathrm{RL\downarrow}}(0-,0+)]%
                                   +J_-[\psi_{\mathrm{LR\downarrow}}(0+,0-)+\psi_{\mathrm{LR\downarrow}}(0-,0+)]\}, \label{eqJ221a} \\
&\psi_{\mathrm{LL\uparrow}}(0+,0+)-(\psi_0)_{\mathrm{LL\uparrow}}(0+,0+)=0, \label{eqJ221b}
\end{align}
\end{subequations}
\begin{subequations}\label{eqJ222}
\begin{align}
&\psi_{\mathrm{RL\uparrow}}(0+,0-)-(\psi_0)_{\mathrm{RL\uparrow}}(0+,0-) \notag \\
=&\frac{\ii}{\epsilon}[-J_z\psi_{\mathrm{RL\uparrow}}(0,(0+)+(0-))%
                                   +J_+\psi_{\mathrm{LL\downarrow}}(0,(0+)+(0-))\notag \\
          &\hspace{2mm}+J_z\psi_{\mathrm{RL\uparrow}}((0+)+(0-),0)%
                                   +J_-\psi_{\mathrm{RR\downarrow}}((0+)+(0-),0)] \notag \\
=&\frac{\ii}{\epsilon}[-J_z\psi_{\mathrm{RL\uparrow}}(0,0)%
                                   +J_+\psi_{\mathrm{LL\downarrow}}(0,0)
                                   +J_z\psi_{\mathrm{RL\uparrow}}(0,0)%
                                   +J_-\psi_{\mathrm{RR\downarrow}}(0,0)] \notag \\
=&\frac{\ii}{2\epsilon}\{J_+[\psi_{\mathrm{LL\downarrow}}(0+,0+)+\psi_{\mathrm{LL\downarrow}}(0-,0-)]%
                                    +J_-[\psi_{\mathrm{RR\downarrow}}(0+,0+)+\psi_{\mathrm{RR\downarrow}}(0-,0-)]\}, \label{eqJ222a} \\
&\psi_{\mathrm{RL\uparrow}}(0-,0+)-(\psi_0)_{\mathrm{RL\uparrow}}(0-,0+)=0, \label{eqJ222b}
\end{align}
\end{subequations}
\begin{subequations}\label{eqJ223}
\begin{align}
&\psi_{\mathrm{LR\uparrow}}(0-,0+)-(\psi_0)_{\mathrm{LR\uparrow}}(0-,0+) \notag \\
=&\frac{\ii}{\epsilon}[+J_z\psi_{\mathrm{LR\uparrow}}(0,(0-)+(0+))%
                                   +J_-\psi_{\mathrm{RR\downarrow}}(0,(0-)+(0+))\notag \\
          &\hspace{2mm}-J_z\psi_{\mathrm{LR\uparrow}}((0-)+(0+),0)%
                                   +J_+\psi_{\mathrm{LL\downarrow}}((0-)+(0+),0)] \notag \\
=&\frac{\ii}{\epsilon}[J_z\psi_{\mathrm{LR\uparrow}}(0,0)%
                                   +J_-\psi_{\mathrm{RR\downarrow}}(0,0)
                                   -J_z\psi_{\mathrm{LR\uparrow}}(0,0)%
                                   +J_+\psi_{\mathrm{LL\downarrow}}(0,0)] \notag \\
=&\frac{\ii}{2\epsilon}\{J_+[\psi_{\mathrm{LL\downarrow}}(0+,0+)+\psi_{\mathrm{LL\downarrow}}(0-,0-)]%
                                    +J_-[\psi_{\mathrm{RR\downarrow}}(0+,0+)+\psi_{\mathrm{RR\downarrow}}(0-,0-)]\}, \label{eqJ223a} \\
&\psi_{\mathrm{LR\uparrow}}(0+,0-)-(\psi_0)_{\mathrm{LR\uparrow}}(0+,0-)=0, \label{eqJ223b}
\end{align}
\end{subequations}
\begin{subequations}\label{eqJ224}
\begin{align}
&\psi_{\mathrm{RR\uparrow}}(0+,0+)-(\psi_0)_{\mathrm{RR\uparrow}}(0+,0+) \notag \\
=&\frac{\ii}{\epsilon}[-J_z\psi_{\mathrm{RR\uparrow}}(0,-(0+)+(0+))%
                                   +J_+\psi_{\mathrm{LR\downarrow}}(0,-(0+)+(0+))\notag \\
          &\hspace{2mm}-J_z\psi_{\mathrm{RR\uparrow}}((0+)-(0+),0)%
                                   +J_+\psi_{\mathrm{RL\downarrow}}((0+)-(0+),0)] \notag \\
=&\frac{\ii}{\epsilon}[-J_z\psi_{\mathrm{RR\uparrow}}(0,0)%
                                   +J_+\psi_{\mathrm{LR\downarrow}}(0,0)
                                   -J_z\psi_{\mathrm{RR\uparrow}}(0,0)%
                                   +J_+\psi_{\mathrm{RL\downarrow}}(0,0)] \notag \\
=&\frac{\ii}{2\epsilon}\{-2J_z[\psi_{\mathrm{RR\uparrow}}(0+,0+)+\psi_{\mathrm{RR\uparrow}}(0-,0-)]\notag \\
          &\hspace{4mm}+J_+[\psi_{\mathrm{RL\downarrow}}(0+,0-)+\psi_{\mathrm{RL\downarrow}}(0-,0+)]%
                                   +J_+[\psi_{\mathrm{LR\downarrow}}(0+,0-)+\psi_{\mathrm{LR\downarrow}}(0-,0+)]\}, \label{eqJ224a} \\
&\psi_{\mathrm{RR\uparrow}}(0-,0-)-(\psi_0)_{\mathrm{RR\uparrow}}(0-,0-)=0, \label{eqJ224b}
\end{align}
\end{subequations}
\begin{subequations}\label{eqJ225}
\begin{align}
&\psi_{\mathrm{LL\downarrow}}(0-,0-)-(\psi_0)_{\mathrm{LL\downarrow}}(0-,0-) \notag \\
=&\frac{\ii}{\epsilon}[-J_z\psi_{\mathrm{LL\downarrow}}(0,-(0-)+(0-))%
                                   +J_+\psi_{\mathrm{RL\uparrow}}(0,-(0-)+(0-))\notag \\
          &\hspace{2mm}-J_z\psi_{\mathrm{LL\downarrow}}((0-)-(0-),0)%
                                   +J_+\psi_{\mathrm{LR\uparrow}}((0-)-(0-),0)] \notag \\
=&\frac{\ii}{\epsilon}[-J_z\psi_{\mathrm{LL\downarrow}}(0,0)%
                                   +J_+\psi_{\mathrm{RL\uparrow}}(0,0)
                                   -J_z\psi_{\mathrm{LL\downarrow}}(0,0)%
                                   +J_+\psi_{\mathrm{LR\uparrow}}(0,0)] \notag \\
=&\frac{\ii}{2\epsilon}\{-2J_z[\psi_{\mathrm{LL\downarrow}}(0+,0+)+\psi_{\mathrm{LL\downarrow}}(0-,0-)]\notag \\
          &\hspace{4mm}+J_+[\psi_{\mathrm{RL\uparrow}}(0+,0-)+\psi_{\mathrm{RL\uparrow}}(0-,0+)]%
                                   +J_+[\psi_{\mathrm{LR\uparrow}}(0+,0-)+\psi_{\mathrm{LR\uparrow}}(0-,0+)]\}, \label{eqJ225a} \\
&\psi_{\mathrm{LL\downarrow}}(0+,0+)-(\psi_0)_{\mathrm{LL\downarrow}}(0+,0+)=0, \label{eqJ225b}
\end{align}
\end{subequations}
\begin{subequations}\label{eqJ226}
\begin{align}
&\psi_{\mathrm{RL\downarrow}}(0+,0-)-(\psi_0)_{\mathrm{RL\downarrow}}(0+,0-) \notag \\
=&\frac{\ii}{\epsilon}[+J_z\psi_{\mathrm{RL\downarrow}}(0,(0+)+(0-))%
                                   +J_-\psi_{\mathrm{LL\uparrow}}(0,(0+)+(0-))\notag \\
          &\hspace{2mm}-J_z\psi_{\mathrm{RL\downarrow}}((0+)+(0-),0)%
                                   +J_+\psi_{\mathrm{RR\uparrow}}((0+)+(0-),0)] \notag \\
=&\frac{\ii}{\epsilon}[+J_z\psi_{\mathrm{RL\downarrow}}(0,0)%
                                   +J_-\psi_{\mathrm{LL\uparrow}}(0,0)
                                   -J_z\psi_{\mathrm{RL\downarrow}}(0,0)%
                                   +J_+\psi_{\mathrm{RR\uparrow}}(0,0)] \notag \\
=&\frac{\ii}{2\epsilon}\{J_-[\psi_{\mathrm{LL\uparrow}}(0+,0+)+\psi_{\mathrm{LL\uparrow}}(0-,0-)]%
                                    +J_+[\psi_{\mathrm{RR\uparrow}}(0+,0+)+\psi_{\mathrm{RR\uparrow}}(0-,0-)]\}, \label{eqJ226a} \\
&\psi_{\mathrm{RL\downarrow}}(0-,0+)-(\psi_0)_{\mathrm{RL\downarrow}}(0-,0+)=0, \label{eqJ226b}
\end{align}
\end{subequations}
\begin{subequations}\label{eqJ227}
\begin{align}
&\psi_{\mathrm{LR\downarrow}}(0-,0+)-(\psi_0)_{\mathrm{LR\downarrow}}(0-,0+) \notag \\
=&\frac{\ii}{\epsilon}[-J_z\psi_{\mathrm{LR\downarrow}}(0,(0-)+(0+))%
                                   +J_+\psi_{\mathrm{RR\uparrow}}(0,(0-)+(0+))\notag \\
          &\hspace{2mm}+J_z\psi_{\mathrm{LR\downarrow}}((0-)+(0+),0)%
                                   +J_-\psi_{\mathrm{LL\uparrow}}((0-)+(0+),0)] \notag \\
=&\frac{\ii}{\epsilon}[-J_z\psi_{\mathrm{LR\downarrow}}(0,0)%
                                   +J_+\psi_{\mathrm{RR\uparrow}}(0,0)
                                   +J_z\psi_{\mathrm{LR\downarrow}}(0,0)%
                                   +J_-\psi_{\mathrm{LL\uparrow}}(0,0)] \notag \\
=&\frac{\ii}{2\epsilon}\{J_-[\psi_{\mathrm{LL\uparrow}}(0+,0+)+\psi_{\mathrm{LL\uparrow}}(0-,0-)]%
                                    +J_+[\psi_{\mathrm{RR\uparrow}}(0+,0+)+\psi_{\mathrm{RR\uparrow}}(0-,0-)]\}, \label{eqJ227a} \\
&\psi_{\mathrm{LR\downarrow}}(0+,0-)-(\psi_0)_{\mathrm{LR\downarrow}}(0+,0-)=0, \label{eqJ227b}
\end{align}
\end{subequations}
\begin{subequations}\label{eqJ228}
\begin{align}
&\psi_{\mathrm{RR\downarrow}}(0+,0+)-(\psi_0)_{\mathrm{RR\downarrow}}(0+,0+) \notag \\
=&\frac{\ii}{\epsilon}[+J_z\psi_{\mathrm{RR\downarrow}}(0,-(0+)+(0+))%
                                   +J_-\psi_{\mathrm{LR\uparrow}}(0,-(0+)+(0+))\notag \\
          &\hspace{2mm}+J_z\psi_{\mathrm{RR\downarrow}}((0+)-(0+),0)%
                                   +J_-\psi_{\mathrm{RL\uparrow}}((0+)-(0+),0)] \notag \\
=&\frac{\ii}{\epsilon}[+J_z\psi_{\mathrm{RR\downarrow}}(0,0)%
                                   +J_-\psi_{\mathrm{LR\uparrow}}(0,0)
                                   +J_z\psi_{\mathrm{RR\downarrow}}(0,0)%
                                   +J_-\psi_{\mathrm{RL\uparrow}}(0,0)] \notag \\
=&\frac{\ii}{2\epsilon}\{+2J_z[\psi_{\mathrm{RR\downarrow}}(0+,0+)+\psi_{\mathrm{RR\downarrow}}(0-,0-)]\notag \\
          &\hspace{4mm}+J_-[\psi_{\mathrm{RL\uparrow}}(0+,0-)+\psi_{\mathrm{RL\uparrow}}(0-,0+)]%
                                   +J_-[\psi_{\mathrm{LR\uparrow}}(0+,0-)+\psi_{\mathrm{LR\uparrow}}(0-,0+)]\}, \label{eqJ228a} \\
&\psi_{\mathrm{RR\downarrow}}(0-,0-)-(\psi_0)_{\mathrm{RR\downarrow}}(0-,0-)=0. \label{eqJ228b}
\end{align}
\end{subequations}
Here, we assumed
\begin{subequations}\label{eqJ229}
\begin{align}
&(0+)+(0-)=0,\quad(0\pm)-(0\pm)=0, \label{eqJ229a}\\
&\psi_{\mathrm{LL\uparrow,\downarrow}}(0,0)%
=\frac{1}{2}[\psi_{\mathrm{LL\uparrow,\downarrow}}(0+,0+)+\psi_{\mathrm{LL\uparrow,\downarrow}}(0-,0-)], \label{eqJ229b}\\
&\psi_{\mathrm{RR\uparrow,\downarrow}}(0,0)%
=\frac{1}{2}[\psi_{\mathrm{RR\uparrow,\downarrow}}(0+,0+)+\psi_{\mathrm{RR\uparrow,\downarrow}}(0-,0-)], \label{eqJ229c}\\
&\psi_{\mathrm{RL\uparrow,\downarrow}}(0,0)%
=\frac{1}{2}[\psi_{\mathrm{RL\uparrow,\downarrow}}(0+,0-)+\psi_{\mathrm{RL\uparrow,\downarrow}}(0-,0+)], \label{eqJ229d}\\
&\psi_{\mathrm{LR\uparrow,\downarrow}}(0,0)%
=\frac{1}{2}[\psi_{\mathrm{LR\uparrow,\downarrow}}(0+,0-)+\psi_{\mathrm{LR\uparrow,\downarrow}}(0-,0+)],
\label{eqJ229e}
\end{align}
\end{subequations}
and so on.
Based on Table~\ref{appA_tab2}, we obtain the $S$ matrix $S_{\mathrm{imp}}^{\mathrm{2w}}$ which is defined by
\begin{equation}\label{eqJ230}
\begin{pmatrix}
\psi_{\mathrm{LL\uparrow}}(0-,0-) \\
\psi_{\mathrm{RL\uparrow}}(0+,0-) \\
\psi_{\mathrm{LR\uparrow}}(0-,0+) \\
\psi_{\mathrm{RR\uparrow}}(0+,0+) \\
\psi_{\mathrm{LL\downarrow}}(0-,0-) \\
\psi_{\mathrm{RL\downarrow}}(0+,0-) \\
\psi_{\mathrm{LR\downarrow}}(0-,0+) \\
\psi_{\mathrm{RR\downarrow}}(0+,0+)
\end{pmatrix}
=S_{\mathrm{imp}}^{\mathrm{2w}}
\begin{pmatrix}
\psi_{\mathrm{LL\uparrow}}(0+,0+) \\
\psi_{\mathrm{RL\uparrow}}(0-,0+) \\
\psi_{\mathrm{LR\uparrow}}(0+,0-) \\
\psi_{\mathrm{RR\uparrow}}(0-,0-) \\
\psi_{\mathrm{LL\downarrow}}(0+,0+) \\
\psi_{\mathrm{RL\downarrow}}(0-,0+) \\
\psi_{\mathrm{LR\downarrow}}(0+,0-) \\
\psi_{\mathrm{RR\downarrow}}(0-,0-)
\end{pmatrix}.
\end{equation}
Subtracting the bottom equation from the top equation in each pair of Eqs.~\eqref{eqJ221a}--\eqref{eqJ228b}, we eliminate
all terms of the eigenfunction $\psi_0$ of the kinetic term $H_0$, and obtain the following:
\begin{subequations}\label{eqJ231}
\begin{align}
&\ii\epsilon[\psi_{\mathrm{LL\uparrow}}(0-,0-)-\psi_{\mathrm{LL\uparrow}}(0+,0+)] \notag \\
&\qquad+\frac{1}{2}\left\{+2J_z\qty[\psi_{\mathrm{LL\uparrow}}(0+,0+)+\psi_{\mathrm{LL\uparrow}}(0-,0-)]\right.\notag \\
   &\hspace{1.7cm}\left.+J_-\qty[\psi_{\mathrm{RL\downarrow}}(0+,0-)+\psi_{\mathrm{RL\downarrow}}(0-,0+)]%
                                     +J_-\qty[\psi_{\mathrm{LR\downarrow}}(0+,0-)+\psi_{\mathrm{LR\downarrow}}(0-,0+)]\right\}=0, \label{eqJ231a}\\
&\ii\epsilon[\psi_{\mathrm{RL\uparrow}}(0+,0-)-\psi_{\mathrm{RL\uparrow}}(0-,0+)] \notag \\
&\qquad+\frac{1}{2}\left\{J_+\qty[\psi_{\mathrm{LL\downarrow}}(0+,0+)+\psi_{\mathrm{LL\downarrow}}(0-,0-)]%
                                   +J_-\qty[\psi_{\mathrm{RR\downarrow}}(0+,0+)+\psi_{\mathrm{RR\downarrow}}(0-,0-)]\right\}=0, \label{eqJ231b}\\
&\ii\epsilon[\psi_{\mathrm{LR\uparrow}}(0-,0+)-\psi_{\mathrm{LR\uparrow}}(0+,0-)] \notag \\
&\qquad+\frac{1}{2}\left\{J_+\qty[\psi_{\mathrm{LL\downarrow}}(0+,0+)+\psi_{\mathrm{LL\downarrow}}(0-,0-)]%
                                   +J_-\qty[\psi_{\mathrm{RR\downarrow}}(0+,0+)+\psi_{\mathrm{RR\downarrow}}(0-,0-)]\right\}=0, \label{eqJ231c}\\
&\ii\epsilon[\psi_{\mathrm{RR\uparrow}}(0+,0+)-\psi_{\mathrm{RR\uparrow}}(0-,0-)] \notag \\
&\qquad+\frac{1}{2}\left\{-2J_z\qty[\psi_{\mathrm{RR\uparrow}}(0+,0+)+\psi_{\mathrm{RR\uparrow}}(0-,0-)]\right.\notag \\
   &\hspace{1.7cm}\left.+J_+\qty[\psi_{\mathrm{RL\downarrow}}(0+,0-)+\psi_{\mathrm{RL\downarrow}}(0-,0+)]%
                                     +J_+\qty[\psi_{\mathrm{LR\downarrow}}(0+,0-)+\psi_{\mathrm{LR\downarrow}}(0-,0+)]\right\}=0, \label{eqJ231d}\\
&\ii\epsilon[\psi_{\mathrm{LL\downarrow}}(0-,0-)-\psi_{\mathrm{LL\downarrow}}(0+,0+)] \notag \\
&\qquad+\frac{1}{2}\left\{-2J_z\qty[\psi_{\mathrm{LL\downarrow}}(0+,0+)+\psi_{\mathrm{LL\downarrow}}(0-,0-)]\right.\notag \\
   &\hspace{1.7cm}\left.+J_+\qty[\psi_{\mathrm{RL\uparrow}}(0+,0-)+\psi_{\mathrm{RL\uparrow}}(0-,0+)]%
                                     +J_+\qty[\psi_{\mathrm{LR\uparrow}}(0+,0-)+\psi_{\mathrm{LR\uparrow}}(0-,0+)]\right\}=0, \label{eqJ231e}\\
&\ii\epsilon[\psi_{\mathrm{RL\downarrow}}(0+,0-)-\psi_{\mathrm{RL\downarrow}}(0-,0+)] \notag \\
&\qquad+\frac{1}{2}\left\{J_-\qty[\psi_{\mathrm{LL\uparrow}}(0+,0+)+\psi_{\mathrm{LL\uparrow}}(0-,0-)]%
                                   +J_+\qty[\psi_{\mathrm{RR\uparrow}}(0+,0+)+\psi_{\mathrm{RR\uparrow}}(0-,0-)]\right\}=0, \label{eqJ231f}\\
&\ii\epsilon[\psi_{\mathrm{LR\downarrow}}(0-,0+)-\psi_{\mathrm{LR\downarrow}}(0+,0-)] \notag \\
&\qquad+\frac{1}{2}\left\{J_-\qty[\psi_{\mathrm{LL\uparrow}}(0+,0+)+\psi_{\mathrm{LL\uparrow}}(0-,0-)]%
                                   +J_+\qty[\psi_{\mathrm{RR\uparrow}}(0+,0+)+\psi_{\mathrm{RR\uparrow}}(0-,0-)]\right\}=0, \label{eqJ231g}\\
&\ii\epsilon[\psi_{\mathrm{RR\downarrow}}(0+,0+)-\psi_{\mathrm{RR\downarrow}}(0-,0-)] \notag \\
&\qquad+\frac{1}{2}\left\{+2J_z\qty[\psi_{\mathrm{RR\downarrow}}(0+,0+)+\psi_{\mathrm{RR\downarrow}}(0-,0-)]\right.\notag \\
   &\hspace{1.7cm}\left.+J_-\qty[\psi_{\mathrm{RL\uparrow}}(0+,0-)+\psi_{\mathrm{RL\uparrow}}(0-,0+)]%
                                     +J_-\qty[\psi_{\mathrm{LR\uparrow}}(0+,0-)+\psi_{\mathrm{LR\uparrow}}(0-,0+)]\right\}=0, \label{eqJ231h}
\end{align}
\end{subequations}
The above equations are cast into the matrix form
\begin{equation}
\qty(\ii\epsilon\mathbb{I}_{8\times8}+\frac{\tilde{M}}{2})
\begin{pmatrix}
\psi_{\mathrm{LL\uparrow}}(0-,0-) \\
\psi_{\mathrm{RL\uparrow}}(0+,0-) \\
\psi_{\mathrm{LR\uparrow}}(0-,0+) \\
\psi_{\mathrm{RR\uparrow}}(0+,0+) \\
\psi_{\mathrm{LL\downarrow}}(0-,0-) \\
\psi_{\mathrm{RL\downarrow}}(0+,0-) \\
\psi_{\mathrm{LR\downarrow}}(0-,0+) \\
\psi_{\mathrm{RR\downarrow}}(0+,0+)
\end{pmatrix}
=\qty(\ii\epsilon\mathbb{I}_{8\times8}-\frac{\tilde{M}}{2})
\begin{pmatrix}
\psi_{\mathrm{LL\uparrow}}(0+,0+) \\
\psi_{\mathrm{RL\uparrow}}(0-,0+) \\
\psi_{\mathrm{LR\uparrow}}(0+,0-) \\
\psi_{\mathrm{RR\uparrow}}(0-,0-) \\
\psi_{\mathrm{LL\downarrow}}(0+,0+) \\
\psi_{\mathrm{RL\downarrow}}(0-,0+) \\
\psi_{\mathrm{LR\downarrow}}(0+,0-) \\
\psi_{\mathrm{RR\downarrow}}(0-,0-)
\end{pmatrix}, \label{eqJ232}
\end{equation}
with
\begin{equation}
\tilde{M}\coloneqq \qty(\begin{array}{cccc|cccc}
+2J_z & & & & & J_- & J_- & \\
& & & & J_+ & & & J_- \\
& & & & J_+ & & & J_- \\
& & & -2J_z & & J_+ & J_+ & \\ \hline
& J_+ & J_+ & & -2J_z & & & \\
J_- & & & J_+ & & & & \\
J_- & & & J_+ & & & & \\
& J_- & J_- & & & & & +2J_z
\end{array}).\label{eqJ233}
\end{equation}
Thus, we obtain the $S$ matrix for the two quantum walkers with an impurity as
\begin{align}
S_{\mathrm{imp}}^{\mathrm{2w}}%
&=\qty(\ii\epsilon\mathbb{I}_{8\times8}+\frac{\tilde{M}}{2})^{-1}\qty(\ii\epsilon\mathbb{I}_{8\times8}-\frac{\tilde{M}}{2})\notag \\
&=\qty(\begin{array}{cccc|cccc}
\alpha_+^{\mathrm{2w}} & & & \beta^{\mathrm{2w}} & & \epsilon_-^{\mathrm{2w}} & \epsilon_-^{\mathrm{2w}} & \\
& \gamma^{\mathrm{2w}} & \delta^{\mathrm{2w}} & & \epsilon_+^{\mathrm{2w}} & & & \epsilon_-^{\mathrm{2w}} \\
& \delta^{\mathrm{2w}} & \gamma^{\mathrm{2w}} & & \epsilon_+^{\mathrm{2w}} & & & \epsilon_-^{\mathrm{2w}} \\
\beta^{\mathrm{2w}} & & & \alpha_-^{\mathrm{2w}} & & \epsilon_+^{\mathrm{2w}} & \epsilon_+^{\mathrm{2w}} & \\ \hline
& \epsilon_+^{\mathrm{2w}} & \epsilon_+^{\mathrm{2w}} & & \alpha_-^{\mathrm{2w}} & & & \beta^{\mathrm{2w}} \\
\epsilon_-^{\mathrm{2w}} & & & \epsilon_+^{\mathrm{2w}} & & \gamma^{\mathrm{2w}} & \delta^{\mathrm{2w}} & \\
\epsilon_-^{\mathrm{2w}} & & & \epsilon_+^{\mathrm{2w}} & & \delta^{\mathrm{2w}} & \gamma^{\mathrm{2w}} & \\
& \epsilon_-^{\mathrm{2w}} & \epsilon_- ^{\mathrm{2w}}& & \beta^{\mathrm{2w}} & & & \alpha_+^{\mathrm{2w}}
\end{array})\label{eqJ234}
\end{align}
with
\begin{subequations}\label{eqJ235}
\begin{align}
&\alpha_\pm^{\mathrm{2w}}=\frac{2\ii J_xJ_yJ_z-{J_z}^2\epsilon+\epsilon^3\pm(2\ii J_z\epsilon^2+2J_xJ_y\epsilon)}%
                             {-2\ii J_xJ_yJ_z+({J_x}^2+{J_y}^2+{J_z}^2)\epsilon+\epsilon^3}, \label{eqJ235a}\\
&\beta^{\mathrm{2w}}=\frac{(-{J_x}^2+{J_y}^2)\epsilon}{-2\ii J_xJ_yJ_z+({J_x}^2+{J_y}^2+{J_z}^2)\epsilon+\epsilon^3}, \label{eqJ235b}\\
&\gamma^{\mathrm{2w}}=\frac{{J_z}^2\epsilon+\epsilon^3}{-2\ii J_xJ_yJ_z+({J_x}^2+{J_y}^2+{J_z}^2)\epsilon+\epsilon^3}, \label{eqJ235c}\\
&\delta^{\mathrm{2w}}=\frac{2\ii J_xJ_yJ_z-({J_x}^2+{J_y}^2)\epsilon}{-2\ii J_xJ_yJ_z+({J_x}^2+{J_y}^2+{J_z}^2)\epsilon+\epsilon^3}, \label{eqJ235d}\\
&\epsilon_\pm^{\mathrm{2w}}=\pm\frac{(J_x\pm J_y)(J_z\pm\ii\epsilon)\epsilon}{-2\ii J_xJ_yJ_z+({J_x}^2+{J_y}^2+{J_z}^2)\epsilon+\epsilon^3}.\label{eqJ235e}
\end{align}
\end{subequations}
This is used in Eq.~\eqref{eq4105} in Sec.~\ref{sec4}. 

\bibliographystyle{elsarticle-num}
\bibliography{yamagishi}

\end{document}